\documentclass[runningheads,envcountsame,orivec]{llncs}
\usepackage[headings]{fullpage}
\usepackage[T1]{fontenc}
\usepackage{lmodern}
\usepackage{amsmath,amssymb}
\usepackage{booktabs,tabularx}
\usepackage{microtype}
\usepackage[hidelinks]{hyperref}
\hypersetup{
  unicode=true,
  pdftitle={Higher Realizability Toposes},
  pdfauthor={Yoshihiro Maruyama},
  pdfkeywords={Realizability \textinfty-topos, Higher tripos,
    Function realizability, Modified realizability, Krivine realizability,
    Lawvere\textendash{}Tierney topology, Contextual choice, Descriptive set theory,
    Univalent universe}
}
\DeclareMathOperator{\Ex}{Ex}
\DeclareMathOperator{\Sh}{Sh}
\DeclareMathOperator{\Asm}{Asm}
\DeclareMathOperator{\RT}{RT}
\DeclareMathOperator{\Map}{Map}
\DeclareMathOperator{\Sub}{Sub}
\DeclareMathOperator{\Loc}{Loc}
\DeclareMathOperator{\isContr}{isContr}
\DeclareMathOperator{\Ab}{Ab}
\DeclareMathOperator{\Hom}{Hom}
\newcommand{\E}{\mathcal E}
\newcommand{\C}{\mathcal C}
\newcommand{\Sp}{\mathcal S}
\newcommand{\HH}{\mathsf H}
\newcommand{\one}{\mathbf 1}
\newcommand{\zero}{\mathbf 0}
\newcommand{\cover}{\twoheadrightarrow}
\newcommand{\into}{\hookrightarrow}
\allowdisplaybreaks[1]

\begin{document}
\title{Higher Realizability Toposes}
\titlerunning{Higher Realizability Toposes}
\author{Yoshihiro Maruyama}
\authorrunning{Y. Maruyama}
\institute{Kyoto University\\
\href{mailto:maruyama@i.h.kyoto-u.ac.jp}{\email{maruyama@i.h.kyoto-u.ac.jp}}}
\maketitle

\begin{abstract}
Building on $\infty$-tripos theory, we develop a theory of 
realizability $\infty$-toposes. We apply $\infty$-tripos theory
to construct the function, modified, and Krivine realizability $\infty$-toposes. 
We recover the Lawvere--Tierney nucleus $j$ that specifies local truth
in each construction by exhibiting, for any $n\geq1$, a single family of
Eilenberg--MacLane $n$-types whose fibre at an ambient truth value $p$
is contractible after sheafification precisely when $j(p)$ holds.
In function realizability, we relate identity types to descriptive set theory.
For each $n\geq 2$, we construct a univalent classifier $U_n$ of exact
truncation height $n+1$. As maps from Baire space into $U_n$ vary, the
sets of inputs admitting a global path from the assigned code to the
terminal code are exactly the $G_\delta$ subsets of Baire space.
After exponentiation by Baire space, the analogous test against the
constant map at the terminal code yields exactly the coanalytic sets,
at the same truncation height.
In modified realizability, a closed proof of local existence under a
stable numerical precondition yields a total recursive function together
with a coherent proof that its output satisfies the specification
whenever the precondition holds. 
In Krivine realizability, contextual choice for $(n-1)$-truncated families 
over suitable $n$-types recovers the subterminal defining the Boolean
localization. Binary cohomology distinguishes slices of 
Anel--Awodey--Barton's effective $\infty$-topos 
even when they share the same ordinary Hyland effective 
topos and their bases have equivalent spaces of global points.
Hyland's effective topos thus occurs as the category of zero-types
in infinitely many pairwise inequivalent realizability $\infty$-toposes 
(recovering Hyland's effective topos as the category of zero-types 
does not uniquely determine an effective $\infty$-topos).
\keywords{Realizability $\infty$-topos \and Higher tripos \and
Function realizability \and Modified realizability \and Krivine realizability \and
Lawvere--Tierney topology \and Contextual choice \and Descriptive set theory \and
Univalent universe}
\end{abstract}

\section{Introduction}

We develop a theory of realizability $\infty$-toposes building on
$\infty$-tripos theory~\cite{Maruyama2026}. In an ordinary realizability
tripos, an entailment is witnessed by a realizer that takes witnesses of
the premise to witnesses of the conclusion, uniformly in the parameters.
The tripos-to-topos construction gives this logic a category of sets
and maps~\cite{HylandJohnstonePitts1980}. The higher tripos theory 
admits type-valued families and their spaces of sections. In a univalent
universe, a path between two codes witnesses an equivalence between the
types they name. When can a realizer supply such a path uniformly as
the parameter varies? More generally, when can a realizer choose a
section of a family of higher types?

To study these questions, we construct the function, modified, and
Krivine realizability \(\infty\)-toposes. Each construction starts
from assemblies, whose maps are tracked by realizers.
Higher exact completion adjoins effective
quotients of Segal groupoids~\cite{Stefanich2023Exact}. If $Q$ is the quotient of $R_\bullet$,
then $R_1\simeq R_0\times_Q R_0$, so the arrows of the groupoid
describe paths between the images of its objects in $Q$. The
$\infty$-tripos structure supplies dependent products and univalent
universes for the resulting families~\cite{Maruyama2026}.

Two more choices determine the interpretation of a family. A
Lawvere--Tierney nucleus $j$ specifies the logic on propositions, and
a type $B$ specifies the context over which the family varies. For an
admissible assembly presentation $\mathcal C$ in the sense of
Definition~\ref{def:admissible}, we form
\begin{equation}
 \mathfrak R(\mathcal C,j;B)
   :=\operatorname{Sh}_j(\operatorname{Ex}_\infty\mathcal C)_{/B},
 \qquad B\in\operatorname{Sh}_j(\operatorname{Ex}_\infty\mathcal C).
 \label{eq:common-form}
\end{equation}
Write $\mathcal T=\operatorname{Ex}_{1,\mathrm{reg}}(\mathcal C)$ for
the ordinary realizability topos and $\mathsf H(\mathcal T)$ for its
bounded higher exact completion. We extend $j$ to a sheafification
one truncation degree at a time and prove that it preserves finite
limits. This uses the singleton construction~\cite{QuirinTabareau2016}
and the path criterion for left exactness~\cite{RijkeShulmanSpitters2020}.
For our models, we obtain
\begin{equation}
 \operatorname{Ex}_\infty\mathcal C\simeq\mathsf H(\mathcal T),
 \qquad
 \operatorname{Sh}_j\mathsf H(\mathcal T)
       \simeq\mathsf H(\operatorname{Sh}_j\mathcal T).
 \label{eq:two-comparisons}
\end{equation}
Thus the ordinary topos, the nucleus, and the chosen higher context $B$ 
determine the higher model up to equivalence (Theorem~\ref{thm:unified}).

How much of the action of sheafification on higher types suffices to
recover the logic? Fix $n\geq1$ and a predicate $P$. Following Diaconescu's
idea of encoding a proposition by identifying two elements~\cite{Diaconescu1975},
we form the quotient $H_P$ of the cyclic group $C_2$ of order two by the relation
$(a=b)\vee P$. The two elements become equal precisely where $P$ holds. Consequently its
Eilenberg--MacLane type $K(H_P,n)$ is contractible precisely there,
and after fibrewise sheafification it is contractible on exactly the
$j$-closure of $P$. Taking the generic predicate on the truth-value
object $\Omega$ gives one family $W_n\to\Omega$ that works for every
nucleus. The locus of fibres made contractible by sheafification has
characteristic map $j:\Omega\to\Omega$
(Corollary~\ref{cor:one-generic-probe}).

\paragraph{Function realizability.}
In Kleene's second algebra, a realizer reads a stream in Baire space
$\mathbb B$. We use the entries of a stream to specify the transition
maps in a sequence of two-element groups. A zero entry leaves the
transition equal to the identity, and a nonzero entry sets it to zero.
The direct limit group is trivial exactly when the stream has
infinitely many nonzero entries. For $n\geq2$, the Eilenberg--MacLane
types of these groups form a family with a univalent classifier $U_n$.
Let $u_*$ be its code for the terminal type. As $f:\mathbb B\to U_n$
varies, the sets of inputs $x$ at which a global path joins $f(x)$ to
$u_*$ are exactly the $G_\delta$ subsets of Baire space.

A map into $U_n^{\mathbb B}$ introduces a second stream. For this
family, a global path to the constant map at $u_*$ exists exactly when
the corresponding groups are trivial for every value of the second
stream. Their zero sections then give a uniform trivialization. This
universal quantification gives exactly the coanalytic sets as equality
loci. For example, one such path exists precisely when the tree coded
by the input has no infinite branch. Both classifiers have exact
truncation height $n+1$ (Theorem~\ref{thm:fn-recognition}).

\paragraph{Modified realizability.}
We obtain the modified model by lifting the closed localization of
nested realizability~\cite{VanOosten1997}. Here the question is what
program can be extracted from a proof that witnesses exist locally.
The natural numbers are projective in the modified model
(Section~\ref{sec:modified-semantics}), so they admit contextual
choice by Corollary~\ref{cor:projective-choice}. Thus, for a bounded
higher family over the natural numbers, pointwise local existence
gives local existence of a section.
The natural numbers also support dependent induction in every parameter
context, where the zero and successor data determine a contractible
space of sections equipped with the prescribed computation paths
(Theorem~\ref{thm:mod-induction}).

Let $P(k)$ be a stable numerical precondition, meaning that
$\neg\neg P(k)$ implies $P(k)$, and let $C(k,m)$ be a bounded family
at an admitted universe level. Suppose a closed proof establishes
that for every $k$ satisfying $P(k)$ there locally exists an $m$ with
$C(k,m)$. We extract one program $e$ that halts on every input $k$.
Whenever $P(k)$ holds, the output comes with a certificate of
$C(k,\varphi_e(k))$ (Theorem~\ref{thm:mod-extraction}).

\paragraph{Krivine realizability.}
Krivine interprets classical proofs by programs~\cite{Krivine2009}.
In the relative Boolean presentation of this semantics, a
subterminal $U$ of the ambient relative topos determines the nucleus
$b_U(p)=((p\Rightarrow U)\Rightarrow U)$~\cite{VanOostenZou2016}.
We recover this logical pole by asking when local witnesses can be
chosen coherently. Fix $n\geq1$. For each ambient truth value $V$ we
construct a pointed $n$-type in the Boolean localization. On this
index, choice for $k$-truncated families means that every pointwise
locally inhabited such family locally admits a section. For
$-1\leq k\leq n-2$ the test holds for every $V$. At $k=n-1$ the
paths from the basepoint form a test family, and choosing a path to
each point contracts the index. At a universe level containing this
family, choice holds precisely for the truth values $V\leq U$.
Their greatest element is the pole (Theorem~\ref{thm:kr-choice-threshold}).
At a single universe code, the loop assigned to a standard program
index $e$ admits a path to the constant loop precisely when
$\varphi_e$ is total. Theorem~\ref{thm:kr-second-identity} establishes
this for every proper pole in the numerical presentations it covers.

\paragraph{Effective realizability.}
We now fix the logic and vary the context in the effective
$\infty$-topos of Anel, Awodey and Barton~\cite{AAB2025}. With matching
size bounds, applying Stefanich's recognition
theorem~\cite{Stefanich2023Exact} gives an equivalence
of exact $\infty$-categories between this model and $\mathsf H(\mathsf{Eff})$,
where $\mathsf{Eff}$ is Hyland's ordinary effective topos. To see what a
higher context can change, take $p:K(A,n)\to1$ in a localized
completion $\mathcal F$, with $A$ an abelian zero-type and $n\geq2$.
Pullback along $p$ gives an equivalence on $(n-2)$-truncated
objects, so in particular the slice has the same ordinary topos.
If $A$ is nonzero, $n$ is the least $m$ for which this slice is the
higher exact completion of a regular $(m,1)$-category
(Theorem~\ref{thm:general-context-dimension}). For every abelian
zero-type $M$, its degree-$n$ cohomology splits as
\[
 H^n(\mathcal F_{/K(A,n)};p^*M)
 \cong H^n(\mathcal F;M)\oplus\Hom_{\Ab(\mathcal F_{\leq0})}(A,M).
\]
Restriction to the basepoint projects onto the first summand. The
additional classes are therefore indexed by homomorphisms in the
ordinary topos (Theorem~\ref{thm:relative-context-cohomology}).

In the effective model, write $D$ for the two-element group with
numerical names for its bits, and give every element of the group
$V=\bigoplus_{\mathbb N}\mathbb F_2$ every number as a name in the
assembly $\nabla V$. For $r\geq0$ put $A_r=D^r\times\nabla V$.
A tracked homomorphism from $A_r$ to $D$ vanishes on $\nabla V$,
since a tracker presented with a common name must return the same
bit. Its restriction to $D^r$ can be any binary linear functional.
Consequently, the slice over $K(A_r,n)$ has binary cohomology
$\mathbb F_2^r$ in degree $n$. The analogous construction in sheaves of
spaces gives Grothendieck slices of every finite localic rank $n\geq2$
with the same ordinary sheaf topos (Theorem~\ref{thm:grothendieck-rank}).
The localic rank is the least $q$ for which the topos has a site of
definition on a small $q$-category with finite limits.

Fix $n\geq2$. The effective slices over $K(A_r,n)$ share Hyland's
effective topos as their ordinary core, and their bases all have global
point space equivalent to $K(V,n)$. The rank of degree-$n$ binary
cohomology recovers $r$ from the slice itself
(Proposition~\ref{prop:separation-recovered}). As $r$ ranges over the
nonnegative integers, this gives infinitely many pairwise inequivalent
realizability $\infty$-toposes.

\subsection{Relation with prior work}
\label{sec:related-work}

This work builds upon Stefanich's recognition theorem for bounded exact
categories~\cite{Stefanich2023Exact} and higher tripos
theory~\cite{Maruyama2026} to show that for admissible presentations, 
sheafification commutes with bounded higher exact completion. The
ordinary topos, the nucleus and the chosen higher context thus determine the
resulting slice. This gives a common construction of the function,
modified and Krivine realizability $\infty$-toposes
(Theorem~\ref{thm:unified}).

Higher sheaves give the left-exact localization generated by a set of
maps in a Grothendieck $\infty$-topos~\cite{AnelEtAl2022}. In this setting,
topological localizations correspond to extended Grothendieck
topologies~\cite{AnelEtAl2024}. Quirin and
Tabareau construct sheafification from a nucleus $j$ one truncation
degree at a time~\cite{QuirinTabareau2016}. We recover $j$ from its
action on a single family. Delooping the quotient group $H_P$ at the generic predicate gives,
for each $n\geq1$, a family $W_n\to\Omega$ independent of $j$. After
fibrewise sheafification, the locus of its contractible fibres has
characteristic map $j$ itself (Corollary~\ref{cor:one-generic-probe}).
The order on local operators studied by Lee and van Oosten
\cite{LeeVanOosten2013} and the interpretation of Turing reducibility
by oracle modalities~\cite{Swan2026Oracle} suggest comparing which
problems each localization makes solvable. For $n\geq2$, we show that $j\leq k$ exactly when, in every
parameter context, each basepoint torsor obtained from $W_n$ that
acquires a section under $j$ also acquires one under $k$ (Theorem~\ref{thm:torsor-order}).

Dubey and Buchholtz proposed a simplicial model of function realizability
over $\mathcal K_2$~\cite{DubeyBuchholtz2022}. For $n\geq2$, write
$c_n(\alpha)$ for the code of the fibre at a stream $\alpha$ in our
bounded higher completion of the ordinary function realizability
topos $\RT(\mathcal K_2)$, and $v_n$ for the
code of $K(C_2,n)$. Over the eventually zero streams, the support of
the identity $c_n(\alpha)=v_n$ is the internal proposition that the
nonzero positions of $\alpha$ have a bound
(Lemma~\ref{lem:fn-support}). This connects code identity to finite-set
bounding~\cite{PaulyFoucheDavie2018}.
For the terminal code $u_*$, the sets of inputs admitting a global identity
witness range over exactly the $G_\delta$ subsets of Baire space.
Exponentiating the classifier by Baire space gives exactly the
coanalytic sets at the same truncation height. In particular, identity
with the constant terminal code recognizes well-foundedness of trees
(Theorem~\ref{thm:fn-recognition}).

Awodey, Gambino and Sojakova relate induction to homotopy-initiality of
algebras~\cite{AwodeyGambinoSojakova2017}. In modified realizability we
prove that the ordinary natural numbers form a homotopy-initial successor
algebra in every parameter context, giving induction with coherent
computation paths (Theorem~\ref{thm:mod-induction}). From a closed proof
of local existence under a stable numerical premise, we extract a program
total on all inputs with a coherent certificate wherever the premise holds
(Theorem~\ref{thm:mod-extraction}).

In Krivine realizability, contextual choice detects the subterminal $U$
that defines the relative Booleanization of van Oosten and
Zou~\cite{VanOostenZou2016}. Choice at an index is a section problem,
equivalent to internal projectivity~\cite{HoTTProjective}. For each
$n\geq1$ and each ambient truth value $V$, we construct an $n$-type
index. At any test level containing its based
path family, choice for $(n-1)$-truncated families on that index holds
precisely when $V\leq U$
(Theorem~\ref{thm:kr-choice-threshold}).

How much higher structure can vary over one ordinary topos?
Hyland's effective topos arises by exact completion of partitioned
assemblies~\cite{Hyland1982,RobinsonRosolini1990}. Van den Berg realizes
it as the homotopy category of a path category~\cite{VanDenBerg2020}.
Awodey and Emmenegger identify Hyland's effective topos as the category
of zero-types in their proposed effective $2$-topos, and Anel, Awodey
and Barton give the same identification in their higher effective topos~\cite{AwodeyEmmenegger2025Effective2,AAB2025}. Speight models
$1$-truncated intensional type theory in partitioned groupoidal
assemblies~\cite{Speight2024}. For any partial combinatory algebra $A$,
external choice identifies $\RT(A)$ with the ordinary exact completion of
its partitioned assemblies~\cite{Frey2019Partitioned}.
Agwu recovers $\RT(A)$ as the homotopy
category of the zero-types in partitioned groupoid assemblies
\cite{Agwu2025}. In cubical assemblies, Uemura constructs a univalent
impredicative universe~\cite{Uemura2019}. In the same model, Swan and
Uemura construct a reflective subuniverse in which Church's thesis
holds~\cite{SwanUemura2021}.

The distinction between global points and computational observations
also appears for type codes. Escard\'o and Streicher's Rice theorem relates
extensional Boolean observations of a universe to weak limited
omniscience~\cite{EscardoStreicher2016}. Higher tripos theory separates
two classifiers with equivalent spaces of global points by their maps
to the natural numbers~\cite{Maruyama2026}. For every finite numerical simple type $\sigma$, we
identify the entire object of observations with $\sigma$ for the
program-indexed classifier and with $\sigma\times\sigma$ for the
tagged classifier (Theorem~\ref{thm:effective-observers}).

Swan uses cohomology to study countable choice in
realizability~\cite{Swan2026Cohomology}. For families with
Eilenberg--MacLane fibre, Shulman's criterion relates principality
to trivial monodromy~\cite{Shulman2014EM}. In our effective examples,
the corresponding cohomology classes determine the least number of
binary torsor coordinates needed to make two such families equivalent
after pullback
(Theorem~\ref{thm:effective-coordinate-cost}). Abelian extensions also
give pointed maps between Eilenberg--MacLane types~\cite{ChristensenFlaten2025}.
We construct a nonzero abelian group $Q$ in the effective topos whose
only global element is zero. For $n\geq2$, a nonsplit extension of $Q$
by the two-element group gives a binary cohomology class over $K(Q,n)$
in degree $n+1$, the first positive degree with nonzero binary cohomology
(Theorem~\ref{thm:effective-invisible-cohomology}).

Cohomology also distinguishes the slices themselves. Zoethout presents
slices over ordinary assemblies by internal partial combinatory
algebras~\cite{Zoethout2020Slices}. For a higher context $K(A,n)$, with
$A$ an abelian zero-type and $n\geq2$, we use maps between
Eilenberg--MacLane types~\cite{LicataFinster2014,Warn2023} to compute
what slicing adds to degree-$n$ cohomology with coefficients pulled back
from an abelian zero-type $M$. The additional summand is the group of
homomorphisms $A\to M$ in the ordinary core
(Theorem~\ref{thm:relative-context-cohomology}). In the effective model
the base has vanishing positive-degree cohomology, so for $A=A_r$ and
the two-element coefficient $M=D$ this summand gives the entire
degree-$n$ group, $\mathbb F_2^r$. At fixed $n$, every finite rank
$r\geq0$ occurs while the bases $K(A_r,n)$ have equivalent spaces of
global points
(Proposition~\ref{prop:separation-recovered}). Cohomology thus
distinguishes infinitely many higher realizability toposes sharing
Hyland's effective topos as their ordinary core.

\subsection{Organization of the paper}

Section~\ref{sec:input} gives conditions on an assembly presentation
that yield dependent products and univalent universes after higher exact
completion. Appendix~\ref{app:source-presentations} verifies these
conditions for relative and nested assemblies.
Sections~\ref{sec:bounded-localization} and~\ref{sec:operations}
construct sheafification in finite homotopy degrees and show how it
interacts with higher exact completion and slicing. We use these results
in Section~\ref{sec:unified-theorem} to construct the realizability
models and determine which indices admit contextual choice.

Section~\ref{sec:universal-probe} recovers a nucleus from its action on
higher families. Section~\ref{sec:higher-context-geometry} studies the
source dimension and cohomology of slices over Eilenberg--MacLane types.
These calculations distinguish the effective slices in
Section~\ref{sec:separation-specialization}. The next three sections
treat recognition in function realizability
(Section~\ref{sec:function-semantics}), induction and extraction in
modified realizability (Section~\ref{sec:modified-semantics}), and
choice and second identities in Krivine realizability
(Section~\ref{sec:krivine-semantics}). Section~\ref{sec:grothendieck-ranks}
classifies the Grothendieck counterparts of the effective slices by
localic rank and binary cohomology.

\section{Admissible presentations and higher exact completion}
\label{sec:input}

A function between assemblies is a morphism if a single tracker
sends every name of each input to a name of its output.
Higher exact completion adjoins quotients of groupoids to these
ordinary objects. Tracked sections and codes for small families in the
source give two ingredients for interpreting dependent types in the
completion. The sections present dependent products, and the codes
present univalent universes.

\subsection{Size and elementary conventions}

We use external choice and a sequence of nested Grothendieck universes
with strongly inaccessible cardinals $\kappa_0<\kappa_1<\cdots$.
Each carrier has cardinality less than $\kappa_i$ for some $i$.
Passing to a larger universe makes the categories and sites small. This common
size convention lets us compare ordinary realizability objects and
spaces at each level.

A cover is an effective epimorphism. An \emph{atlas} of $X$ is a
cover $A\cover X$ from an object of the specified class, for example
an assembly or a zero-type. A regular $\infty$-category
has finite limits and pullback-stable realizations of \v{C}ech nerves.
In an exact $\infty$-category all internal Segal groupoids have
effective realizations stable under base change. A regular functor
preserves finite limits and covers. Between exact categories it
therefore preserves effective groupoid realizations.
An object $P$ is \emph{regular-projective} if every map $P\to Y$
lifts through every cover $X\cover Y$, up to homotopy. Equivalently,
$\pi_0\Map(P,X)\to\pi_0\Map(P,Y)$ is surjective for every such cover.

An $n$-type is an $n$-truncated object. In the internal language,
a $(-1)$-type is a proposition and a zero-type is a set. For
$n\geq0$ an $(n+1)$-type has $n$-truncated identity types.
The zero-types form the \emph{ordinary core}.
For $q\geq-1$, a map $X\to B$ has \emph{relative height at most $q$} if it is a
$q$-truncated object of the slice over $B$, or equivalently if its
homotopy fibres are $q$-types in their parameter contexts.
For example, the fibre of a point $b:\one\to B$ over $b$ is the
loop object $\Omega_b B=\one\times_B\one$.
Write $\tau_{\leq q}^{B}X$, or $\|X\|_q$ when the base is understood,
for the $q$-truncation of $X\to B$ in that slice. It is the universal
$q$-truncated object over $B$ receiving a map from $X$ over $B$.
The map $X\to B$ is \emph{$q$-connected} if
$\tau_{\leq q}^{B}X\simeq\one_B$, where $\one_B$ is the identity of $B$,
and \emph{infinitely connective} if this holds for every finite $q\geq-1$.
An exact $\infty$-category is \emph{$q$-complicial}
if every object admits a cover by a $q$-truncated object
\cite{Stefanich2023Exact}.
The term \emph{bounded} means that each object has a finite
truncation bound, depending on the object.

We use the universe-based definition of elementary $\infty$-toposes 
due to Apol and Petrowitsch \cite{ApolPetrowitsch2026}, which 
have finite limits, local cartesian closure, and 
enough closed univalent universal families. 
The higher realizability models constructed here also have 
effective groupoid quotients stable under base change, 
disjoint finite coproducts, and a distinguished classifier
$\one\into\Omega$ for all monomorphisms. 

We note that Rasekh's axiomatization of elementary $\infty$-toposes 
uses enough complete Segal universes~\cite{RasekhElementary}; 
Rasekh also proves the existence theorem for natural-number objects 
in elementary $\infty$-toposes in the setting of \cite{Rasekh2021NNO}.

\subsection{Admissible computational presentations}

For a regular source $\mathcal C$, let
\[
 \mathcal H_{\mathcal C}
 =\operatorname{Sh}_{\mathrm{reg}}(\mathcal C,\mathcal S),\qquad
 \mathcal E_{\mathcal C}
 =\operatorname{Ex}_\infty(\mathcal C)\subseteq\mathcal H_{\mathcal C}.
\]
Here $\mathcal S$ is the $\infty$-category of spaces in the chosen
ambient universe. The regular topology is generated by covers consisting of a
single effective epimorphism. The completion is the full closure of
representables under effective groupoid realizations. Its unit is
fully faithful and regular. We call an object in the image of
$\mathcal C$ \emph{original}. Every completed object has an
original atlas \cite{Stefanich2023Exact}.

For a family $p:E\to V$ in the completion, write $\mathcal M(p)$
for its cover-local pullback class. A map $Y\to X$ belongs to
$\mathcal M(p)$ if there is a cover $X'\cover X$ and a map $X'\to V$
such that $Y\times_X X'\to X'$ is equivalent over $X'$ to the
pullback of $p$. Source covers suffice when testing on original
bases, since completed covers admit further original atlases.
In a presentable locally cartesian closed $\infty$-category, univalent
families correspond to bounded local classes of maps~\cite{GepnerKock2017}.

\begin{definition}[Admissible presentation]\label{def:admissible}
A computational presentation is admissible when it has the following data.
\begin{enumerate}
\item An ordinary regular category $\mathcal C$ with stable disjoint
finite sums, presenting the ordinary realizability topos
$\mathcal T$ by exact/regular completion.
\item A full replete, finite-limit-closed covering class
$\mathcal P\subseteq\mathcal C$ of regular projectives, and weak
section families in $\mathcal C$. For each $Z\to Y\to X$ in
$\mathcal C$ there is $W\to X$ in $\mathcal C$ and an evaluation
$W\times_XY\to Z$ over $Y$. Given $T\to X$ and a map
$T\times_XY\to Z$ over $Y$ in $\mathcal C$, there are a regular
epimorphism $T'\to T$ in $\mathcal C$ and a map $T'\to W$ over
$X$ whose evaluation is the restriction of the given map.
\item A collection of presented families
$p_\lambda:E_\lambda\to V_\lambda$ in
$\operatorname{Ex}_\infty\mathcal C$, with
$V_\lambda\in\mathcal C$, of finite relative truncation degrees.
The classes $\mathcal M_\lambda:=\mathcal M(p_\lambda)$ are
directed by inclusion, together contain every map of the completion,
and include a distinguished level consisting of all monos.
Choose a cofinal directed subsystem, including the distinguished mono
level, in which each class contains all monos. Its members are the
\emph{working levels}. Thus every presented class is contained in a
working level. We require each working level to be closed under
identities, image inclusions, relative diagonals, composition, and
dependent products of composable maps in that level. These products
are computed in $\mathcal H_{\mathcal C}$. Henceforth $\lambda$
indexes the working levels.
The source-code and bounded-quotient conditions of higher tripos
theory provide a way to verify these closures in the source.
\end{enumerate}
The last item supplies a hierarchy of type-valued families over the
ordinary propositional tripos. Each family has its code base in the
original source and its total space in the completion.
\end{definition}

By Lemma~\ref{lem:coherent-sections}, conditions (1) and (2) ensure that
these dependent products belong to $\mathcal E_{\mathcal C}$.
Condition (3) requires each working level
to be closed under those products.

The inaccessible cardinal bounds the size of the codes. The truncation
degree determines the depth of the following atlas construction. Let
$X\to V$ be $n$-truncated with $V$ original, and choose an original
atlas $A\cover X$. For $n=-1$, the subobject $X\hookrightarrow V$
is the regular image of $A\to V$. For $n\geq0$, consider the kernel
pair $A\times_X A$ of $A\cover X$. Its comparison map
\[
 A\times_X A\longrightarrow A\times_V A
\]
is a pullback of the diagonal of $X\to V$. Its base is original,
its total object belongs to the completion by finite-limit closure,
and its relative height is at most $n-1$. Recursion thus presents the
kernel pair over $A\times_V A$. Its iterated fibre products over $A$
form the nerve of $A\cover X$, whose effective quotient recovers $X$ over $V$.
Regularity of the source gives the same construction for the higher
sources below \cite{Maruyama2026}.

\subsection{Dependent products and univalent universes in the completion}

By the Carboni--Rosolini criterion, a finitely complete source has weak
dependent products precisely when its exact/lex completion is locally
cartesian closed~\cite{CarboniRosolini2000,Emmenegger2020LCC}.
Emmenegger also derives local cartesian closure from a categorical form
of the fullness axiom~\cite{Emmenegger2020Fullness}.
We use dependent type notation in each slice. A type over $X$ is a
map to $X$, and a term of that type is a section. For
$Z\to Y\to X$, the fibre of $\Pi_{Y/X}Z$ at $x$ is the space of
sections of $Z_x\to Y_x$. The relative diagonal interprets identity.
The proposition $\operatorname{isContr}(T)=
\sum_{t:T}\prod_{u:T}(t=u)$ asserts that $T$ has a centre of
contraction. The image of $T\to1$ is the propositional truncation
$\|T\|$, which asserts local existence of an element of $T$.

For the applications to relative and nested assemblies, we describe
dependent products explicitly in terms of section codes. Write
$\mathcal H$ for the ambient category of regular sheaves and $\mathcal E$
for the completion. By the projective-cover criterion of higher tripos
theory~\cite{Maruyama2026}, dependent products in $\mathcal E$ are
computed in $\mathcal H$. Over an original projective base, we construct
an atlas of section codes for the dependent product. The fibre of its
kernel pair over two codes is the space of paths between the sections
they represent.

\begin{lemma}[Finite descent for section codes]
\label{lem:coherent-sections}
Suppose that the first two conditions of Definition~\ref{def:admissible}
hold. Let $Z\to Y\to X$ be in $\mathcal E$, with $Z\to Y$
$n$-truncated. If $X$ is an original projective, there is an original
$W\to X$ and an evaluation over $Y$ inducing a cover
\[
 W\longrightarrow S=\Pi^{\mathcal H}_{Y/X}Z .
\]
This holds for every finite $n$. The section object $S$ belongs to
$\mathcal E$ for every base $X$ in the completion.
\end{lemma}

\begin{proof}
First consider source diagrams over an original base. Equations between
maps of source objects can be imposed on section codes. Indeed, for a
source mono $D\hookrightarrow C$ and $C\to A$, a weak section object
$W_0\to A$ has image precisely $\Pi_{C/A}D\hookrightarrow A$ in
regular sheaves. Evaluation gives one inclusion. For the other, test
on a representable and use the local lifting property of the section
codes. Images of maps of source objects are their source images by
regularity. Finite intersections and these products of monos thus
define, in the source itself, the subobject of codes satisfying any
finite list of universally quantified equations.

Now take $X$ in the chosen covering class $\mathcal P$ and
use members of $\mathcal P$ for the projective tests $T$ and all
the projective covers below. Take $m=n+1$, so $m=0$ for $n=-1$.
In the slice over $X$
choose a partial augmented semisimplicial hypercover $U_{\leq m}\to Y$:
at each degree take a projective source cover of the matching object,
the finite limit specifying all compatible faces over $Y$. Such covers
exist by the atlas property and the source projective cover. Since
$X,T,U_k$ lie in $\mathcal P$, finite-limit closure puts each
$T\times_X U_k$ in the class of source projectives. Every such
projective $P$ remains projective in the completion. Indeed, given
a cover $E\cover B$ and a map $P\to B$, choose an original atlas
$A\cover P\times_B E$. The composite $A\to P$ is a cover in the
source, so it has a section. Composing this section with $A\to E$
gives the required lift.

Put $D_k=U_k\times_Y Z$. For a
semisimplicial diagram $L_\bullet$ in $\mathcal E/X$, let $M_kL$ be
the limit of its degree-$k$ boundary diagram in that slice, with
$M_0L=X$. Choose a relative projective resolution
$V_{\le m}\to D_{\le m}$ by taking a projective cover
\[
 V_k\cover D_k\times_{M_kD}M_kV
\]
at each step. The composites
$V_k\to U_k$ form a finite diagram in the ordinary source.
Take weak section objects for these maps over $X$, take their product,
and impose all face equations by the preceding construction. This
gives an original parameter object $W$ for compatible levelwise
sections. A coherent section over $T\times_XY$ with $T$ original
projective gives a point of $W$ over $T$: lift successively along the
relative matching covers from each projective $T\times_XU_k$. The
matching objects enforce the previously chosen
faces. Mapping spaces between the resulting source
objects are discrete, so their face compatibilities
are precisely the equations used in defining $W$. The weak codes at
each degree may be chosen over $T$ since $T$ splits source covers.

For the converse, the universal compatible sections give a map to
$Z$ over $Y$ from the partial realization of $W\times_X U_{\le m}$
in $\mathcal H$. Restriction along this partial resolution is an
equivalence on sections of $n$-truncated families. To see this,
continue the semisimplicial resolution by matching covers. In the
Grothendieck $\infty$-topos $\mathcal H/W$ the resulting hypercover
has infinitely connective augmentation
\cite{Stefanich2023Exact}. Over $W\times_XY$,
orthogonality to truncated maps therefore identifies the section spaces.
In degrees $k>m$, the space of extensions across a $k$-simplex
with prescribed boundary is contractible. Indeed, for an
$n$-type $D$, the map $D\to D^{S^{k-1}}$ sending each point to
the corresponding constant map is an equivalence when $k-1>n$.
This argument also applies when $n=-1$, where a
proposition has at most one inhabitant. Realizations and the matching
cover conditions commute with base change in $\mathcal H$. This gives
the asserted evaluation over $W$ and identifies the evaluation of each
lift just constructed with the original section. Original projectives
cover the representable generators of $\mathcal H$, so $W\to S$ is an
effective epimorphism. An arbitrary original regular-projective $X$
is a retract of some $P\in\mathcal P$ by splitting a cover. Compose
$Y\to X$ with the resulting mono $X\hookrightarrow P$. Restricting
the section code just obtained over $P$ back to $X$ gives the asserted
original code by base change for dependent products and covers.

We now prove membership of $S$ in $\mathcal E$ by induction on $n$,
allowing all bases at each stage. Over an original projective base in
degree $-1$, $S$ is the image of $W\to X$. For $n\geq0$, the relation
$W\times_S W\to W\times_X W$ is the dependent product over $Y$ of
the family of identities between the two evaluated sections. That
family is $(n-1)$-truncated. Induction places the relation in
$\mathcal E$. Its higher nerve terms are finite pullbacks, so its
effective quotient $S$ also lies in $\mathcal E$.

Finally choose a projective cover $P\to X$ for an arbitrary completed
base. The preceding argument represents $S_P=P\times_XS$. All higher
terms of its descent diagram are represented, since in simplicial
degree $k$ the term is
$S_P\times_P P^{\times_X(k+1)}$, using its first projection to $P$.
Their effective quotient is $S$. This establishes the induction for
all bases.\qed
\end{proof}

\begin{proposition}[Finite-height logical completion]
\label{prop:input-completion}
An admissible presentation gives a bounded exact elementary
$\infty$-topos $\mathcal E_{\mathcal C}$, and
\[
 (\mathcal E_{\mathcal C})_{\leq0}\simeq\mathcal T.
\]
The unit identifies original mapping spaces and subobject lattices
and preserves existing dependent products. Every object has a cover
by an original zero-type. For each level there is a univalent family
$u_\lambda:\widetilde U_\lambda\to U_\lambda$ with
\[
 \operatorname{Map}(X,U_\lambda)
 \simeq
 \bigl((\mathcal E_{\mathcal C}/X)_{\mathcal M_\lambda}\bigr)^{\simeq}.
\]
Here $(-)^{\simeq}$ denotes the maximal $\infty$-groupoid. Its objects
in this formula are the $\mathcal M_\lambda$-families over $X$, its
paths are equivalences over $X$, and its higher paths are their homotopies.
If the class has relative bound $n$, its classifier is an
$(n+1)$-type. The hierarchy is cumulative up to coherent equivalence.
\end{proposition}

\begin{proof}
The regular-sheaf construction identifies the completion with the
full exact closure of its source. For an ordinary source it is bounded:
the bounded objects contain the source and are closed under effective
groupoid realizations, since the degree-zero and arrow objects bound
the degree of the quotient by the atlas and diagonal criteria
\cite{Stefanich2023Exact}.
Lemma~\ref{lem:coherent-sections}
gives all dependent products and computes them in regular sheaves.

Apply the individual-family reconstruction
\cite{Maruyama2026} at the truncation degree of each
presented family. Lemma~\ref{lem:coherent-sections} supplies the
section codes required at that degree.
For a presentation $E\to V$ form the internal object of equivalences
between its two pullbacks to $V\times V$. Dependent products give
the fibrewise mapping objects. For a map $f:A\to B$, take the
product of the space of pairs $(g,gf\simeq\mathrm{id}_A)$ and
the space of pairs $(h,fh\simeq\mathrm{id}_B)$ with $g,h:B\to A$.
These are homotopy fibres of composition. Over any test object, a
point of this product supplies both a left and a right homotopy
inverse to $f$, so $f$ is an equivalence in that slice. In this
case both spaces of inverses are contractible
since composition by $f$ is an equivalence on mapping spaces.
This constructs the required subobject. Coherently associative composition
gives a groupoid with object space $V$ and these
equivalences as arrows. More explicitly, its $k$-simplices are
composable strings of $k$ equivalences, with the usual composition
maps in the core of the slice category. Its quotient $U$ has this
groupoid as the nerve of $V\cover U$.

The groupoid acts on $E$ by evaluation. Realizing the action nerve
gives a family $\widetilde U\to U$ whose pullback to $V$ is $E\to V$.
After pullback along $V\times V\to U\times U$, effectiveness identifies
paths between codes in $U$ with equivalences between the corresponding
fibres of $E\to V$. Univalence then follows by descent over $U\times U$.

A local presentation of a family now gives a map to $U$, since the
equivalences on overlaps and all their compatibilities are the
descent data for this map. Conversely a map to $U$ has local lifts to $V$, and its
pullback belongs to the presented class. This proves the equivalence
of spaces in the proposition. For fibres of height $n$, equivalences
between them are $n$-truncated, so $U$ has height at most $n+1$.
Inclusions of classes induce the cumulative comparisons by the
representing property, including their composition coherences.

For an original object $A$, a completed subobject has an original
atlas $P$ and is the regular image of $P\to A$, hence original.
A zero-type $X$ has an original atlas $A\cover X$. Its kernel pair
is therefore an ordinary equivalence relation $R\hookrightarrow A\times A$.
Conversely, the effective quotient of such a relation is a zero-type
because its diagonal is a mono after pullback along the square of the atlas.
For a second presentation $B\cover Y$ with kernel pair $S$, pulling
a graph back to $A\times B$ gives an original relation, invariant
under $R$ and $S$, total on $A$ and single-valued modulo $S$.
Conversely such a relation descends to a subobject of $X\times Y$.
Its projection to $X$ is both a cover by totality and a mono by
single-valuedness, so it is the graph of a unique map. This is the
ordinary exact/regular completion, on objects and on morphisms.
The unit is fully faithful by the Yoneda embedding
into regular sheaves. An existing source
dependent product represents the ambient one on representable
tests by the adjunction, so these agree by density. The ordinary
subobject classifier also classifies higher-context monos: classify
on an original atlas, descend the characteristic map on its overlaps
(using original atlases there as well), and use that the target
classifier is a set. This realizes the distinguished mono level.

We construct finite sums intrinsically in the completion, starting
with the source objects $0$ and $2$. As shown in the proof of
Lemma~\ref{lem:coherent-sections}, original regular-projective
objects remain regular-projective in the completion. Mapping out of such
a projective preserves covers and finite limits, and hence effective
groupoid quotients. The source objects $0$ and $2=1\amalg1$ are
projective. For an original object $X$, the space of maps from $0$ to $X$
is contractible, and a map from $2$ to $X$ amounts to a pair of global
sections of $X$. Those identities
extend through groupoid quotients, making $0$ initial and
$2=1\amalg1$ intrinsically in the completion. For arbitrary $X,Y$,
choose a common universe classifying them; their two codes give a
family $F\to2$. In the slice over $2$, the coproduct of the two
point inclusions is the identity on $2$. Pullback to $F$ preserves this coproduct
because it has a right adjoint, and composition to the terminal
context preserves it as a left adjoint. Hence $F=X\amalg Y$.
Disjointness follows by pullback from that of the two source
inclusions into $2$.\qed
\end{proof}

\paragraph{The codomain doctrine.}
Take
$\mathbb D_{\mathcal C}(X)=\mathcal C/X$, with reindexing by
pullback and dependent sums by composition. For a completed section
problem over an original base, cover its dependent product by an
original object and restrict evaluation to that cover. Pulling the
cover back along the transpose of a parameterized section gives a
local code and a homotopy over the family from its evaluation to the
given section. When the parameter is original, refining by an original
atlas gives a cover in the source. This local representation
of sections, including the lower-height problems of representing
identities between them, is \emph{coherent fullness} at the given
height. The weak universes have original code
bases by hypothesis, and their five class closures give the
source-local codes for the logical operations
\cite{Maruyama2026}. These data give the higher tripos structure
on $\mathbb D_{\mathcal C}$. Applying the higher tripos-to-topos
construction $\mathsf T_\infty$ gives
$\mathsf T_\infty\mathbb D_{\mathcal C}\simeq\mathcal E_{\mathcal C}$.
For the assembly sources, let $\nabla I$ be the codiscrete assembly
on a set $I$, in which every point has the full set of names at each
level. The doctrine $I\mapsto\mathcal C/\nabla I$ records type-valued
fibres and their coherent maps. Taking subobjects of the terminal
object in each fibre gives
$I\mapsto\operatorname{Sub}_{\mathcal C}(\nabla I)$, the ordinary
realizability tripos of the assembly presentation.

\paragraph{Trackers for parameterized sections.}
Following Hofstra~\cite{Hofstra2006}, take a filtered ordered partial
combinatory algebra (OPCA) $(A,\leq,\cdot,A')$. Assembly names lie in
$A$ and global trackers lie in the filter $A'$. A partitioned
assembly has a specified name $\nu(x)$ and realizes $x$ by
$\downarrow\nu(x)$. Such assemblies form a finite-limit-closed
projective cover. The section-object carrier over $x$ consists of
sections admitting a tracker in $A$. The names of a section form the
downward closure of the pairs $\langle s,a\rangle$ with $s$ a name
of $x$ and $a$ a tracker of that section.
Suppose $h$ tracks the map from the parameter object to the base
and $d$ tracks evaluation on paired inputs. Fix a bracket abstraction
$c\in A'$ for $\lambda d\,r\,b.\,d\langle r,b\rangle$ and put
$\sigma(d,r)=c d r$. Then $\sigma(d,r)b\leq d\langle r,b\rangle$
whenever the right-hand side is defined. Thus $\sigma(d,r)$ tracks
the fibre section at the parameter name $r$, and
$r\mapsto\langle hr,\sigma(d,r)\rangle$ tracks the transpose.

Theorem~\ref{thm:source-verification} in Appendix~\ref{app:source-presentations}
constructs the corresponding univalent hierarchies for relative and
nested assemblies. The inaccessible cardinal bounds the sizes of the name
sets and finite diagrams. The fibre height determines how many matching levels suffice to represent
a section. In nested realizability one actual tracker acts uniformly on
both potential and actual names.

\section{Lawvere--Tierney sheafification at finite truncation heights}
\label{sec:bounded-localization}

The completion has the same classifier of propositions as its
ordinary core. A nucleus on this classifier therefore already
specifies which subobjects of any higher context are dense. We
construct the reflection of a higher type from the reflections of
its identity types, starting with the action of the nucleus on
propositions.

Let $\mathcal E$ be a bounded exact elementary $\infty$-topos with
the closed hierarchy above. A Lawvere--Tierney nucleus is a map
$j:\Omega\to\Omega$ such that all internal truth values $P,Q$ satisfy
\[
 P\leq jP,\qquad j(jP)=jP,\qquad
 j(P\wedge Q)=jP\wedge jQ,\qquad j\top=\top.
\]
These laws hold in every context. For a subobject $P\into B$ with
characteristic map $\chi_P:B\to\Omega$, its $j$-closure
$\overline{P}^{\,j}$ has characteristic map $j\circ\chi_P$.
We write $j\leq k$ when $jP\Rightarrow kP$ for every truth value
$P$ in every context. The subobject $P\into B$ is \emph{$j$-closed}
when $\overline{P}^{\,j}=P$ and \emph{$j$-dense} when
$\overline{P}^{\,j}=B$ as subobjects of $B$. A type is $j$-local when it is
right orthogonal to dense monos, with an equivalence of full
mapping spaces. Internally this is expressed by the proposition
\[
 \Loc_j(X)
 =\prod_{P:\Omega}
    \bigl(jP\to\operatorname{isEquiv}(X\to X^P)\bigr).
\]
All definitions are also made in slices, using the induced nucleus.
Indeed, if $D\hookrightarrow B$ is dense, its fibre $P_b$ at each
$b:B$ satisfies $jP_b$. The displayed condition extends a map
$D\to X$ by taking the inverse of $X\to X^{P_b}$ at each $b$.
Conversely, for a proposition $P$ in any context, restrict to the
subcontext $jP$. There $P$ is dense, and remains dense after every
further pullback. Orthogonality of $X$ on all these tests makes the
constant-map comparison an equivalence over $jP$ by Yoneda.

For a code $v$ in a univalent level $U_\lambda$, write
$\operatorname{El}(v)$ for the fibre of its universal family at
$v$. The subtype
$U_{\lambda,\le n}=\sum_{v:U_\lambda}
\operatorname{isTrunc}_n(\operatorname{El}(v))$ classifies its
$n$-truncated families. Here $\operatorname{isTrunc}_{-1}(X)$ is
$\prod_{x,y:X}(x=y)$ and
$\operatorname{isTrunc}_{r+1}(X)=
\prod_{x,y:X}\operatorname{isTrunc}_r(x=y)$ for $r\ge-1$.
These assertions are propositions: a witness at the first stage
makes $X$ a proposition and its identities contractible wherever
there are endpoints; the inductive step uses function
extensionality. Thus the inclusion into $U_\lambda$ is monic.
Univalence identifies identities in the subtype with equivalences
between $n$-types, themselves $n$-truncated, giving the bound $n+1$.

\begin{lemma}[Local univalent subuniverses]
\label{lem:local-universe}
Let $U_{\lambda,\leq n}$ classify the $n$-truncated families in a
closed level containing all monos. The subuniverse
\[
 U^j_{\lambda,\leq n}
   =\sum_{v:U_{\lambda,\leq n}}
       \Loc_j(\operatorname{El}(v))
\]
is $j$-local. Its tautological family is a family of local types.
\end{lemma}

\begin{proof}
Work in any context, and let $P$ be a dense proposition. A family
$F:P\to U^j_{\lambda,\leq n}$ extends to the type
$\prod_{p:P}\operatorname{El}(F(p))$. It belongs to the same
level by dependent-product closure and the inclusion of all monos.
It is an $n$-type and is local, since local families are closed
under dependent products: orthogonality is tested using the product
adjunction and pullback stability of dense monos. Over $P$ the
extension is equivalent to the original family. Conversely the
extension of the constant family $X$ is $X^P\simeq X$, by
locality. Univalence turns these equivalences into inverse maps of
classifiers, including all identity data. This proves orthogonality
of the displayed subuniverse in every context.\qed
\end{proof}

\begin{theorem}[Bounded Lawvere--Tierney sheafification]
\label{thm:bounded-reflection}
Every nucleus $j$ above has a left-exact reflection
\[
 a_j:\mathcal E\rightleftarrows\mathcal E_j:i_j
\]
onto its local types. It preserves each finite truncation bound,
acts on propositions by $j$, and is stable in slices. The category
$\mathcal E_j$ is bounded, exact, and locally cartesian closed,
and has an exhaustive closed hierarchy of univalent universes.
For each family at level $\lambda$, a local universe $U^j_\mu$
classifies its reflection, with
$\mathcal M_\lambda\subseteq\mathcal M_\mu$.
\end{theorem}

\begin{proof}
Using the singleton construction of Quirin and Tabareau~\cite{QuirinTabareau2016},
we construct the reflector in every slice by induction on truncation degree.
At each successor stage we take the $j$-closure of the image of the
localized singleton map in a univalent level. For each $n$-type $X$,
the unit $X\to a_nX$ into the resulting local $n$-type induces an equivalence
\begin{equation}
 \Map(a_nX,Q)\simeq\Map(X,Q)
 \quad\text{for every bounded local }Q.
 \label{eq:strong-reflection-invariant}
\end{equation}
The local target $Q$ may have any finite truncation degree, independently
of $n$. In the same induction we prove stability under base change and
preservation of finite products.

If $S\hookrightarrow Z$ is $j$-closed and $Z$ is local, then $S$
is local. Indeed, given a dense mono $D\hookrightarrow T$ and a map $D\to
S$, extend the composite $D\to S\hookrightarrow Z$ to $T$ by locality of
$Z$. The inverse image of $S$ under its extension
is closed and contains $D$, so is all of $T$. The space of factorizations
through the mono is contractible. This holds in every slice.
In particular $jP$ is local, being a closed subobject of $\one$.
For $n=-1$ take $a_{-1}P=jP$. The unit is a dense mono, so
orthogonality gives \eqref{eq:strong-reflection-invariant}; the
other assertions are the nucleus laws.

\emph{Localized singletons.} Suppose the construction
has been made through $n$, and let $T$ be an $(n+1)$-type.
Exhaustiveness puts the entire family $(x,t)\mapsto a_n(x=t)$ over
$T\times T$ in one level. Its classifying
map lands in the local subuniverse of Lemma~\ref{lem:local-universe}.
Form
\[
 \sigma:T\longrightarrow Z=(U^j_{\lambda,\leq n})^T,
 \qquad \operatorname{El}(\sigma(x)(t))\simeq a_n(x=t).
\]
The type $Z$ is local and $(n+1)$-truncated. Factor $\sigma$
as $T\cover S\into Z$, and define $a_{n+1}T$ as the
$j$-closure of $S$ in $Z$. Let $\eta_T:T\to a_{n+1}T$ be the
composite $T\cover S\into a_{n+1}T$. A closed subobject of a local type
is local, and a subobject retains its truncation bound. Thus
$a_{n+1}T$ is a local $(n+1)$-type and
$S\into a_{n+1}T$ is dense.

\emph{Paths between singletons.} To calculate the path type from
$\sigma(x)$ to $\sigma(y)$, we use the fact that the type of
equivalences between two local types is local. Represent an
equivalence by a function equipped with a left inverse and a right
inverse, each with its witnessing homotopy. This total
space is a finite limit of function types with local codomain, so
is local.
Write $F_x(t)=a_n(x=t)$. The relative induction hypothesis and
dependent Yoneda give
\[
 \prod_{t:T}(F_x(t)\to F_y(t))
 \simeq\prod_{t:T}((x=t)\to F_y(t))
 \simeq a_n(y=x).
\]
We claim that each family of maps $F_x(t)\to
F_y(t)$ on the left consists of equivalences. For
$p:y=x$, precomposition with $p$ gives the fibre equivalences
$F_x(t)\simeq F_y(t)$. Their type $\prod_t\operatorname{Equiv}
(F_x(t),F_y(t))$ is local, so the induction adjunction extends
this assignment to $a_n(y=x)$. The Yoneda comparison evaluates
the underlying transformation at $t=x$ on the unit of
$\operatorname{refl}_x$. Its value on the given transformation is
the unit of $p$, by path induction. Thus the extension is a right
inverse to forgetting equivalence data and applying Yoneda. Since
being an equivalence is a proposition, this forgetful map is monic,
so it is an equivalence. After reversing the path $p$, univalence
and function extensionality give, uniformly in $x,y$,
\begin{equation}
 (\eta_T x=\eta_T y)\simeq a_n(x=y).
 \label{eq:localization-paths}
\end{equation}
The map $\operatorname{ap}_{\eta_T}$ sends a path in $T$ to its image
under $\eta_T$. Under this comparison it is the stage-$n$ unit, as path
induction reduces the assertion to the evaluation at reflexivity just
computed. The same calculation holds for the two images in $S$,
since both $S$ and $a_{n+1}T$ are subobjects of $Z$.

\emph{Descent to the image.} Let $R_\bullet$ be the \v{C}ech nerve
of $T\cover S$. The diagonals
with repeated vertices give a simplicial map
$d_\bullet:\operatorname{const}(T)\to R_\bullet$. Over
$T^{k+1}$, its degree-$k$ map has
fibres
\[
 \prod_{r=1}^k(t_0=t_r)\longrightarrow
 \prod_{r=1}^k(\eta_T t_0=\eta_T t_r).
\]
For $k\geq1$ this is the stage-$n$ unit, by
\eqref{eq:localization-paths}, product preservation, and base
change. For $k=0$ it is the identity. For any bounded local $Q$,
use $T^{k+1}\times Q$ as the local target over $T^{k+1}$.
The relative induction adjunction identifies restriction along $d_k$
with an equivalence $\Map(R_k,Q)\to\Map(T,Q)$. Since the restrictions
are induced by $d_\bullet$, they give a map of cosimplicial spaces
to the constant diagram on $\Map(T,Q)$. Effectiveness now yields
\[
 \Map(S,Q)\simeq\lim_{[k]\in\Delta}\Map(R_k,Q)
             \simeq\Map(T,Q).
\]
Restriction along the dense mono $S\into a_{n+1}T$ is an
equivalence for local $Q$, proving
\eqref{eq:strong-reflection-invariant} at $n+1$.

The construction was made in arbitrary slices. For $g:B\to C$,
pullback preserves local families: its left adjoint, composition
with $g$, carries a dense mono over $B$ to the same dense mono over
$C$. Also, if $Q$ is local over $B$, then $\Pi_gQ$ is local over $C$
by orthogonality to pulled-back dense monos. The product adjunction
therefore shows that $g^*a_{n+1}X$ has the same reflection
universal property as $a_{n+1}g^*X$. Finite-product preservation
follows by currying twice against local targets, using locality
of exponentials with local codomain. Uniqueness of the adjoints
provides all substitution and product coherences.

\emph{Preservation of pullbacks.} The operations at successive stages agree on their common
domains by \eqref{eq:strong-reflection-invariant}. Since every
object is finitely truncated, they assemble to one reflection.
Local families are closed under dependent sums when their bases
are local: extend a map to the local base across a dense mono,
then extend its section of the resulting local fibre family;
both extensions have contractible spaces of choices. Thus this
is a modality. For a modality on a fixed universe, the path
comparison is equivalent to preservation of pullbacks
\cite{RijkeShulmanSpitters2020}. The following argument allows the
reflected family to lie in a larger working level.
Write $L$ for the reflection. For a local family $D\to LA$, both
$D$ and $LA$ are local. Taking the fibres at $\mathrm{id}_{LA}$
and $\eta_A$ of the restriction square for maps to $D\to LA$
shows that sections of $D$ over $LA$ restrict equivalently to
sections over $A$. The argument holds in each parameter slice.

For $X\xrightarrow{f} Z\xleftarrow{g}Y$ put
$R=LX\times_{LZ}LY$. For a local $Q$, a map $R\to Q$ is a term of
\[
 \prod_{x':LX}\prod_{y':LY}
       \bigl((Lf(x')=Lg(y'))\longrightarrow Q\bigr).
\]
The function types into $Q$ are local, and dependent products of
local families are local. Restrict sections along $\eta_X$ and
$\eta_Y$ in turn by the preceding argument. For $x:X$ and $y:Y$,
\eqref{eq:localization-paths} identifies the remaining path type
with $L(f(x)=g(y))$ in that context, compatibly with the unit.
By the relative adjunction, precomposition with this unit identifies
functions from the reflected path type into $Q$ with functions from
$f(x)=g(y)$ into $Q$.
These equivalences identify restriction along
$X\times_ZY\to R$ with an equivalence of mapping spaces into $Q$.
Since $R$ is local by finite-limit closure, it is the reflection
of $X\times_ZY$. The terminal object is local as well, proving
left exactness.

\emph{The local category and its universes.} Reflecting ambient
groupoid quotients gives their quotients in $\mathcal E_j$.
Left exactness proves their effectiveness and base-change
stability. Dependent products between local objects are computed
ambiently, by the product adjunction and left exactness of $a_j$.
Finite coproducts are reflected coproducts. Left exactness gives
disjointness, with the initial object of the local category used
in the pullback calculation.

Let $n_\lambda$ bound the truncation degree of the fibres of $u_\lambda$.
Then $U_{\lambda,\leq n_\lambda}\to U_\lambda$ is an equivalence.
Thus the subtype of codes for local types in this level is
\[
 U^j_\lambda:=\sum_{v:U_\lambda}\Loc_j(\operatorname{El}(v))
       \simeq U^j_{\lambda,\leq n_\lambda}.
\]
Restrict $u_\lambda$ to $U^j_\lambda$. Its base is local by
Lemma~\ref{lem:local-universe}, and its total space is local
because its fibres and base are local. For local bases it
classifies precisely ambient $\mathcal M_\lambda$-maps with
local total spaces. Indeed, if both endpoints of a map are local,
extension of sections across a dense mono is the equivalence on
fibres of the two endpoint mapping-space equivalences. Thus its
family is local in the slice. Classification and univalence
give the asserted equivalence on full spaces of families.

To see cover locality in the local category, suppose a map has
small pullback along a local cover $B'\cover B$. Its restrictions
to the \v{C}ech nerve of this cover give compatible codes in the
local classifier, by univalence. Effectiveness in the local category
descends them to a code on $B$. Its decoded family is equivalent to
the original: their pullbacks and all transition equivalences agree
on the nerve, and realizing the pulled-back nerves recovers their
total spaces over $B$. Thus covers intrinsic to the local category
give descent for small families.
Composition and relative diagonals are computed ambiently, as are
dependent products between local objects. An image inclusion in the
local category is a mono and hence belongs to every working level.
Concretely it is the closure of the ambient image. The
classes are directed and exhaustive, and their cumulative maps
are supplied coherently by classification. Thus they give the
asserted closed hierarchy. At the mono level, a proposition $P$ is
local precisely when $jP=P$. Locality extends its identity across the
dense mono $P\hookrightarrow jP$, and the converse follows from
closedness. Thus the local subobject classifier is
$\Omega_j=\sum_{P:\Omega}(jP=P)$, with truth as its universal mono.
Its tautological total space is contractible by propositional
extensionality.\qed
\end{proof}

\begin{corollary}[Ordinary recovery under localization]
\label{cor:ordinary-localization}
Suppose $\mathcal E$ is zero-complicial and
$\mathcal E_{\leq0}=\mathcal T$. Then
\[
 (\mathcal E_j)_{\leq0}\simeq\operatorname{Sh}_j(\mathcal T).
\]
\end{corollary}

\begin{proof}
The inclusion $i_j$ preserves and reflects truncation. A local
zero-type is therefore an ordinary $j$-sheaf: test orthogonality
on ordinary dense monos. Conversely let $Q$ be an ordinary
$j$-sheaf. Finite truncations in an exact category exist and are
computed in its regular-sheaf presentation
\cite{Stefanich2023Exact}.
Since $\Omega$ is zero-truncated, subobjects of $X$ correspond
to subobjects of $\tau_{\leq0}X$; a mono and its closure are
obtained by pulling back the corresponding mono and closure on
$\tau_{\leq0}X$. The map from that pullback to the ordinary subobject
is a base change of the $0$-connected truncation unit, hence is
$0$-connected. Its target is therefore the zero-truncation of the
pullback. Thus a dense mono $A\into B$ induces an
ordinary dense mono $\tau_{\leq0}A\into\tau_{\leq0}B$.
Maps into $Q$ factor through these zero-truncations. Ordinary
sheaf orthogonality therefore gives
$\Map(B,Q)\simeq\Map(A,Q)$ for every dense mono in the higher
category. Hence the local zero-types are exactly the ordinary
sheaves.\qed
\end{proof}

\section{Presentation independence and change of context}
\label{sec:operations}

Sheafifying the ordinary realizability topos and then taking its higher
completion gives the same higher category, up to canonical equivalence,
as sheafifying the completion. Localization preserves the covers by
zero-types in Proposition~\ref{prop:input-completion}.
Slicing gives a further way to change the model. The homotopy fibres
over the new base determine the relative height of its atlases.

By Stefanich's recognition theorem
\cite{Stefanich2023Exact}, for a bounded,
zero-complicial exact $\infty$-category $\mathcal E$ the canonical
regular functor
\[
 \operatorname{Ex}_{\infty}(\mathcal E_{\leq 0})
       \longrightarrow \mathcal E
\]
is an equivalence. Conversely, the same result gives boundedness
and zero-compliciality for the higher exact completion of an ordinary
regular category. Its zero-types form the ordinary exact/regular
completion \cite{Stefanich2023Exact}.

\subsection{Independence of the assembly presentation}

\begin{proposition}[Ordinary exact completion can be performed first]
\label{prop:operations-source-independence}
Let $\mathcal C$ be an ordinary regular category and put
$\mathcal T=\operatorname{Ex}_{1,\mathrm{reg}}(\mathcal C)$. The canonical
regular functor induces an equivalence
\[
 \operatorname{Ex}_{\infty}(\mathcal C)
       \simeq \operatorname{Ex}_{\infty}(\mathcal T).
\]
This equivalence is natural in regular functors of ordinary sources.
Consequently, when an assembly category presents an ordinary
realizability topos by exact/regular completion, either source gives
the same higher exact category.
\end{proposition}

\begin{proof}
Let $\mathcal D$ be any exact $\infty$-category. Its full subcategory
$\mathcal D_{\leq0}$ is an ordinary Barr-exact category.
In particular, for an ordinary equivalence relation on a zero-type
$A$, the diagonal of its effective quotient pulls back along the
cover from $A\times A$ to the relation mono. Thus the diagonal is
monic and the quotient is again a zero-type. Images of maps between
zero-types are subobjects of zero-types. It follows that the inclusion
preserves and reflects finite limits and regular epimorphisms between
its objects. A finite-limit-preserving functor from an ordinary category
to $\mathcal D$ takes values in $\mathcal D_{\leq0}$ because it
preserves the iterated diagonal condition that characterizes being
zero-truncated. Fullness of the inclusion also identifies the spaces of
natural transformations. Thus the ordinary exact/regular universal
property gives natural equivalences
\[
\begin{aligned}
 \operatorname{Fun}_{\mathrm{reg}}(\mathcal C,\mathcal D)
 &\simeq
 \operatorname{Fun}_{\mathrm{reg}}(\mathcal C,\mathcal D_{\leq0})\\
 &\simeq
 \operatorname{Fun}_{\mathrm{reg}}(\mathcal T,\mathcal D_{\leq0})\\
 &\simeq
 \operatorname{Fun}_{\mathrm{reg}}(\mathcal T,\mathcal D).
\end{aligned}
\]
Applying the higher exact-completion universal property to the two
endpoints proves the claim. Naturality follows from the naturality of
restriction and of both universal properties.\qed
\end{proof}

\begin{remark}[Transport of size hierarchies]
A universe level is determined by the class of small maps it
classifies. The equivalence transports each such class and its
classifying family together. This identifies the corresponding
levels across carrier presentations whenever their size conventions
select the same classes.
\end{remark}

\subsection{Recognizing bounded reflective localizations}

\begin{proposition}[Reflection is reconstructed from its zero-types]
\label{prop:operations-reflection-reconstruction}
Let $\mathcal E$ be bounded, exact, and zero-complicial, and let
\[
 a:\mathcal E\rightleftarrows\mathcal D:i
\]
be a reflective inclusion whose left adjoint is left exact. Then
$\mathcal D$ is bounded, exact, and zero-complicial, and there is a
canonical equivalence
\[
 \mathcal D\simeq
 \operatorname{Ex}_{\infty}(\mathcal D_{\leq0}).
\]
In particular, let $\mathcal T$ be an ordinary elementary topos, let
$j$ be a Lawvere--Tierney topology on $\mathcal T$, and suppose a
left-exact reflective $j$-sheafification on
$\mathcal E=\operatorname{Ex}_{\infty}(\mathcal T)$ has been constructed,
with zero-types precisely $\operatorname{Sh}_{j}(\mathcal T)$. Then
\[
 \operatorname{Sh}_{j}(\operatorname{Ex}_{\infty}(\mathcal T))
       \simeq
 \operatorname{Ex}_{\infty}(\operatorname{Sh}_{j}(\mathcal T)).
\]
For an admissible presentation, Theorem~\ref{thm:bounded-reflection}
constructs this reflector and Corollary~\ref{cor:ordinary-localization}
identifies its zero-types.
\end{proposition}

\begin{proof}
Finite limits in $\mathcal D$ are computed by its fully faithful right
adjoint. For a groupoid object $R_\bullet$ of $\mathcal D$, realize
$iR_\bullet$ in $\mathcal E$ and apply $a$. Since $a$ is a left adjoint,
this computes the realization in $\mathcal D$. Since $a$ preserves
finite limits, it carries the equivalence between the groupoid and the
\v{C}ech nerve of its quotient to the corresponding equivalence in
$\mathcal D$. For stability write
$S=|iR_\bullet|$ and take $Z\to aS$ in $\mathcal D$. In $\mathcal E$
the pulled-back diagram realizes to $S\times_{iaS}iZ$, by base
change of $iR_\bullet$ along $S\times_{iaS}iZ\to S$. Reflecting its
realization gives $aS\times_{aS}Z\simeq Z$. Thus $\mathcal D$ is exact, and $a$
preserves effective epimorphisms.

Both $a$ and $i$ preserve finite limits and therefore truncation.
Boundedness of $\mathcal E$ implies boundedness of $\mathcal D$.
For $X\in\mathcal D$, choose a cover $A\twoheadrightarrow iX$
by a zero-type $A$ in $\mathcal E$. Applying $a$ gives a cover
$aA\twoheadrightarrow X$ by the zero-type $aA$ in $\mathcal D$.
Hence $\mathcal D$ is zero-complicial. Recognition applies to both
$\mathcal E$ and $\mathcal D$. Under the resulting equivalences, the
reflector is the regular extension of
$a_{\leq0}:\mathcal E_{\leq0}\to\mathcal D_{\leq0}$, since $a$ is
regular and the exact-completion universal property determines that
extension. Its fully faithful right adjoint is then determined by
adjunction. Taking the ordinary reflection to be $j$-sheafification
gives the second equivalence.\qed
\end{proof}

\begin{corollary}[Slicing over an ordinary base]
\label{cor:operations-ordinary-slice}
Let $\mathcal T$ be an ordinary exact category and let $B\in\mathcal T$.
Then
\[
 \operatorname{Ex}_{\infty}(\mathcal T/B)
       \simeq\operatorname{Ex}_{\infty}(\mathcal T)/B.
\]
\end{corollary}

\begin{proof}
The slice on the right is bounded and exact. An original atlas
$A\twoheadrightarrow X$ of an object over $B$ is an atlas in the slice;
because $A$ and $B$ are zero-truncated, $A\to B$ is zero-truncated as an
object of the slice. Its zero-type category is $\mathcal T/B$.
Apply the recognition theorem.\qed
\end{proof}

\begin{remark}[Atlas dimension over a higher base]
\label{rem:operations-higher-base}
For a nonzero abelian group $G$, consider the basepoint object
$*\to K(G,2)$ in bounded spaces over $K(G,2)$. Since the base is
connected and simply connected, every zero-truncated family is a
constant local system $S\times K(G,2)\to K(G,2)$. A map from this
family to the basepoint object assigns to each $s\in S$ a
nullhomotopy of the identity of $K(G,2)$. Since the identity induces
the nonzero map $\mathrm{id}_G$ on $\pi_2$, this forces $S=\varnothing$.
The basepoint object has fibre $K(G,1)$ at the basepoint and supplies
its own atlas of height one. Thus one is the minimum atlas height.
Theorem~\ref{thm:general-context-dimension} extends this loop-space
calculation to higher degrees.
\end{remark}

\subsection{Exchange of localization and slicing}

\begin{theorem}[Slice localization over an arbitrary base]
\label{thm:operations-exchange}
Let $a:\mathcal E\rightleftarrows\mathcal D:i$ be a reflective
inclusion of finitely complete $\infty$-categories with $a$ left exact,
and let $B\in\mathcal E$. There is a left-exact reflective localization
\[
 a_B:\mathcal E/B\rightleftarrows\mathcal D/aB:r_B
\]
given by
\[
 a_B(X\to B)=(aX\to aB),\qquad
 r_B(Y\to aB)=(B\times_{iaB}iY\to B).
\]
The local maps $p:X\to B$ are exactly those for which the naturality
square of the unit is cartesian, equivalently
\[
 X\longrightarrow B\times_{iaB}iaX
\]
is an equivalence.

If $\mathcal E$ is bounded and $a$ is a topological $j$-sheafification,
write $j_B$ for its induced
slice topology. A monomorphism over $B$ is $j_B$-dense precisely when
its underlying monomorphism is $j$-dense. Then
\[
 \operatorname{Sh}_{j_B}(\mathcal E/B)
       \simeq\operatorname{Sh}_{j}(\mathcal E)/aB.
\]
For $B\in\mathcal D$ this specializes to
\[
 (\operatorname{Sh}_{j}(\mathcal E))/B
       \simeq\operatorname{Sh}_{j_B}(\mathcal E/iB),
\]
and the local objects on the right are precisely the maps to $iB$ with
local total space. For an ordinary Lawvere--Tierney operator, the induced
operator on the slice subobject classifier is
$j\times\operatorname{id}_B:\Omega\times B\to\Omega\times B$.
\end{theorem}

\begin{proof}
For $X\to B$ and $Y\to aB$, the adjunction $a\dashv i$ and the
universal property of pullback give
\[
\begin{aligned}
 \operatorname{Map}_{\mathcal D/aB}(aX,Y)
 &\simeq
 \operatorname{Map}_{\mathcal E/iaB}(X,iY)\\
 &\simeq
 \operatorname{Map}_{\mathcal E/B}
       (X,B\times_{iaB}iY).
\end{aligned}
\]
Here $X\to iaB$ is the composite of $X\to B$ with the unit of $B$.
This proves the adjunction. Its counit is an equivalence, since
left exactness gives
\[
 a(B\times_{iaB}iY)
       \simeq aB\times_{aB}Y\simeq Y.
\]
Thus $r_B$ is fully faithful. The unit is the displayed comparison,
so its invertibility gives the stated characterization of local maps.
Left exactness of $a_B$ follows from that of $a$ and the slice formulas
for finite limits.

For the topological assertion call a slice object local if it is
orthogonal to the monos whose underlying monos are $j$-dense.
Every $r_BY$ has this property by adjunction, and these orthogonality
conditions are closed under finite limits. We claim that a map $f$
between local slice objects inverted by $a_B$ is already invertible.
Its relative truncation degree is finite by boundedness. If $f$ is
monic, its underlying mono is inverted by $a$, so $f$ is dense.
Restriction along $f$ is then an equivalence on maps into either
endpoint, which gives an inverse. At a higher degree, $\Delta f$ is
between local objects, has lower relative degree, and is inverted by
left exactness. Induction makes $\Delta f$ invertible, and the mono
case applies to $f$. For a local $X$ apply the claim to the unit
$X\to r_Ba_BX$. This identifies intrinsic slice locality with the
image of $r_B$. When $B$ is local its unit is an equivalence and
$r_B$ reduces to inclusion.

Finally, the characteristic predicate of a monomorphism in a slice
is its ordinary characteristic predicate together with the map to
$B$, so the slice closure operator is $j\times\operatorname{id}_B$.\qed
\end{proof}

\subsection{A normal form for slicing and localization by ambient topologies}

\begin{theorem}[Normal form for descended localizations and slices]
\label{thm:operations-normal-form}
Let $\mathcal E$ be a bounded finitely complete $\infty$-category.
Consider a finite sequence of the following operations, starting
from $\mathcal E$. Assume each localization in the sequence has a
topological reflector, as provided for the nuclei of
Theorem~\ref{thm:bounded-reflection}.
\begin{enumerate}
\item slicing by an arbitrary object of the current category;
\item localizing an unsliced ambient category by a topology;
\item after slicing, localizing by a topology pulled back from the
      current unsliced ambient category.
\end{enumerate}
The sequence determines a closure operator $j$, a context $B$, and
a canonical equivalence from the resulting category to
\[
 \operatorname{Sh}_{j}(\mathcal E)/B,
 \qquad B\in\operatorname{Sh}_{j}(\mathcal E).
\]
Here $j$ is the pullback-stable closure operator on subobjects
induced by the composite ambient localization. If $\mathcal E$ has
a subobject classifier $\Omega$, this operator is represented by
a Lawvere--Tierney nucleus $j:\Omega\to\Omega$.
For $\mathcal E=\operatorname{Ex}_{\infty}(\mathcal T)$, if the
bounded $j$-sheafification has the prescribed ordinary zero-types,
the same category is
\[
 \operatorname{Ex}_{\infty}
       (\operatorname{Sh}_{j}(\mathcal T))/B.
\]
\end{theorem}

\begin{proof}
Maintain a presentation $\mathcal E_j/B$, with
$\mathcal E_j=\operatorname{Sh}_j(\mathcal E)$. Initially take the
identity topology and $B=1$. Slicing again by an object $C\to B$
replaces this presentation by $\mathcal E_j/C$, by associativity of
slicing. If $k$ is a topology on $\mathcal E_j$, localizing
$\mathcal E_j/B$ by its induced slice topology gives
\[
 \operatorname{Sh}_{k}(\mathcal E_j)/a_kB
\]
by Theorem~\ref{thm:operations-exchange}. The composite of the two
reflectors has as local objects those orthogonal to the $j$-dense
monos and to the images under $i_j$ of the $k$-dense monos of
$\mathcal E_j$. The first condition puts an object in $\mathcal E_j$,
where full faithfulness identifies the second with $k$-locality.
Both classes consist of monos, so this is a topological localization.
Explicitly its closure of $P\hookrightarrow C$ is the pullback along
$C\to i_ja_jC$ of $i_j$ applied to the $k$-closure of
$a_jP\hookrightarrow a_jC$. This formula follows by pulling back the
composite unit square for the mono and is stable under substitution.
Write $j'$ for this closure operator. When $\Omega$ exists, the
characteristic map of the $j'$-closure of the generic mono
$1\hookrightarrow\Omega$ represents $j'$ by pullback stability.
The updated pair is $(j',a_kB)$. An unsliced localization
is the special case $B=1$. Induction proves the first assertion.
Every comparison used is the natural comparison supplied by a
pullback universal property or a reflective adjunction; hence the
comparisons respect compositions and identities with their canonical
coherences. Proposition~\ref{prop:operations-reflection-reconstruction}
gives the second displayed description.\qed
\end{proof}

\begin{remark}[Topologies on a two-component slice]
In $\mathbf{Set}/2\simeq\mathbf{Set}\times\mathbf{Set}$ a reflective
localization can act independently on the two components. For example,
one localization fixes the first factor and sends the second to the
terminal category. The topologies induced from $\mathbf{Set}$ in
Theorem~\ref{thm:operations-normal-form} act identically on the two
components. The identity topology fixes both factors, and the chaotic
topology sends both to the terminal category.
\end{remark}

\subsection{Closed restriction and open observations}

In nested realizability a single subterminal determines the open
and closed subtoposes. To compare observations in these two models,
consider, in any of our elementary models, a subterminal $U$ and
the nuclei $o_U(P)=(U\to P)$ and $c_U(P)=U\vee P$.
Testing open locality at $P=U$ gives $X\simeq X^U$ by the
constant-map comparison. Conversely, whenever $U\to P$ the projection
$U\times P\to U$ is an equivalence. Exponential adjunction then
makes the constant-map comparison $X^U\to(X^U)^P$ an equivalence.
Thus this condition characterizes open-local objects.

Testing closed locality at $P=\bot$ gives $U\to\isContr(X)$,
equivalently $X\times U\to U$ is an equivalence. Conversely, given
this condition and $U\vee P$, it suffices to prove the proposition
that $X\to X^P$ is an equivalence. Under $P$ the proposition $P$
is contractible, so evaluation inverts the comparison.
Under $U$ both $X$ and $X^P$ are contractible.
Disjunction elimination gives closed locality. Both criteria hold
after any change of parameter.

\begin{proposition}[Closed-to-open constancy]
\label{prop:operations-open-closed}
Let $\mathcal E$ be locally cartesian closed and let $U\hookrightarrow1$
be subterminal. Suppose $C\times U\simeq U$ and the canonical map
$O\to O^U$ is an equivalence. The constant-map comparison is an
equivalence
\[
 O\longrightarrow O^C.
\]
In particular,
\[
 \operatorname{Map}_{\mathcal E}(C,O)
       \simeq\operatorname{Map}_{\mathcal E}(1,O).
\]
The assertion holds in every slice with its corresponding
subterminal.
\end{proposition}

\begin{proof}
Exponential adjunction and the two hypotheses give natural
equivalences
\[
 O^C\simeq(O^U)^C\simeq O^{U\times C}\simeq O^U\simeq O.
\]
The inverse composite is induced by $C\to1$ and is therefore the
constant-map comparison. Apply global sections, or repeat the same
calculation in a slice.\qed
\end{proof}

\begin{remark}[Exponentials under closed reflection]
The arrow topos gives an explicit calculation. Take
$\mathcal E=\mathbf{Set}^{[1]}$ and $U=(\varnothing\to1)$. Its
closed subcategory consists of arrows $S\to1$, and the reflector is
\[
 a(A\to B)=(A\to1).
\]
Take $X=(\varnothing\to1)$ and $Y=(\varnothing\to2)$. The domain
component of $Y^X$ is the two-element set of commuting squares from
$X$ to $Y$, so $a(Y^X)$ corresponds to $2$. Computing the exponential
after reflection gives $(aY)^{aX}=\varnothing^{\varnothing}=1$ under
the same identification. For maps between closed objects we can
compute the dependent product in the ambient category. After pullback
to $U$ these maps are equivalences, and so are their dependent
products. The ambient product is therefore closed and represents the
dependent product in the closed subcategory.
\end{remark}

\section{The unified models and contextual choice}
\label{sec:unified-theorem}

An ordinary realizability topos, a nucleus on its propositions, and
a context of parameters together specify a higher model. Its dependent
products characterize the indices for which local witnesses in each
fibre assemble, after a cover of the parameter, into a coherent section.

\begin{definition}[Localized higher realizability in a context]
For an admissible presentation $\mathcal C$, a nucleus $j$ on
$\mathcal T=\operatorname{Ex}_{1,\mathrm{reg}}\mathcal C$, and a
bounded object $B$ of the localized completion, define
\[
 \mathfrak R(\mathcal C,j;B)
 =\operatorname{Sh}_j(\operatorname{Ex}_\infty\mathcal C)_{/B}.
\]
We also write
$\mathfrak R(\mathcal T,j;B)=\mathsf H(\operatorname{Sh}_j\mathcal T)/B$,
where $\mathsf H(\mathcal T)=\operatorname{Ex}_\infty\mathcal T$.
This expression describes the same slice directly in terms of the
ordinary topos.
\end{definition}

The ordinary classifier $\Omega$ also classifies subobjects in higher
contexts, since characteristic maps descend along covers by zero-types.
Thus the same nucleus acts in every context.

\begin{theorem}[Unified construction and higher-tripos presentation]
\label{thm:unified}
Let $(\mathcal C,j;B)$ be as above, put
$\mathcal T_j=\operatorname{Sh}_j\mathcal T$, and suppose $B$ is
$s$-truncated, with $s\geq0$. Set $d=\max\{0,s-1\}$. Then:
\begin{enumerate}
\item The category $\mathfrak R(\mathcal C,j;B)$ is a bounded
exact elementary $\infty$-topos with a cumulative closed hierarchy
of univalent universes. Its universe at level $\lambda$ is the
pullback of the local universe to $B$.
\item There are canonical equivalences
\[
 \mathfrak R(\mathcal C,j;B)
 \simeq\mathsf H(\mathcal T_j)/B
 \simeq\operatorname{Ex}_\infty(\mathcal C^{(d)}_{j,B}),
 \quad
 \mathcal C^{(d)}_{j,B}:=(\mathsf H(\mathcal T_j)/B)_{\leq d}.
\]
The category $\mathcal C^{(d)}_{j,B}$ is a regular $(d+1,1)$-category
and its codomain doctrine admits higher-tripos structure with this
completion.
\item Equivalent ordinary presentations, corresponding nuclei,
and transported contexts give equivalent higher models. Selected
universe classes are transported along the equivalence.
\item Pullback along maps of contexts supplies substitution,
with dependent sums and products and their Beck--Chevalley
equivalences. The exchange comparisons of
Theorem~\ref{thm:operations-exchange} and the normalization of
Theorem~\ref{thm:operations-normal-form} are compatible with
these substitutions.
\end{enumerate}
\end{theorem}

\begin{proof}
Proposition~\ref{prop:input-completion} constructs the base model;
Theorem~\ref{thm:bounded-reflection} constructs its localization
with local universes. Corollary~\ref{cor:ordinary-localization}
and Proposition~\ref{prop:operations-reflection-reconstruction}
identify it with $\mathsf H(\mathcal T_j)$. Slices inherit finite
limits, local cartesian closure, exactness, and disjoint finite
coproducts. If $u^j_\lambda$ is a local universal family, then
\[
 B\times\widetilde U^j_\lambda\longrightarrow
 B\times U^j_\lambda
\]
over $B$ classifies the corresponding families in the slice.
Classification, univalence, and closure of the levels follow by
pullback. Finite truncation of total objects and of $B$ bounds
every object of the slice.

Write $\mathcal F=\mathsf H(\mathcal T_j)$. It is
zero-complicial. For $Y\to B$ choose a cover $P\cover Y$ by a
zero-type $P$. For the composite $f:P\to B$, the fibre over $b:B$ is
$\sum_{x:P}(f(x)=b)$. Its base $P$ is a zero-type and its path fibres
are $(s-1)$-truncated, so $f$ is $d$-truncated.
Thus $\mathcal F/B$ is $d$-complicial. The $d$-types in the slice
are closed under finite limits. The image of a map between $d$-types
is a subobject of its target, so it too is a $d$-type. The subcategory
therefore inherits pullback-stable image factorizations from
$\mathcal F/B$ and is regular. Boundedness and $d$-compliciality now give
the asserted completion \cite{Stefanich2023Exact}.

For a section problem $Z\to Y\xrightarrow{f}T$ in this completion
over $T\in\mathcal C^{(d)}_{j,B}$, take a cover $W\cover\Pi_fZ$
by a $d$-type $W$ and restrict evaluation to $W\times_TY$.
The transpose of any parameterized section pulls this cover back
to a cover on which the section is represented. For a parameter
in the source, cover this pullback by a $d$-type to obtain
a source cover. This supplies coherent fullness at every finite
height \cite{Maruyama2026}. For each chosen
universe in the slice choose a cover $V_\lambda\cover U_\lambda$
by a $d$-type and pull back its universal family to $V_\lambda$.
Its relative height stays finite. The recursion on kernel pairs in
Section~\ref{sec:input} therefore gives a finite atlas presentation
of its total space over this original code base.
Given a classifying map from another source object, its pullback
of this code cover admits a further cover by a $d$-type. The composite
is a source cover. Hence the generated local class is exactly
the desired universe class. These presentations are directed and
exhaustive by slice classification. The five closures in
Definition~\ref{def:admissible} hold for the slice classes, so
the generic identity, image, diagonal, composite and dependent
product admit source-local codes
\cite{Maruyama2026}. The codomain doctrine with these
data therefore gives the stated higher-tripos presentation.

Presentation independence is
Proposition~\ref{prop:operations-source-independence}, followed
by the natural equivalence of localized ordinary toposes and
their completions. An equivalence preserves dependent products by
their universal property and transports universes by their classes
of small maps.
Substitution is ordinary pullback in slices. Its adjoints are
composition and dependent product. Their Beck--Chevalley
comparisons are induced by the pullback universal property.
The relative localization comparisons are induced by the same
pullbacks and adjunction units, so they agree coherently.\qed
\end{proof}

Section~\ref{sec:higher-context-geometry} identifies ordinary sources for
slices over ordinary parameters and over group classifying types.
For a slice over $K(A,n)$ with $A\ne 0$ and $n\ge 2$, the minimum
regular source dimension is exactly $n$
(Corollary~\ref{cor:context-transition}).

\subsection{Computational presentations and their localizations}

For a discrete partial combinatory algebra $A$ and a subalgebra $A_\#$
with common combinators and inherited defined application, write
$\mathcal N(A,A_\#)$ for the nested realizability topos,
$\mathcal T^r(A,A_\#)$ for its relative open subtopos, and
$\mathcal T^m(A,A_\#)$ for its modified relative closed subtopos.
For ordered inputs write $\operatorname{RT}(A',A)$ for the relative
topos of a filtered OPCA. Table~\ref{tab:examples} gives the
specializations used below. Here $\mathcal K_1$ and $\mathcal K_2$
are Kleene's first and second algebras, respectively.

\begin{table}[!hb]
\caption{Realizability models in the form $\mathfrak R(\mathcal T,j;B)$.
The first six models use the terminal context. The effective slices use $K(A_r,n)$.}
\label{tab:examples}
\begin{tabularx}{\textwidth}{@{}p{0.22\textwidth}X X@{}}
\toprule
Model & Ordinary topos and nucleus & Higher model \\
\midrule
Effective & $\mathcal T=\mathsf{Eff}$, $j=\mathrm{id}$
 & $\mathsf H(\mathsf{Eff})$ \\
Function & $\mathcal T=\operatorname{RT}(\mathcal K_2)$,
 $j=\mathrm{id}$ & $\mathsf H(\operatorname{RT}(\mathcal K_2))$ \\
Relative & $\mathcal T=\operatorname{RT}(A',A)$,
 $j=\mathrm{id}$ & $\mathsf H(\operatorname{RT}(A',A))$ \\
Nested & $\mathcal T=\mathcal N(A,A_\#)$,
 $j=\mathrm{id}$ & $\mathsf H(\mathcal N(A,A_\#))$ \\
Modified & $\mathcal T=\mathsf{Eff}_2$,
 $j=c_u$, $c_u(\Phi)=u\vee\Phi$
 & $\mathsf H(\mathsf{Mod})$ \\
Krivine & $\mathcal T=\operatorname{RT}(A',A)$,
 $j=b_U$\newline $b_U(P)=(P\Rightarrow U)\Rightarrow U$
 & $\mathsf H(\operatorname{Sh}_{b_U}\mathcal T)$ \\
Effective slice & $\mathcal T=\mathsf{Eff}$,
 $j=\mathrm{id}$, $B=K(A_r,n)$
 & $\mathsf H(\mathsf{Eff})/K(A_r,n)$ \\
\bottomrule
\end{tabularx}
\end{table}

\paragraph{Function realizability.}
Set $A=A'=\mathcal K_2$ with discrete order in the relative
assembly construction and choose $\kappa_0>|\mathbb N^{\mathbb N}|$.
The identity-localized case of Theorem~\ref{thm:unified}
gives the higher function model. Its original Baire and numerical
objects retain all their maps by full faithfulness. With the usual
Baire assembly, these are exactly the continuous maps
$\mathbb N^{\mathbb N}\to\mathbb N$ and
$\mathbb N^{\mathbb N}\to\mathbb N^{\mathbb N}$.
In this absolute model every Baire function is available as a
realizer. Section~\ref{sec:function-semantics} uses continuity to
determine the subsets of Baire space recognized by identity with a
fixed code.

\paragraph{Nested and modified realizability.}
Write a nested truth value as $(P,P_a)$, with potential realizers
$P\subseteq A$ and actual realizers $P_a\subseteq A_\#\cap P$.
For subsets $P,Q\subseteq A$, write $P\Rightarrow Q$ for the set
of $r\in A$ such that $r\cdot a$ is defined and belongs to $Q$ for
every $a\in P$. Implication between nested truth values is then
\[
 (P,P_a)\Rightarrow(Q,Q_a)
 =\bigl(P\Rightarrow Q,
 A_\#\cap(P\Rightarrow Q)\cap(P_a\Rightarrow Q_a)\bigr),
\]
and entailment uses one actual realizer uniformly over the index
set. On a nested truth value $\Phi$, the subterminal $u=(A,\varnothing)$
determines the open nucleus $o_u(\Phi)=u\Rightarrow\Phi$ and the closed
nucleus $c_u(\Phi)=u\vee\Phi$. Their ordinary sheaf toposes are respectively
$\mathcal T^r$ and $\mathcal T^m$
\cite{BirkedalVanOosten2002}.
The displayed formulas use the potential--actual presentation of Maschio and Streicher
\cite{MaschioStreicher2015}. Applying our comparison between sheafification
and higher exact completion (Proposition~\ref{prop:operations-reflection-reconstruction})
to these two ordinary localizations gives
\[
 \operatorname{Sh}_{o_u}\mathsf H(\mathcal N)
 \simeq\mathsf H(\mathcal T^r),\qquad
 \operatorname{Sh}_{c_u}\mathsf H(\mathcal N)
 \simeq\mathsf H(\mathcal T^m).
\]
For $A=A_\#=\mathcal K_1$, write $\mathsf{Eff}_2$ for the ordinary
nested realizability topos over the Sierpi\'nski base. Its open and
closed subtoposes are $\mathsf{Eff}$ and the modified realizability
topos $\mathsf{Mod}$, respectively. Tarantino recovers this pair of
ordinary toposes using arrow algebras~\cite{Tarantino2025}. The displayed equivalences
identify their higher completions inside $\mathsf H(\mathsf{Eff}_2)$.

By the sheaf criterion preceding
Proposition~\ref{prop:operations-open-closed}, a closed object is
characterized by $X\times u\simeq u$. Its universe formula becomes
\begin{equation}
 U^c_\lambda
 =\sum_{v:U_\lambda}
      \prod_{z:u}\operatorname{isContr}(\operatorname{El}(v)).
 \label{eq:closed-universe-specialization}
\end{equation}
Over $u$, the code of $\one$ supplies a centre for the type of
contractible small types. Univalence identifies paths to this centre
with equivalences to $\one$, contractible for each such type.
The tautological total space is also contractible there, being a
contractible family over a contractible base. Thus both are closed,
and this restricted family
classifies the ambient-small maps between closed objects.
Proposition~\ref{prop:operations-open-closed} supplies the
closed-to-open constancy theorem at every higher degree.

\paragraph{Krivine realizability.}
Fix an abstract Krivine structure in the sense of Streicher~\cite{Streicher2013}
and assume its tripos is nondegenerate. The van Oosten--Zou representation
supplies a filtered OPCA $(A,A')$ and a downset
of $A$ disjoint from $A'$, representing a proper subterminal $U$ in
its relative topos. The Krivine tripos is equivalent to the relative
Booleanization for $b_U(P)=((P\Rightarrow U)\Rightarrow U)$
\cite{VanOostenZou2016}. The fixed propositions
form a Boolean algebra whose bottom is $b_U(\bot)=U$. We call this
subterminal the \emph{logical pole}.
Thus the same higher construction gives
\[
 \operatorname{Sh}_{b_U}\mathsf H(\operatorname{RT}(A',A))
 \simeq
 \mathsf H\bigl(\mathbf{Set}[P_{\mathrm{Krivine}}]\bigr).
\]
Here $P_{\mathrm{Krivine}}$ is the tripos of the chosen Krivine
structure, and $\mathbf{Set}[P_{\mathrm{Krivine}}]$ is the ordinary
topos obtained from it by the tripos-to-topos construction.
In a nondegenerate localized model the canonical two-element
abelian group is nonzero. Its finite deloopings give nontrivial
types in all positive degrees. Section~\ref{sec:universal-probe}
recovers $b_U$, hence $U$, from one higher family.

\paragraph{Uniform computational observations.}
A set-valued observation takes the same value on equivalent small
types. In an assembly presentation, this invariance can be tested
on a single tracked map between the original assemblies.

\begin{proposition}[Observation through a univalent code quotient]
\label{prop:observations}
Let $p:X\to I$ be a family and let $c:I\cover U$ exhibit its
univalent completion, so the \v{C}ech relation of $c$ is the internal
space of fibrewise equivalences of $p$. For every zero-type $D$,
maps $U\to D$ are exactly maps $g:I\to D$ satisfying
\[
 \prod_{i,j:I}\bigl(\operatorname{Equiv}(X_i,X_j)
                     \to(g(i)=g(j))\bigr).
\]
This correspondence is an equivalence of spaces. For original
assemblies $I,D$ in the unlocalized completion, full faithfulness
identifies the maps $g$ with the original tracked maps $I\to D$.
\end{proposition}

\begin{proof}
Mapping out of the effective quotient $c$ gives the totalization
of its \v{C}ech mapping diagram. Since $D$ is a zero-type, equality of
the two restrictions to the arrow object supplies all higher
compatibilities uniquely. This arrow object is the fibre-equivalence
object, so the descent condition is the one displayed above.
Full faithfulness of the assembly unit
identifies the maps between original objects.\qed
\end{proof}

\subsection{Characterizing contextual choice in the ordinary core}

A set-valued observation descends when its outputs agree. A section
of a varying family also carries paths identifying its values under
transport in the index $I$, together with their higher coherences.
Which indices turn local existence in every fibre into local existence
of such a section? In the completion $\mathcal F=\HH(\mathcal T_j)$,
covers by zero-types give a characterization in the ordinary core.
Write $\|Y\|_q$ for relative $q$-truncation, with $\|Y\|_{-1}$ the
image of the structure map. A term of $\|Y\|_{-1}$ expresses local
existence of a section of $Y$. Directedness places each family
and its presenting atlas in a common universe level.

\begin{definition}[Contextual choice]
An index $I\in\mathcal F$ admits \emph{contextual higher choice} if, for
every $C\in\mathcal F$ and every family $Y$ over $C\times I$, the map
\[
 \|\Pi_pY\|_{-1}\longrightarrow\Pi_p\|Y\|_{-1},
 \qquad p:C\times I\to C,
\]
is an equivalence over $C$. A zero-type $I$ admits \emph{ordinary
contextual choice} if, for each zero-type $C$ and each ordinary
regular epimorphism $V\cover C\times I$, the ordinary section object
$\Pi_pV$ covers $C$.
\end{definition}

\begin{theorem}[Choice is determined by the ordinary core]
\label{thm:choice-indices}
In a bounded higher model with covers by zero-types, an index $I$ admits
contextual higher choice if and only if it is a zero-type admitting
ordinary contextual choice in the core. These conditions are also
equivalent to each of the following:
\begin{enumerate}
\item every $\Pi_{C\times I\to C}$ preserves covers;
\item for all $C,Y$ as above and all finite $q\geq-1$,
$\|\Pi_pY\|_q\longrightarrow\Pi_p\|Y\|_q$ is an equivalence.
\end{enumerate}
Here the ordinary covers and dependent products of zero-type families
are computed in the higher model.
\end{theorem}

\begin{proof}
Suppose higher choice holds and let $Y\cover Z$ be a cover over
$C\times I$. Pull it back along the evaluation map
$(\Pi_pZ)\times I\to Z$. Each fibre of the resulting family of lifts
is locally inhabited. Choice in the parameter context $\Pi_pZ$ makes
the section object of this family cover $\Pi_pZ$. This says exactly that $\Pi_pY\to\Pi_pZ$
is a cover. Conversely, apply preservation of covers to
$Y\cover\|Y\|_{-1}$. Its target after $\Pi_p$ is propositional, so
it is the image of $\Pi_pY$.

A right adjoint preserves finite limits. Induction on diagonals
therefore shows that a cover-preserving $\Pi_p$ preserves
$q$-connected maps for finite $q\geq-1$. The base case is preservation of
covers. For $q\geq0$, a map is $q$-connected precisely when it is a cover
and its diagonal is $(q-1)$-connected. The functor $\Pi_p$ preserves
$q$-truncated maps by the diagonal criterion as well. Applying it
to the truncation unit gives a $q$-connected map to a $q$-truncated
object, the universal relative truncation. Taking $q=-1$ gives
the converse implication.

Choose a cover $A\cover I$ by a zero-type. Higher choice makes its section
object cover $\one$. After that cover the universal section exhibits
the pullback of $I$ as a retract of the pullback of $A$ in the
parameter slice. Truncation is stable under retracts and descends
along covers, so $I$ is a zero-type.
Restriction to ordinary parameters gives ordinary contextual choice.

Conversely suppose $I$ is such a zero-type. To test $\Pi_p(Y\cover Z)$
choose a cover $P\cover\Pi_pZ$ by a zero-type. Evaluate on $P\times I$,
pull back the cover and cover the resulting lift object by a zero-type
$R$. The composite $R\cover P\times I$ is a regular epimorphism in
the ordinary core. Ordinary choice gives a cover
$S=\Pi_{P\times I\to P}R\cover P$. Evaluation now supplies a lift
$S\to\Pi_pY$ over $\Pi_pZ$. Since $S\to\Pi_pZ$ is a cover, so is
$\Pi_pY\to\Pi_pZ$. This proves the assertion for every higher $C$.
The hypotheses on the ordinary operations hold in $\mathcal F$:
the full core is closed under the indicated products, images
and finite limits, and a map between sets is a cover
precisely when its image is the whole target.
\qed
\end{proof}

\begin{corollary}[Projective indices and the finite-height test]
\label{cor:projective-choice}
If the ambient model has a finite-product-closed covering class of
regular projectives containing $\one$, then higher choice holds
exactly at its regular-projective indices. More generally, under
the assumptions of Theorem~\ref{thm:choice-indices}, for $k\geq0$
contextual choice for $k$-truncated families over $C\times I$,
in every parameter context $C$, holds
exactly when $\tau_{\leq k+1}I$ is a zero-type and $\tau_{\leq0}I$
has ordinary contextual choice. Choice for proposition-valued
families holds for every $I$.
\end{corollary}
\begin{proof}
If $I$ is regular-projective, it is a retract of a member of the
covering class. Thus $P\times I$ is projective for each member $P$.
The preceding evaluation test for covers can now be solved by
lifting on $P\times I$ itself. Conversely higher choice gives mere
sections of every cover onto $I$. Projectivity of $\one$ turns them
into actual sections. Pulling a cover back along an arbitrary map
out of $I$ proves regular projectivity.

For the finite-height assertion put $J=\tau_{\le k+1}I$.
The classifier of a fixed level of $k$-types is a $(k+1)$-type.
Hence every such family descends along $C\times I\to C\times J$.
By connected--truncated orthogonality in the slice,
pullback preserves the space of sections of the descended
family. The same holds for its pointwise propositional
truncation. It suffices to test $J$. A cover of this
$(k+1)$-type by a zero-type has $k$-truncated fibres, so the choice
test forces $J$ to be a set by the retract argument. Testing ordinary
covering families gives ordinary choice on $J\simeq\tau_{\le0}I$.
Conversely the theorem gives full choice on that index. For a
proposition-valued family both sides of the comparison already
agree by idempotence of propositional truncation.\qed
\end{proof}

\section{Recovering and comparing Lawvere--Tierney nuclei}
\label{sec:universal-probe}

A section of the based path family $x\mapsto(b=x)$ contracts its
base to $b$. We construct a pointed type for each predicate $P$
whose locus of contraction is exactly $P$. Localizing this family
then computes the closure of $P$ in the same parameter context.

Let $\mathcal E$ be an exact locally cartesian closed $\infty$-category
with disjoint finite coproducts and a subobject classifier $\Omega$.
Suppose that a Lawvere--Tierney nucleus $j$ admits a left-exact
reflection $a_j:\mathcal E\rightleftarrows\mathcal F_j:i_j$ whose
induced universal closure operator is $j$.
The bounded elementary models constructed above satisfy these
hypotheses. The construction below encodes an intuitionistic predicate
as equality of two elements of a quotient group.

\paragraph{Relative localization.}
The predicate and its closure are subobjects of the same parameter
$B$. To compare the original and reflected families over $B$, use
the slice adjunction of Theorem~\ref{thm:operations-exchange}. Its
left adjoint is
\[
 a_{j,B}:\mathcal E_{/B}\longrightarrow(\mathcal F_j)_{/a_jB},
 \qquad (X\to B)\longmapsto(a_jX\to a_jB)
\]
and its fully faithful right adjoint is
\[
 r_{j,B}(Y\to a_jB)
   =\bigl(B\times_{i_ja_jB}i_jY\to B\bigr).
\]
Write
\[
 \ell^j_B=r_{j,B}a_{j,B},\qquad
 \ell^j_BX=B\times_{i_ja_jB}i_ja_jX.
\]
Both $a_{j,B}$ and $r_{j,B}$ preserve finite limits. For every
$f:C\to B$, left exactness gives the coherent base-change identity
\begin{equation}
 f^*\ell^j_BX\simeq\ell^j_Cf^*X.
 \label{eq:probe-modal-base-change}
\end{equation}
For a mono $P\hookrightarrow B$, $\ell^j_BP$ is its
$j$-closure $j_B(P)\hookrightarrow B$.

Write $A\dashv I$ for $a_{j,B}\dashv r_{j,B}$. For maps $p:Y\to Z$
and $q:W\to Y$ in its target, form the ambient dependent product
$V=\Pi_{Ip}(IW)\to IZ$. For every $T\to IZ$, left exactness of $A$
and the adjunction give
\begin{align*}
 \operatorname{Map}_{/IZ}(T,V)
 &\simeq\operatorname{Map}_{/IY}(T\times_{IZ}IY,IW)\\
 &\simeq\operatorname{Map}_{/Y}(AT\times_ZY,W).
\end{align*}
Since $A$ inverts the unit $T\to IAT$ over $IZ$, restriction along the
unit induces an equivalence on these mapping spaces.
At $T=V$ the identity therefore extends to a retraction $IAV\to V$
of the unit in the slice. Local objects are closed under retracts,
so $V$ is local. Restricting the same equivalence to local $T$ shows
that $AV$ is $\Pi_pW$, with $I\Pi_pW\simeq V$. This proves local
cartesian closure and preservation of these products by $I$.
Dependent sums between local objects
are compositions, so they are preserved as well. Consequently
\begin{equation}
 r_{j,B}\operatorname{isContr}_{(\mathcal F_j)_{/a_jB}}(Y)
 \simeq
 \operatorname{isContr}_{\mathcal E_{/B}}(r_{j,B}Y).
 \label{eq:probe-contr-inclusion}
\end{equation}

\subsection{Finite deloopings and section spaces}

In the following arguments a type over $C$ means an object of the
slice, and its identity family is its relative diagonal. We write
$\underline{\Map}$ and $\underline{\Hom}$ for internal objects and
$\Map$ and $\Hom$ for external mapping spaces and ordinary homomorphism
sets, respectively. Taking global sections of the internal mapping
object recovers the external mapping space.

\begin{lemma}[Finite bars and their internal maps]\label{lem:finite-bars}
In a parameter slice let $A,M$ be abelian group zero-types.
Put $K(A,0)=A$ and let $K(A,n+1)$ be the effective quotient
of the one-object groupoid with arrow group $K(A,n)$.
For $n\geq1$ this gives a pointed $(n-1)$-connected $n$-type with
$\Omega K(A,n)=K(A,n-1)$. For $n\geq1$, there are natural equivalences
\[
\begin{aligned}
 \underline{\Map}_*(K(A,n),K(M,n))&\simeq
              \underline{\Hom}_{\Ab}(A,M),\\
 \underline{\Map}(K(A,n),K(M,n))&\simeq
              K(M,n)\times\underline{\Hom}_{\Ab}(A,M).
\end{aligned}
\]
For equivalences the second factor is the object of coefficient
isomorphisms. The equivalences are stable under change of parameter.
Left exact functors preserving effective groupoid quotients commute
with the displayed construction of $K$.
\end{lemma}
\begin{proof}
Commutativity supplies the abelian group structure on each successive
bar by passage through finite products. The basepoint map is a
cover, and effectiveness identifies its fibre over the
basepoint with the preceding group. This gives the
loop identity and, by induction using connectivity and diagonals, the
truncation and connectivity assertions. Over a further test $T$, a
pointed map is a pointed map in the slice over $T$. Embed that exact
slice fully and with effective quotients in its category of regular
sheaves of spaces. For $n=1$ the equivalence between groups and
pointed connected one-types identifies the space of pointed maps with
the discrete set of homomorphisms. For $n\geq2$ use the pointed
Eilenberg--MacLane equivalence
\cite{Lurie2009HTT}. Full faithfulness and
finite limits identify the groups, maps and pointed realizations in
this calculation with those in the original slice. Testing all $T$
and applying Yoneda proves the internal assertion. For an unpointed
map $f$ the decomposition is $(f(0),\,f-f(0))$; addition of the
constant restores $f$. Translation is an equivalence, so the same
decomposition restricts to equivalences. Pullback and the stated
functors preserve the nerves termwise and preserve their realizations.
\qed
\end{proof}

\begin{lemma}[Torsors, sections and connecting maps]\label{lem:principal-sections}
Work in any such slice. For a group $G$ with effective bar $BG$ and
$f:X\to BG$, the pullback $T_f=X\times_{BG}\one$ is a principal
$G$-torsor. Its section space is the path space from $f$ to the
constant basepoint in $\Map(X,BG)$. Conversely every principal
$G$-torsor has this form, as in the classification of principal
$\infty$-bundles in an $\infty$-topos~\cite{NikolausSchreiberStevenson2015}.
For a coherent right $G$-action on
$T$ over $X$, principality means that $T\to X$ is a cover and
$(t,g)\mapsto(t,tg)$ identifies $T\times G$ with $T\times_XT$.
For a short exact sequence $0\to D\to E\to A\to0$ of abelian
zero-types there is, for $n\geq1$, a pointed connecting map
\[
 \partial:K(A,n)\longrightarrow K(D,n+1)
 \quad\text{with homotopy fibre }K(E,n).
\]
A nullhomotopy of $\partial$ induces a splitting $A\to E$ of the
original coefficient map.
\end{lemma}
\begin{proof}
The assertion about the action follows by pulling back the
\v{C}ech nerve of $\one\to BG$. A section of its pullback is
a map $X\to\one$ and a homotopy in $BG$; the first choice is
contractible. Conversely the action nerve of a principal torsor
is its \v{C}ech nerve. Projection of the action nerve to the
one-object group nerve defines $X\to BG$ on realizations.
After pullback along $T\cover X$ the two torsors identify with the
trivial torsor by the action, compatibly on overlaps. Descent gives
the asserted equivalence over $X$.

Consider $q:K(E,n)\to K(A,n)$ in the regular-sheaf ambient
topos. The homotopy sequence of its fibre has $D$ in degree $n$,
trivial cokernel in degree $n-1$ (trivial components for $n=1$),
and trivial groups in all remaining degrees. The canonical map from
$K(D,n)$ is therefore an equivalence by finite Postnikov induction.
This calculation restricts to the exact slice by fullness and finite
limits. The basepoint of $K(A,n)$ is a cover because $K(A,n)$ is
connected. It factors through $q$ via the basepoint of $K(E,n)$, so $q$ is
a cover. Translation identifies its kernel pair with
$K(E,n)\times K(D,n)$. Its classifying map is $\partial$; the zero of
$K(E,n)$ supplies its pointing. If it is null, $q$ has a section.
Subtract the value of that section at the basepoint, and normalize the
section homotopy by the same translation. This yields a pointed right
inverse up to pointed homotopy. Applying $\pi_n$ yields $A\to E$
splitting $E\to A$. All arguments commute with change of slice.\qed
\end{proof}

\subsection{Recovering a nucleus from one generic family}
In $\mathcal E_{/B}$ put $C_2=1_B\amalg1_B$, with addition modulo
two. Given $P\hookrightarrow1_B$, define an equivalence relation
$R_P\hookrightarrow C_2\times C_2$ by assigning $1_B$ to its two
diagonal components and $P$ to its two off-diagonal components.
The decomposition of $C_2\times C_2$ into four summands
is preserved by every change of parameter, since
coproducts are disjoint and stable under pullback.
In logical notation, $xR_Py\iff(x=y)\vee P$. Disjunction elimination
gives compatibility with addition for every intuitionistic predicate
$P$, so $R_P$ is a group congruence.
Let
\[
 H_P=C_2/R_P
\]
be its effective quotient group, and let $e,s:1_B\to H_P$ be the
images of the two summands. Effectiveness gives the exact identity
\begin{equation}
 (e=s)\simeq P.
 \label{eq:probe-generator-equality}
\end{equation}
The quotient map $C_2\to H_P$ is an effective epimorphism. Pullback
along its square $C_2^2\to H_P^2$ identifies the diagonal of $H_P$
with $R_P\hookrightarrow C_2^2$. Monomorphisms descend along
effective epimorphisms, so $H_P$ is a zero-type.
If $e=s$, both summands of the cover $C_2\to H_P$ map to $e$.
Equality in $H_P$ is propositional, so this cover shows that every
element equals $e$. Conversely, contractibility implies $e=s$.
Thus
\begin{equation}
 \operatorname{isContr}(H_P)\simeq P.
 \label{eq:probe-group-contractibility}
\end{equation}

By Lemma~\ref{lem:finite-bars}, for $n\geq1$ the family
\[
 Q_n(P)=K(H_P,n)=B^nH_P
\]
is a pointed $(n-1)$-connected
$n$-type with $\Omega^nQ_n(P)\simeq H_P$. If $Q_n(P)$ is
contractible then so is $H_P$; if $H_P$ is trivial then every
iterated bar construction is terminal. Hence
\begin{equation}
 \operatorname{isContr}(Q_n(P))\simeq P.
 \label{eq:probe-type-contractibility}
\end{equation}
All these constructions commute with base change. In particular,
pullback preserves the congruence quotients and the bar quotients
by universality of effective groupoid realizations. The displayed
contractibility equivalences therefore hold as internal subobjects
in every further parameter context.

\begin{theorem}[Contractibility reconstruction of a nucleus]
\label{thm:universal-probe}
Fix $n\geq1$. For every $B\in\mathcal E$ and
$P\hookrightarrow B$, there is a canonical equivalence of
subobjects of $B$
\begin{equation}
 \operatorname{isContr}_{\mathcal E_{/B}}
       \bigl(\ell^j_B Q_n(P)\bigr)
 \simeq j_B(P).
 \label{eq:probe-localized-contractibility}
\end{equation}
Furthermore, for $P,Q\hookrightarrow B$,
\begin{equation}
 \ell^j_B Q_n(P)\simeq\ell^j_B Q_n(Q)
 \quad\Longleftrightarrow\quad j_B(P)=j_B(Q).
 \label{eq:probe-unpointed-classification}
\end{equation}
The equivalence on the left is unpointed.
\end{theorem}

\begin{proof}
The functor $a_{j,B}$ is left exact and a left adjoint. It therefore
preserves finite coproducts, effective congruence quotients, group
nerves, and their realizations. There are consequently canonical
equivalences
\[
 a_{j,B}H_P\simeq H_{a_{j,B}P},\qquad
 a_{j,B}Q_n(P)\simeq Q_n(a_{j,B}P).
\]
Applying \eqref{eq:probe-type-contractibility} in the local slice,
then \eqref{eq:probe-contr-inclusion}, gives
\begin{align*}
 \operatorname{isContr}_{\mathcal E_{/B}}
      (\ell^j_BQ_n(P))
 &\simeq
 r_{j,B}\operatorname{isContr}_{(\mathcal F_j)_{/a_jB}}
      (Q_n(a_{j,B}P))\\
 &\simeq r_{j,B}a_{j,B}P
 \simeq j_B(P).
\end{align*}
An unpointed equivalence in \eqref{eq:probe-unpointed-classification}
preserves this contractibility subobject and hence implies equality
of closures. Conversely, full faithfulness of $r_{j,B}$ turns
equality of the closures into an equivalence
$a_{j,B}P\simeq a_{j,B}Q$; functoriality of the quotient and bar
constructions then gives the required equivalence.\qed
\end{proof}

\begin{corollary}[One generic family detects all nuclei]
\label{cor:one-generic-probe}
Let $t:1\hookrightarrow\Omega$ be the generic mono and form, in
$\mathcal E_{/\Omega}$,
\[
 W_n=Q_n(t)=B^n_\Omega\bigl((C_2)_\Omega/R_t\bigr).
\]
Then
\begin{equation}
 \chi_{\operatorname{isContr}_{\mathcal E_{/\Omega}}
      (\ell^j_\Omega W_n)}=j:\Omega\longrightarrow\Omega.
 \label{eq:generic-probe-nucleus}
\end{equation}
Consequently the equivalence class over $\Omega$ of the single
localized $n$-type family $\ell^j_\Omega W_n$ determines the
nucleus $j$. For two nuclei with the stated reflections,
\begin{equation}
 j\leq k
 \quad\Longleftrightarrow\quad
 \operatorname{isContr}(\ell^j_\Omega W_n)
 \leq\operatorname{isContr}(\ell^k_\Omega W_n)
 \quad\text{in }\operatorname{Sub}_{\mathcal E}(\Omega).
\end{equation}
\end{corollary}

\begin{proof}
The characteristic map of the generic mono $t$ is
$\operatorname{id}_\Omega$, so the characteristic map of its
$j$-closure is $j$. Apply
\eqref{eq:probe-localized-contractibility}. The order assertion is
the pointwise order on characteristic maps. All-context recovery
follows by pulling this identity back along characteristic maps and
using \eqref{eq:probe-modal-base-change}.\qed
\end{proof}

In the local slice the reflected basepoint covers the reflected
family, so each fibre is connected. Its locus of contractibility,
viewed back over the original parameter, recovers the closure of the
predicate.

\subsection{Identities between codes and the order on nuclei}

Contractibility tells us when the probe for $P$ has the same code
as the terminal type. Comparing the probes for two predicates also
records a translation in the target Eilenberg--MacLane fibre.
Fix $n\geq2$ and a parameter slice over $C$. Choose a univalent
level classifying the probes in this slice and a univalent level
classifying their reflections in the target slice.
Proposition~\ref{prop:input-completion} and
Theorem~\ref{thm:bounded-reflection} supply these levels in the bounded
elementary models. Write $X_P=Q_n(P)$ and use the same symbol for its code.

\begin{theorem}[Identity types between probe codes]\label{thm:probe-identities}
There is an internal equivalence
\[
 (X_P=_{U}X_Q)\simeq(P\leftrightarrow Q)\times X_Q.
\]
In particular $(X_P=_{U}\one)\simeq P$. Reclassify the reflected
families in a univalent universe of the target slice and apply
$r_{j,C}$ to the identity between those codes. The resulting identity
family over $C$ is
\[
 (j_C(P)\leftrightarrow j_C(Q))\times\ell_C^j X_Q .
\]
On self-identities the induced map from the first calculation to the
second is the modal unit $X_P\to\ell_C^j X_P$.
\end{theorem}
\begin{proof}
Write $g$ for the image of the nonzero binary generator in $H_P$.
The quotient cover gives the disjunction $x=0\vee x=g$ for every
$x:H_P$. Apply it to $x=u(g)$ for an automorphism $u$, with the
propositional goal $u(g)=g$. The case $u(g)=g$ is immediate, and in
the case $u(g)=0$, applying $u^{-1}$ gives $g=0$. Quotient induction
now shows $u=\mathrm{id}$. Given two
isomorphisms $v,w:H_P\to H_Q$, apply this to $w^{-1}v$.
The isomorphism object is a set, so this proves it is a proposition
in every parameter context. Any isomorphism
preserves contractibility and hence gives $P\leftrightarrow Q$.
Conversely that logical equivalence identifies the defining
congruences on $C_2$ and induces mutually inverse quotient maps.
By Lemma~\ref{lem:finite-bars}, unpointed equivalences split into
this isomorphism object and a translation in $X_Q$. Univalence
gives the formula. Take $Q=\top$ for the unit case.

In the target, use its intrinsic binary group and the equations
$a_{j,C}H_P\simeq H_{a_{j,C}P}$ and
$a_{j,C}X_P\simeq K(H_{a_{j,C}P},n)$. Repeat the first calculation
there. The right adjoint preserves products and exponentials between
local objects, so it takes the logical factor to the equivalence
of the closures and the translation factor to $\ell_C^jX_Q$.
Point evaluation and translation are natural for the reflector.
On self-identities the logical factor is terminal, and comparison
on the value at zero is exactly the unit.\qed
\end{proof}

The basepoint torsors themselves detect the order of the nuclei.
Indeed, $\one_C\to Q_n(P)$ is a torsor under the loop group
$K(H_P,n-1)$, and a section contracts $Q_n(P)$. Varying $P$
therefore tests when sections pass from one localization to another.

\begin{theorem}[Torsor sections characterize the order on nuclei]\label{thm:torsor-order}
Fix $n\geq2$. For two nuclei $j,k$, the inequality $j\leq k$ holds
exactly when the following implication holds in every parameter slice:
every principal $K(M,n-1)$-torsor, with $M$ an abelian zero-type in
that slice, which has a section after $j$-reflection has a section
after $k$-reflection. Explicitly, for a torsor $T\to X$ in
$\mathcal E_{/C}$ the section is over $a_jX$ in the slice over
$a_jC$, and likewise for $k$. It suffices to test
$\one_C\to K(H_P,n)$ for $P\hookrightarrow C$.
If $j<k$, one such torsor $T\to X$ over a zero-type parameter has an empty
section space after $j$ and a contractible section space after $k$.
The original torsor satisfies $\tau_{\leq q}^{X}T\simeq\one_X$
for every $-1\leq q\leq n-2$.
\end{theorem}
\begin{proof}
Suppose $j\leq k$. We first show in every slice that a map of
finite relative truncation degree inverted by $j$-reflection is
also inverted by $k$-reflection. For a mono this follows from
$j$-density implying $k$-density. In the inductive step the image
inclusion is inverted by $j$, since a left-exact reflector preserves
covers and images. Hence its $k$-reflection is invertible, and
the reflected map is a cover. Its diagonal has one lower degree
and is inverted by $j$. Induction and left exactness make the
reflected diagonal invertible as well. A cover with invertible
diagonal is an equivalence.

Let $T\to X$ be the specified torsor in a parameter slice over $C$.
It has relative degree at most $n-1$, checked after its trivializing
cover. Both $T$ and $\ell_X^jT$ are therefore $(n-1)$-types over
$X$. The unit $T\to\ell_X^jT$ has finite relative degree and is
inverted by $a_{j,X}$, hence by $a_{k,X}$. A section over $a_jX$
corresponds by adjunction to a section of $\ell_X^jT\to X$.
Apply $a_{k,X}$ and use the preceding equivalence to obtain a
section over $a_kX$.
By Lemma~\ref{lem:principal-sections} the basepoint of $K(H_P,n)$
is a principal $K(H_P,n-1)$-torsor. A section contracts its base.
Its reflected section thus exists exactly when the reflected
coefficient is trivial, equivalently when $j_C(P)=\one_C$.
Given the test implication, restrict $P$ to the subobject
$D=j_C(P)$ of $C$. There it is $j$-dense, hence $k$-dense by the
test. Pullback stability of closure gives $D\leq k_C(P)$.
Taking $C=\Omega$ and the generic mono yields $j\leq k$.

For strict order take $t:\one\into\Omega$, $C=k_\Omega(t)$,
and $P=t$ regarded as a subobject of $C$. Its $k$-closure is $C$
and its $j$-closure is the proper subobject $j_\Omega(t)$ of $C$,
since the two characteristic maps are distinct. A section exists
precisely when the closure fills $C$. Thus the $k$-reflected basepoint
is an equivalence, and the proper $j$-closure gives an empty section
space after $j$-reflection. The subobject $j_\Omega(t)\hookrightarrow C$
itself distinguishes the two localizations. Local triviality identifies
the original fibres locally with the $(n-2)$-connected $K(H_P,n-1)$,
giving the truncation assertion. The same parameter and predicate work for every
$n$ in the statement.\qed
\end{proof}

\section{Source dimension and relative cohomology of higher contexts}
\label{sec:higher-context-geometry}

To measure how the context affects a model with fixed logic, consider
a basepoint $b:1\to B$. As an object over $B$, it has fibre $\Omega_b B$ at $b$. The height of this loop object gives
a lower bound on the dimension of a source presenting the slice. For
$B=K(A,n)$, cohomology also detects the coefficient $A$ through maps
from $B$ to other Eilenberg--MacLane types.

\paragraph{Atlas dimension in regular sheaves.}
Let $\mathcal F$ be an exact $\infty$-category with a fully faithful
finite-limit-preserving functor
\[
  i:\mathcal F\hookrightarrow\mathcal H
\]
into a Grothendieck $\infty$-topos, whose essential image is closed
under realizations of internal Segal groupoids.  Such realizations in
$\mathcal F$ are computed in $\mathcal H$. The embedding and the
completions use compatible external size bounds. For $q\geq-1$, say that $\mathcal F$
has \emph{atlas dimension at most $q$} if every object admits a
cover by a $q$-type, that is, if $\mathcal F$ is $q$-complicial.

The embedding $i$ preserves and reflects finite truncation: use the
diagonal characterization, starting with equivalences.  It also
preserves and reflects covers, because it computes the realization
of every \v{C}ech nerve.  These assertions hold in slices.  Consequently
ambient locality of truncation and ambient retract arguments apply
to objects and maps of $\mathcal F$.

For a localized exact category $\mathcal F$, its own regular-sheaf
presentation supplies such an embedding at the chosen outer size.
\[
 \mathcal F\simeq\operatorname{Ex}_\infty(\mathcal F)
 \hookrightarrow\operatorname{Sh}_{\mathrm{reg}}(\mathcal F,\mathcal S).
\]
This embedding computes effective epimorphisms using the quotients
intrinsic to the localized model.

\begin{lemma}[A loop-space lower bound for atlas dimension]
\label{lem:general-basepoint-obstruction}
Let $b:1\to B$ be a pointed object of $\mathcal F$, and let $q\geq-1$.
If the object $b$ of $\mathcal F_{/B}$ admits a cover from a
$q$-truncated object of that slice, then $\Omega_b B$ is
$q$-truncated in $\mathcal F$.
\end{lemma}

\begin{proof}
Write the proposed slice cover as
\[
 (Y_0\xrightarrow{f}B)\longrightarrow(1\xrightarrow{b}B).
\]
It consists of a map $p_0:Y_0\to1$ and a specified homotopy
$f\simeq b p_0$.  Since covers in slices are detected on underlying
maps, $p_0$ is a cover.  By hypothesis $f$ is $q$-truncated.  Pulling
back $f$ along $b$ therefore gives a $q$-truncated object
\[
  T=Y_0\times_B1\simeq Y_0\times\Omega_bB.
\]
The homotopy $f\simeq b p_0$ gives the product decomposition over $Y_0$.

Now base-change $T\to1$ along the cover $p_0$.  Its pullback
$T\times Y_0\to Y_0$ is $q$-truncated.  In the slice over $Y_0$,
the object $\Omega_b B\times Y_0\to Y_0$ is a retract of this
pullback.  In product coordinates, the section and retraction are
\[
 \begin{split}
  s(\omega,y)&=(y,\omega,y),\\
  r(y',\omega,y)&=(\omega,y),
 \end{split}
 \qquad r s=\operatorname{id}.
\]
The diagonal of $Y_0$ and the product projections define $s$ and $r$
in the slice. Truncated maps are closed under retracts, so the pullback of $\Omega_b B\to1$
along $Y_0\to1$ is $q$-truncated.  Truncation is local for effective
epimorphisms in $\mathcal H$, hence $i\Omega_bB$ is $q$-truncated.
Since $i$ reflects truncation, $\Omega_b B$
is $q$-truncated in $\mathcal F$.\qed
\end{proof}

\begin{theorem}[Sharp dimension of a higher context]
\label{thm:general-context-dimension}
Suppose that $\mathcal F$ is bounded and has covers by zero-types.
Let $n\geq2$ and let $b:1\to B$ be a pointed $n$-type whose loop
object $\Omega_bB$ has exact truncation height $n-1$. Then
\begin{enumerate}
\item $\mathcal F_{/B}$ is bounded and has atlas dimension exactly
      $n-1$;
\item the category
      $\mathcal R_B=(\mathcal F_{/B})_{\leq n-1}$ is a regular
      $(n,1)$-category, and the canonical functor induces an equivalence
      \[
        \operatorname{Ex}_{\infty}(\mathcal R_B)
        \simeq\mathcal F_{/B};
      \]
\item if a regular $(m,1)$-category has higher exact completion
      equivalent to $\mathcal F_{/B}$, then $m\geq n$.
\end{enumerate}
\end{theorem}

\begin{proof}
For an object $Y\to B$ choose a cover $P\to Y$ in $\mathcal F$
with $P$ zero-truncated.  The composite $P\to B$ is
$(n-1)$-truncated. More generally, for a map $f:X\to B$ from an
$a$-type, compute its fibre in $\mathcal H$ over a generalized
point $y:T\to iB$. In the slice over $T$ it is the dependent sum
of the path types $((i f)(x)=y)$ over $x:(iX)_T$. The base is
$a$-truncated and the paths are $(n-1)$-truncated, so closure of
truncated maps under composition gives the bound
$\max\{a,n-1\}$. Applying this with $a=0$ shows that $P\to B$ is an
$(n-1)$-truncated object of the slice and gives a cover of
$Y\to B$.  This proves the upper atlas bound.

By Lemma~\ref{lem:general-basepoint-obstruction}, a $q$-truncated
object covering $b$ gives a $q$-truncation bound for $\Omega_bB$.
Its exact height is $n-1$, so every such bound satisfies $q\geq n-1$.
Together with the upper bound this gives atlas dimension $n-1$.

If $Y$ is $a$-truncated in $\mathcal F$, the same estimate makes
$Y\to B$ $\max\{a,n-1\}$-truncated.  Boundedness of
$\mathcal F$ therefore implies boundedness of its slice.
The $(n-1)$-truncated objects are closed under finite limits and
regular images: an image is a subobject of its truncated target.
Thus $\mathcal R_B$ is regular, and its mapping spaces are
$(n-1)$-truncated, making it an $(n,1)$-category.
The exact-completion recognition theorem
\cite{Stefanich2023Exact} now gives the
displayed equivalence, using boundedness and $(n-1)$-compliciality.

Finally, the higher exact completion of a regular $(m,1)$-category
is $(m-1)$-complicial
\cite{Stefanich2023Exact}.
Thus a regular source for the slice satisfies $m-1\geq n-1$ by the
atlas dimension just computed. The source $\mathcal R_B$ attains equality.\qed
\end{proof}

\begin{remark}[The top homotopy group and relative truncation]
For a pointed $n$-type the internal homotopy group $\pi_n(B,b)$
is the zero-type $\Omega_b^nB$ with its loop multiplication.
If $\Omega_bB$ is $(n-2)$-truncated, its $(n-1)$-fold loop object
is a pointed proposition and hence contractible. Thus a nonzero
internal $\pi_n(B,b)$ gives exact height $n-1$ for $\Omega_bB$.
For example, the effective $\infty$-topos contains a nonzero abelian group
$Q$ in its ordinary core whose only global element is zero
(Lemma~\ref{lem:effective-invisible-quotient}).
Yet $\pi_n(K(Q,n),0)\cong Q\ne 0$.
The basepoint map $b:1\to B$ shows why the dimension is measured in
the slice. Its domain is a zero-type in $\mathcal F$, and its fibre
at $b$ is the loop object of height $n-1$.
\end{remark}

\subsection{Cohomology of an Eilenberg--MacLane context}

Under the hypotheses of Theorem~\ref{thm:general-context-dimension},
the least source dimension for a slice over $K(A,n)$ is $n$ for
every nonzero abelian zero-type $A$ and $n\geq2$. To distinguish
groups $A$ at a fixed $n$, consider the maps $K(A,n)\to K(M,n)$
as $M$ varies.
Assume in addition that $\mathcal F_{\leq0}$ is an elementary topos.
For $A\in\Ab(\mathcal F_{\leq0})$, the finite bar construction of
Lemma~\ref{lem:finite-bars} gives $K_{\mathcal F}(A,k)$.
The embedding into $\mathcal H$ preserves this construction and
its truncation and connectivity bounds.

For $k\geq1$ put
\[
 H^k(\mathcal F;M)=
 \pi_0\operatorname{Map}_{\mathcal F}
       (1,K_{\mathcal F}(M,k)).
\]
Here $M$ is an abelian zero-type. Restriction to the basepoint
splits off $H^k(\mathcal F;M)$ from the cohomology of the slice.
Its kernel measures the contribution of the context.

\begin{theorem}[Cohomology added by a higher context]
\label{thm:relative-context-cohomology}
Let $A,M\in\operatorname{Ab}(\mathcal F_{\leq0})$, let $n\geq2$,
and put $B=K_{\mathcal F}(A,n)$, $p:B\to1$, with canonical
basepoint $b:1\to B$.  Then
\[
 p^*:\mathcal F_{\leq q}\simeq
       (\mathcal F_{/B})_{\leq q}
 \qquad(0\leq q\leq n-2),
\]
with inverse $b^*$.  Moreover there are natural isomorphisms
\begin{align*}
 H^k(\mathcal F_{/B};p^*M)
   &\cong H^k(\mathcal F;M)
       &&(0<k<n),\\
 H^n(\mathcal F_{/B};p^*M)
   &\cong H^n(\mathcal F;M)\oplus
       \operatorname{Hom}_{\operatorname{Ab}(\mathcal F_{\leq0})}(A,M).
\end{align*}
In degree $n$, the first inclusion is $p^*$, its retraction is
$b^*$, and the second summand is the kernel of $b^*$.
\end{theorem}

\begin{proof}
First work in $\mathcal H$.  The object $iB$ is
$(n-1)$-connected.  The classifier of $q$-truncated small families
is $(q+1)$-truncated.  If $q\leq n-2$, every classifying map
from $iB$ therefore factors, with a contractible space of choices,
through $iB\to1$.  Classifiers can be taken in a sufficiently
large ambient universe for each family under discussion.
For $q$-truncated $X,Y$ in $\mathcal H$, their exponential
$Y^X$ is $q$-truncated. With $\bar p:iB\to1$, we have
\[
 \Map_{\mathcal H_{/iB}}(\bar p^*X,\bar p^*Y)
 \simeq \Map_{\mathcal H}(iB,Y^X)
 \simeq \Map_{\mathcal H}(1,Y^X).
\]
The second equivalence is inverse to pullback along $\bar p$ and
gives full faithfulness on mapping spaces. Thus
pullback from $1$ induces an equivalence between $q$-truncated
objects of $\mathcal H$ and of $\mathcal H_{/iB}$.

For a $q$-truncated map $Y\to B$ in $\mathcal F$, the fibre $b^*Y$ belongs
to $\mathcal F$ by finite-limit closure.  The ambient
equivalence $Y\simeq p^*b^*Y$ is an equivalence in
$\mathcal F_{/B}$ by fullness.  The same argument restricts the
ambient mapping-space equivalences.  This proves the first assertion.

Base change commutes with the bar construction, so
\[
 K_{\mathcal F_{/B}}(p^*M,k)
 \simeq (B\times K_{\mathcal F}(M,k)\longrightarrow B).
\]
Its space of sections is
$\operatorname{Map}_{\mathcal F}(B,K_{\mathcal F}(M,k))$.
When $k<n$, connectivity of $iB$ and truncation of the target
give an equivalence
\[
 \operatorname{Map}_{\mathcal F}(B,K_{\mathcal F}(M,k))
 \simeq
 \operatorname{Map}_{\mathcal F}(1,K_{\mathcal F}(M,k)).
\]
Full faithfulness restricts the ambient equivalence to
$\mathcal F$.  Taking components proves the lower-degree formula.

For $k=n$ apply the pointed and unpointed mapping formulas of
Lemma~\ref{lem:finite-bars} in $\mathcal H$ to $iA$ and $iM$.
The embedding is full and preserves the finite limits that express
the abelian group laws. Thus its external pointed maps are the
discrete set $\Hom_{\Ab(\mathcal F_{\leq0})}(A,M)$. The unpointed
mapping space is the product of this set with
$\Map_{\mathcal F}(1,K_{\mathcal F}(M,n))$. Evaluation at $b$ is
the first-factor projection when the factors are ordered as in the
statement, and its section takes a global point to the constant map.
On components the splitting sends $[f]$ to
\[
 \bigl([f(b)],\,\pi_n(f-\operatorname{const}_{f(b)})\bigr).
\]
The difference carries its canonical pointing at $b$. Evaluation
preserves addition, and $\pi_n$ carries sums of pointed maps to
sums of homomorphisms since it preserves products. The inverse
takes $([u],h)$ to $[\operatorname{const}_u+K(h,n)]$. These group
isomorphisms commute with precomposition by the maps $K(g,n)$
induced on the source and postcomposition by those induced on the
coefficient object.\qed
\end{proof}

\begin{corollary}[Dimension of a localized Eilenberg--MacLane context]
\label{cor:localized-context-survival}
Let $\mathcal E$ be bounded, exact, and zero-complicial, and let
$L:\mathcal E\rightleftarrows\mathcal F:i$ be a left exact
reflective localization.  Suppose $\mathcal F$ satisfies the
logical and ambient hypotheses above.  For
$A\in\operatorname{Ab}(\mathcal E_{\leq0})$ and $n\geq2$,
\[
 L K_{\mathcal E}(A,n)\simeq K_{\mathcal F}(LA,n).
\]
If $LA=0$, the corresponding slice is $\mathcal F$.
If $LA\neq0$, the slice
$\mathcal F_{/K_{\mathcal F}(LA,n)}$ has atlas dimension exactly
$n-1$ and least finite exact-completion source dimension $n$.
Its cohomology is given by
Theorem~\ref{thm:relative-context-cohomology}, with $A$ replaced
by $LA$.
\end{corollary}

\begin{proof}
The reflector preserves finite limits and every existing
colimit, in particular effective groupoid realizations.
Applying it through the iterated bar construction proves the
delooping comparison. A cover $P\to iX$ by a zero-type in
$\mathcal E$ becomes a cover $LP\to X$ by the zero-type $LP$ in
$\mathcal F$. Thus $\mathcal F$ is zero-complicial.
Boundedness follows from the fully faithful finite-limit
inclusion: an object of $\mathcal F$ has the same finite
truncation bounds in $\mathcal E$.  If $LA=0$, its delooping
is terminal.  Otherwise
$\Omega K_{\mathcal F}(LA,n)=K_{\mathcal F}(LA,n-1)$ has
nonzero $(n-1)$st homotopy group $LA$, so
Theorem~\ref{thm:general-context-dimension} applies.\qed
\end{proof}

\begin{remark}[Recovering the coefficient from the marked slice]
As $M$ varies, the kernels of $b^*$ in degree $n$ form the functor
$\Hom_{\Ab(\mathcal F_{\leq0})}(A,-)$. Yoneda therefore recovers $A$
from the slice together with its projection and basepoint. The kernel
singles out the coefficient homomorphisms alongside the classes
pulled back from $\mathcal F$. If $1$ is projective in $\mathcal F$,
global sections preserve effective groupoid
realizations, giving
$\operatorname{Map}_{\mathcal F}(1,K_{\mathcal F}(M,k))
\simeq K(\operatorname{Map}_{\mathcal F}(1,M),k)$.
Then the base cohomology vanishes in positive degrees, so the entire
degree-$n$ group consists of coefficient homomorphisms.
\end{remark}

\begin{corollary}[The exact transition from ordinary to higher contexts]
\label{cor:context-transition}
Let $\mathcal F=\mathsf H(\mathcal T_j)$.
\begin{enumerate}
\item For $B\in\mathcal T_j$,
$\mathcal F/B\simeq\mathsf H(\mathcal T_j/B)$.
\item For a group object $G$ of $\mathcal T_j$,
$\mathcal F/BG\simeq\mathsf H(\mathcal T_j^G)$, where
$\mathcal T_j^G$ is the ordinary topos of internal $G$-actions.
\item For a nonzero abelian $A\in\mathcal T_j$ and $n\geq2$,
$\mathcal F/K(A,n)$ has zero-type category $\mathcal T_j$ and minimum
regular source dimension $n$.
\end{enumerate}
\end{corollary}

\begin{proof}
The first assertion is
Corollary~\ref{cor:operations-ordinary-slice}. A map from a
zero-type to a one-type is zero-truncated; hence
$\mathcal F/BG$ is zero-complicial. A zero-truncated family over
$BG$ pulls back along $\one\cover BG$ to a zero-type $X$.
Descent in $\mathcal H$ along the \v{C}ech nerve $G^\bullet$
gives an ordinary $G$-action on $X$: coherence reduces to the
action equations since $X$ is a zero-type. Conversely the action
nerve $X\times G^\bullet$ lies in $\mathcal F$, and its effective
quotient $X/\!/G\to BG$ is computed by the embedding into
$\mathcal H$. Its pullback to $\one$ is $X$, so the family is
zero-truncated by descent. The two constructions are inverse by
ambient descent and fullness. Under the same descent equivalence
maps of families are precisely equivariant maps. This identifies
the zero-type category with $\mathcal T_j^G$.
Recognition gives the second assertion. The third assertion follows from
Theorems~\ref{thm:general-context-dimension}
and~\ref{thm:relative-context-cohomology}.\qed
\end{proof}

For example, take $\mathcal T_j$ to be the modified or a Krivine
realizability topos of Section~\ref{sec:unified-theorem}. For
each nonzero abelian $A$ in that topos and each $n\geq2$, the
slice $\mathsf H(\mathcal T_j)/K(A,n)$ has ordinary core
$\mathcal T_j$ and least regular source dimension $n$. Its
degree-$n$ cohomology includes the ambient cohomology as the
summand split off by the basepoint.

\section{Slice cohomology and numerical observations in effective realizability}
\label{sec:separation-specialization}

In the higher effective topos $\mathcal E=\mathsf H(\mathsf{Eff})$,
we use assemblies over Kleene's first algebra. An assembly gives each
element a nonempty set of numerical names. A single partial recursive
function on names tracks each morphism. We first count the tracked
homomorphisms that contribute binary cohomology classes to a slice.
We then compute the numerical observations of a type code. In both
calculations $D=\one\amalg\one$ has addition modulo two and numerical
names for its elements.

\subsection{Classification of the effective slices over \texorpdfstring{\(K(A_r,n)\)}{K(A\_r,n)}}

In the degree-$n$ splitting of
Theorem~\ref{thm:relative-context-cohomology}, a homomorphism from
the coefficient of the base to $D$ gives a binary cohomology class.
To vary how many such homomorphisms exist, put
$V=\bigoplus_{\mathbb N}\mathbb F_2$ and give every element of $V$
every natural number as a name. Write $\nabla V$ for this codiscrete
assembly, and set
\[
 A_r=D^r\times\nabla V,\qquad
 \mathcal E_{r,n}=\mathcal E/K(A_r,n),
 \quad r\geq0,\ n\geq2.
\]
The factor $\nabla V$ makes every $A_r$ nonzero, including $A_0$.
The projection to $D^r$ identifies the tracked homomorphisms
$A_r\to D$ with the linear functionals on $D^r$.

\begin{proposition}[Unmarked classification of the effective slices]
\label{prop:separation-recovered}
Let $p_{r,n}:K(A_r,n)\to\one$ be the projection. The categories
$\mathcal E_{r,n}$ have
\[
 (\mathcal E_{r,n})_{\leq q}\simeq\mathcal E_{\leq q}
 \quad(0\leq q\leq n-2),\qquad
 H^n(\mathcal E_{r,n};p_{r,n}^*D)\cong\mathbb F_2^r.
\]
Their first nonzero positive cohomological degree, allowing all
abelian zero-type coefficients, is $n$. Consequently
\[
 \mathcal E_{r,n}\simeq\mathcal E_{s,m}
 \quad\Longleftrightarrow\quad(r,n)=(s,m).
\]
For fixed $n$, the defining bases share the global point space $K(V,n)$
up to equivalence.
\end{proposition}

\begin{proof}
A tracked map from $\nabla V$ to $D$ is constant. A single input
realizer names every element of $V$, so a tracker sends it to a
common name for all the images. In $D$ each name specifies one bit.
If the map is a homomorphism, that bit is
zero. Restriction to $D^r\times\{0\}$ therefore identifies every
homomorphism $A_r\to D$ with a binary linear functional on $D^r$.
Conversely each such functional is tracked by its finite table
after projection to $D^r$. Thus
\[
 \Hom_{\Ab(\mathsf{Eff})}(A_r,D)\cong\mathbb F_2^r.
\]
In the effective higher completion the terminal object is
regular-projective: original partitioned projectives retain their
lifting property through original atlases. The global sections functor
$\Gamma=\Map_{\mathcal E}(\one,-)$ is therefore regular and preserves
effective groupoid quotients.
Positive-degree base cohomology vanishes. Apply
Theorem~\ref{thm:relative-context-cohomology} for the displayed
cohomology formula and the low-type equivalences.

Cohomology vanishes in degrees $0<k<n$ for every coefficient. In
degree $n$, the identity of $A_r\ne0$ gives a nonzero class with
coefficient $p_{r,n}^*A_r$. Every equivalence
preserves zero-types, abelian group objects, their iterated bar realizations and global
sections. It therefore preserves this first degree. It also carries
$p_{r,n}^*D$ to the corresponding group: its underlying object is
$\one\amalg\one$, its unit is the first summand, and addition
on the four product summands is prescribed by addition modulo two.
Thus the dimension of degree-$n$ cohomology with this coefficient
recovers $r$, including $r=0$. Equality of the parameters
gives the converse equivalence by construction.

Externally $\Gamma A_r\cong\mathbb F_2^r\oplus V\cong V$.
Preservation of the iterated bar construction by $\Gamma$
therefore gives $\Gamma K(A_r,n)\simeq K(V,n)$.\qed
\end{proof}

The identity of $A_r$ has a concrete torsor representative. Its
delooping is the identity of $K(A_r,n)$, whose fibre over zero gives
the basepoint cover $\one\to K(A_r,n)$. A section of this cover is
a contraction of $K(A_r,n)$. Thus $\pi_n K(A_r,n)\cong A_r\ne0$
witnesses the nontriviality of the torsor in the slice.

\subsection{The minimum number of binary torsor coordinates}

Each binary cohomology class determines a torsor whose total space
trivializes the class. Given two families over the same context, we
compute the minimum number of such coordinates that make them equivalent.

Fix $r\geq0$, $n\geq2$, and write $B_{r,n}=K(A_r,n)$. Choose
$s\geq0$ and put $G_s=K(D^s,n-1)$. By a family with fibre $G_s$ we mean a
family locally equivalent to the constant $G_s$-family. We
classify these families up to equivalence over the fixed base,
allowing changes of coordinates on the fibre.
For a binary matrix $L$ of size $d\times r$, set
\[
 \lambda_L(a,v)=La,\qquad
 f_L=K(\lambda_L,n): B_{r,n}\longrightarrow K(D^d,n).
\]
Let $P_L$ be the homotopy fibre over zero, with map
$t_L:P_L\to B_{r,n}$. Extension of context along $t_L$ adjoins a
witness that $f_L$ is zero at the current point. Its \emph{number
of binary coordinates} is $d$.

\begin{lemma}[Matrix coordinates for families]
\label{lem:effective-matrix-families}
The equivalence classes of $G_s$-families over $B_{r,n}$ are
in bijection with subspaces of $(\mathbb F_2^r)^*$ of dimension at most
$s$. The family $P_M$ for an $s\times r$ matrix $M$ corresponds to
its row space.
\end{lemma}
\begin{proof}
Internally, homomorphisms $A_r\to D^s$ form the finite numerical
assembly of $s\times r$ matrices. To see the assertion with parameters,
it suffices to test on ordinary objects, since maps to both zero-types
factor through zero-truncation. Take an ordinary object $T$
and choose an assembly cover $C\cover T$. For fixed carrier values of
$c\in C$ and $a\in D^r$, the tuples $(c,a,v)$ have a common realizer
as $v$ ranges over $V$. The tracker of the homomorphism has the same
output name on these inputs, and the distinct outputs in $D^s$ have
disjoint realization sets. Thus its value is independent of $v$.
Evaluation on the standard basis of $D^r$ defines a matrix over
$T$ whose multiplication formula agrees with the homomorphism
after the cover and hence before it. Conversely all matrix operations
on finite binary tuples are tracked. This proves the internal formula,
including zero rows or columns.

Let $H_s$ be the finite numerical group $\mathrm{GL}_s(\mathbb F_2)$.
By Lemma~\ref{lem:finite-bars} in every slice, evaluation
at zero decomposes a self-equivalence of $G_s$ uniquely up to coherent
homotopy as $x\mapsto u+K(h,n-1)(x)$ with $u:G_s$ and $h:H_s$.
Composition is the semidirect product law. These formulas also apply
to $n=2$, when pointed connected one-types classify groups.
The image of the code of $G_s$ in a univalent universe is therefore
the delooping of $G_s\rtimes H_s$: its basepoint is a cover whose
loop group is this automorphism group. Write $Y//H$ for the effective
quotient of an action groupoid. The split extension yields a
projection from that component to $BH_s$, with fibre
$BG_s=K(D^s,n)$ and monodromy the coefficient action of $H_s$.
Indeed pulling back along the point cover of $BH_s$ reduces the
group extension to its kernel, and the change of lift by $h$ acts
by conjugation. Thus the component is $K(D^s,n)//H_s$.

Put $B=B_{r,n}$. Since $B$ is simply connected and $BH_s$ is a
one-type, evaluation gives $(BH_s)^B\simeq BH_s$, stably in
parameters. Exponentiating the projection above and pulling back along
$1\to BH_s$ gives fibre $K(D^s,n)^B$, with the same monodromy.
Evaluation at the basepoint of $B$, followed by subtraction of that
value, identifies this fibre with
\[
 K(D^s,n)\times
 \underline{\Hom}_{\Ab}(A_r,D^s).
\]
For the pointed factor the formula follows by testing pointed maps
in every parameter slice. Descent gives the action quotient of this
product by $H_s$ as the full internal object of families.
By projectivity, global sections preserve this effective quotient.
Externally, its classes are the orbits of the binary matrices under
invertible row operations, since $\Gamma K(D^s,n)$ is connected.
Two matrices with $s$ rows are in the same orbit precisely when
they have the same reduced row echelon form, or equivalently the same
row space. The map $f_M$ represents the matrix $M$ with value zero
at the basepoint. Its fibre is the corresponding translation torsor,
which becomes the stated unframed family upon passage to the action
quotient.\qed
\end{proof}

The row space classifies the family over the original context.
Pullback along $t_L$ makes a row trivial precisely when it lies
in the span of the rows of $L$. Thus we can compute the minimum by
finding a subspace of least dimension modulo which the two classifying
subspaces have the same image.

\begin{theorem}[Minimum number of binary torsor coordinates]
\label{thm:effective-coordinate-cost}
Let two $G_s$-families over $B_{r,n}$ have row subspaces $Y,Z$.
The least number of binary coordinates in an extension $t_L$ after
which the families are globally equivalent is
\[
 d_{\min}(Y,Z)=\max\{\dim Y,\dim Z\}-\dim(Y\cap Z).
\]
A full-row-rank matrix attains this minimum. Equivalently this is the
least number of such coordinates giving an identity of their universe
codes in the extended context.
\end{theorem}
\begin{proof}
Let $\ell=\operatorname{rank}L$, $W=\operatorname{row}L$. In bases
realized by finite recursive changes of
coordinates, the coefficient map has the form
\[
 D^\ell\times A_{r-\ell}\longrightarrow D^\ell\times D^{d-\ell},
 \qquad (u,a)\longmapsto(u,0).
\]
Taking the fibre of its $n$-fold delooping gives
$P_L\simeq B_{r-\ell,n}\times K(D^{d-\ell},n-1)$. Up to homotopy,
$t_L$ is projection onto the first factor followed by inclusion of
the delooped kernel. In these bases the coefficient inclusion
$A_{r-\ell}\to A_r$ sends $(v,w)$ to $(0,v,w)$, with $w$ the
codiscrete coordinate. The projection onto the first factor has a section. Consequently a
row $a$ gives a null map $f_a t_L$ exactly when it vanishes on the
coefficient kernel: by pointed classification this is necessary
(subtract the basepoint value of an unpointed nullhomotopy), and
sufficient by restriction of $K(-,n)$. Equivalently $a\in W$.

Choose matrices $M,N'$ for $Y,Z$. After pullback to $P_L$, both
classifying maps lift to $K(D^s,n)$ inside its action quotient.
An equivalence between their underlying families is a path in that
quotient. Its projection to $BH_s$ is a loop, hence has linear part
a map $P_L\to H_s$. Both factors in the decomposition of $P_L$ are
connected for $n\geq2$, so this map is a single matrix $h$.
For each fixed $h$, path lifting in the action groupoid identifies
the remaining datum with a homotopy from $h f_Mt_L$ to $f_{N'}t_L$.
Such a homotopy exists precisely when each row of $N'-hM$ lies in
$W$, by the preceding kernel test. In the quotient
$(\mathbb F_2^r)^*/W$, such an $h$ exists exactly when the two row
spans coincide. Indeed the two lists of $s$ rows then define
surjections from $\mathbb F_2^s$ onto the same span. Choose a basis
of that span, lift it through each surjection, and complete each
lifted list by a basis of its kernel. The resulting change of basis
is the required row operation, including for a zero span. Thus the
criterion is $Y+W=Z+W$.

Put $a=\dim Y$, $b=\dim Z$, $c=\dim(Y\cap Z)$. If $Y+W=Z+W$,
the image of $Y$ modulo $Z$ lies in that of $W$, and conversely
with $Y,Z$ exchanged. Hence $d\geq\dim W\geq\max(a,b)-c$.
Assume $a\geq b$, interchanging the spaces if necessary. Extend a
basis of $Y\cap Z$ separately by $y_1,\ldots,y_{a-c}$ and
$z_1,\ldots,z_{b-c}$. Take as rows of $L$ the $b-c$ vectors
$y_i+z_i$ and the remaining $a-b$ vectors $y_i$. Their projections
modulo $Z$ are independent. Modulo their span the two bases generate
the same subspace, so this matrix realizes the bound. The conclusion
about codes is univalence.\qed
\end{proof}

The formula identifies this minimum with the injection distance of the two
subspaces \cite{SilvaKschischang}. For example, the planes $\langle e_1,e_2\rangle$ and
$\langle e_2,e_3\rangle$ in $(\mathbb F_2^3)^*$ are identified by
adjoining the single coordinate $e_1+e_3$. Each image has rank two
in the quotient. Thus one coordinate identifies the two families,
and trivializing both uses the three independent coordinates in the
sum of the two planes.

\subsection{Cohomology over a base with contractible global point space}

The calculation for $A_r$ reads off $r$ from binary cohomology in
degree $n$. We now construct a group $Q$ for which the first positive
cohomological degree depends on the coefficient. Write $p:K(Q,n)\to1$
for the projection. The identity of $Q$ gives a nonzero class in degree
$n$ with coefficient $p^*Q$. Every homomorphism $Q\to D$ is zero,
and a nonsplit extension gives a binary class in degree $n+1$, the
first positive degree for $p^*D$. To construct $Q$ and the extension
together, put $C=\nabla\mathbb F_2$ and let $j:D\to C$ be the
identity on carriers. Taking its cokernel in the abelian groups of
$\mathsf{Eff}$ gives
\begin{equation}
 0\longrightarrow D\xrightarrow{j}C\xrightarrow{q}Q
 \longrightarrow0.
 \label{eq:effective-invisible-extension}
\end{equation}
Here $j$ is monic by injectivity of its carrier map. The quotient
can equally be formed by the effective relation given by translation
by $D$, which identifies its kernel with $D$.

\begin{lemma}[A nonzero quotient with trivial global sections]
\label{lem:effective-invisible-quotient}
The group $Q$ is nonzero, $\Gamma Q=0$, and $\Hom(Q,M)=0$ for every
abelian assembly $M$.
\end{lemma}
\begin{proof}
Every tracked map $C\to D$ is constant, since the two elements of
$C$ share a name and each numerical name in $D$ specifies one bit.
Hence $j$ is a proper subobject of $C$ and its cokernel is nonzero.
Every global point of $Q$ lifts to $C$ by projectivity, and every
global point of $C$ comes from $D$. This gives $\Gamma Q=0$.
For $g:Q\to M$ a group homomorphism, $gq$ is a map of assemblies
with $gqj=0$. Since $j$ is surjective on carriers, $gq$ is zero
on carriers and therefore zero as an assembly map. The epimorphism
$q$ now gives $g=0$.\qed
\end{proof}

\begin{theorem}[First degrees of detection for the quotient coefficient]
\label{thm:effective-invisible-cohomology}
Let $n\geq2$, $B_Q=K(Q,n)$, $p:B_Q\to1$, and $\mathcal Z_n=\E/B_Q$.
The global space $\Gamma B_Q$ is contractible, and the ordinary core
of $\mathcal Z_n$ is $\mathsf{Eff}$. Its first nonzero positive
cohomology with abelian zero-type coefficients occurs in degree $n$.
In that degree $H^n(\mathcal Z_n;p^*M)=0$ for every abelian
assembly $M$. The first positive degree with nonzero cohomology
for the intrinsic binary coefficient $p^*D$ is $n+1$.
\end{theorem}
\begin{proof}
Preservation of finite bars by $\Gamma$ gives $\Gamma B_Q\simeq
K(0,n)$. By Theorem~\ref{thm:relative-context-cohomology}, every abelian
zero-type coefficient is a pullback $p^*M$. Since the terminal object of
$\E$ is projective, cohomology with these coefficients vanishes in
positive degrees below $n$ and is $\Hom(Q,M)$ in degree $n$. The identity
of $Q$ and Lemma~\ref{lem:effective-invisible-quotient} give the
asserted detection and vanishing.

For the binary class apply Lemma~\ref{lem:principal-sections} to
\eqref{eq:effective-invisible-extension}. It makes the $n$-fold
delooping $q_n:K(C,n)\to B_Q$ a principal $K(D,n)$-torsor with
classifying Bockstein $\partial_n:B_Q\to K(D,n+1)$.
A nullhomotopy of $\partial_n$ would split $q_n$. Subtracting the
value of the section at zero gives a pointed section up to homotopy,
since the normalization is natural for group homomorphisms and
takes $\operatorname{id}_{B_Q}$ to itself. Taking $\pi_n$ would
split $q$ by a homomorphism $Q\to C$. Since $C$ is an abelian
assembly, $\Hom(Q,C)=0$, so the composite of any such homomorphism
with $q$ is zero. The identity of $Q\ne0$ is nonzero. Hence
$[\partial_n]\ne0$ in degree $n+1$ with coefficient $p^*D$.
Its image on global point spaces is zero, since its source is
contractible and its target is connected.\qed
\end{proof}

\subsection{Numerical observations of two type classifiers}

We turn from the cohomology of slices to maps from a type classifier
into the numerical assembly $N$, in which each numeral names itself.
A program-indexed family and a family with two explicit tags give
classifiers with equivalent spaces of global points and different
observations in $N$~\cite{Maruyama2026}. We compute the internal
objects of observations of both classifiers at every finite numerical
simple type. The internal calculation also describes observations
that depend on a higher parameter.

Fix an acceptable numbering $e\mapsto\varphi_e$ of the partial
recursive functions. For $e,s\in\mathbb N$ let $l_e(s)$ be the largest $k\leq s$ such
that every computation $\varphi_e(i)$ with $i<k$ halts in at most
$s$ steps. This function is computable and nondecreasing in $s$.
Between stages $s$ and $s+1$ take the endomorphism of $D$ that is zero
if $l_e(s+1)>l_e(s)$ and the identity otherwise. Write $d_e[s,t)$
for the composite up to stage $t$. Over $e$, form the group $M_e$ of
classes of pairs $(s,a)\in\mathbb N\times D$, with
\[
 (s,a)\sim_e(t,b)
 \quad\Longleftrightarrow\quad
 \exists u\geq s,t\quad d_e[s,u)(a)=d_e[t,u)(b).
\]
This is a single group assembly $M$ over $N$. A name of
$(e,[s,a])$ is a recursive tuple encoding $e$ and any representative
of the class. To add two names, transport both to their maximum stage
and add their bits. A further common stage verifies independence
of the representatives and the group laws. Zero has name $(e,0,0)$
and inversion fixes every element. These formulas track the group
operations uniformly in $e$ and present the quotient directly as an
assembly over $N$.

If $\varphi_e$ is total, $l_e$ increases infinitely often, every
class eventually maps to zero, and the zero section identifies the
fibre with $1$. Otherwise $l_e$ stabilizes. Fix a stage $S$ after
stabilization. The maps $a\mapsto[S,a]$ and $[t,b]\mapsto
d_e[t,\max(t,S))(b)$ are inverse tracked group maps between $D$
and $M_e$. Thus $M_e$ is isomorphic to the trivial
group when $e$ is total and to $D$ otherwise.

Fix $n\geq2$. Form the relative delooping $K_N(M,n)$ and classify
it in a level containing these countable presentations and their
finite bar constructions. Let $c:N\cover\mathcal M_n$ be the image
factorization of its classifying map into that universe. The pullback
of the universal family to this image is univalent: the inclusion
of the image is monic, so the identity space of two codes remains the
space of equivalences of their fibres. Fix indices $e_0,e_1$ of a
total and a nowhere-defined program, respectively, and write
$u_i=c(e_i)$ for their universe codes. For comparison, let
$\mathcal W_n$ be the image of the map from $1\amalg1$ that tags
the codes for $1$ and $K(D,n)$ in the same universe. Choosing
$e_0,e_1$ gives a map $1\amalg1\to N$ and hence an inclusion of
images $\mathcal W_n\hookrightarrow\mathcal M_n$. The two tagged
fibre types are inequivalent over every noninitial context, so the
tag descends to a map $\mathcal W_n\to1\amalg1$. A factorization of
the program-indexed family through $\mathcal W_n$ would therefore
compute a tag for every program, deciding totality. The same reduction
determines which numerical observations descend to $\mathcal M_n$.

\begin{theorem}[Global homotopy and numerical observation]
\label{thm:effective-observers}
The spaces of global points of $\mathcal M_n$ and $\mathcal W_n$ are
both equivalent to $1\amalg K(\mathbb F_2,n+1)$. For every finite
numerical simple type $\sigma$, generated from $1,N$ by products and
function types, evaluation at $u_0$ and at the two tagged codes gives,
respectively,
\[
 \sigma^{\mathcal M_n}\simeq\sigma,
 \qquad
 \sigma^{\mathcal W_n}\simeq\sigma\times\sigma .
\]
Under evaluation at both $u_0,u_1$, the first comparison gives the
diagonal of $\sigma$. These are internal equivalences and hence
apply with arbitrary additional higher parameters.
\end{theorem}

\begin{proof}
The terminal assembly is projective in $\E$, since it is partitioned
and a lifting problem can be refined by an assembly atlas. Thus
$\Gamma=\Map(1,-)$ preserves covers: projectivity gives surjectivity
on components. It preserves finite limits, so takes the effective
groupoid presenting any quotient to the \v{C}ech groupoid of the
induced cover of spaces. It follows that $\Gamma$ also preserves
finite deloopings. In particular every component of $\Gamma\mathcal M_n$
contains a numeral code. The fibre computation gives exactly two
components. They are distinct since an equivalence between the two
fibres would, on $n$-fold looping, make $D$ trivial. At the terminal
code the loop space is contractible. At the other code the loop group
is $\Gamma K(D,n)$. Translation identifies $K(D,n)$ with the identity
component of its self-equivalences~\cite{BuchholtzEtAl2025}. This is the
whole automorphism group since pointed self-equivalences correspond to
automorphisms of $D$, and $D$ is cyclic of order two. Translation respects
composition. These loop groups determine the claimed global components.

The two tagged codes have empty equivalence object between their
fibres in every noninitial context: looping an equivalence there
would identify the two disjoint points of $D$. Their equivalence
groupoid is therefore the disjoint union of the two automorphism
groupoids. Its quotient gives $\mathcal W_n\simeq1\amalg K(D,n+1)$
internally. Global points of this sum split over its map to $1\amalg1$,
whose global points are the two ordinary tags. This proves the
assertion about $\Gamma\mathcal W_n$ as well.

Let $f:\mathcal M_n\to N$. The composite $fc:N\to N$ is total
recursive by full faithfulness on assemblies. It takes the fixed
values $f(u_0)$ and $f(u_1)$ according to whether $e$ is total,
since equivalent fibres give equal outputs. If the values differed,
$fc$ would decide totality. Indeed a program ignoring its input and
simulating a fixed computation is total exactly when that computation
halts, so such a decision would decide halting. Equality on all
numerals is equality of assembly maps, and $c$ is a cover. Hence
$f$ is constant.

For a parameter $X$, let $F:X\times\mathcal M_n\to N$, and put
$g(x)=F(x,u_0)$. Choose an assembly cover $A\cover X$. Every
carrier element $a$ of $A$ gives a global point by choosing any
one of its names. Applying the preceding result at that point shows
that $F$ and $g\pi_X$ have the same carrier values on
$A\times N$. The maps there are assembly morphisms, so they agree.
Since $A\times N\to X\times\mathcal M_n$ is a cover and $N$ is
a set, their equality descends. Yoneda now identifies evaluation
as inverse to $N\to N^{\mathcal M_n}$. For any object $Z$, if
constant maps give $Y\simeq Y^{\mathcal M_n}$, then exponential
adjunction gives
$(Y^Z)^{\mathcal M_n}\simeq (Y^{\mathcal M_n})^Z\simeq Y^Z$.
The property also passes to products and contains the terminal type.
Induction on $\sigma$ proves the first formula. Every such $\sigma$
is a set. Mapping the connected $K(D,n+1)$ into a set gives exactly
the constants, in every context. The two summands of
$\mathcal W_n$ therefore allow two independent outputs.\qed
\end{proof}

\section{Code identities and classical reflection in function realizability}
\label{sec:function-semantics}

In the program-indexed family of
Theorem~\ref{thm:effective-observers}, progress in a computation
sets a transition in a sequence of binary groups to zero. In function
realizability we can use the entries of a stream to specify those
transitions directly. Infinitely many nonzero entries make the
resulting group trivial. On an eventually zero stream, a bound on the
nonzero positions identifies the group with $C_2$. The trivial fibres
give a recognition theorem for subsets of Baire space. The cyclic fibres
relate a continuous choice of identities to a continuous choice of tail bounds.

Write \(B=\mathbb N^{\mathbb N}\), let \(\mathbb B\) be the assembly
whose unique name for \(\alpha\in B\) is \(\alpha\), and work in
\(\mathcal E=\mathsf H(\RT(\mathcal K_2))\). Choose an admitted
inaccessible cardinal greater than \(|B|\). In \(\mathcal K_2\), the
\(j\)-th output of \(r\cdot\alpha\) is one less than the first positive
value \(r(\langle j,\alpha|k\rangle)\), in order of increasing prefix
length \(k\). The application \(r\cdot\alpha\) is defined as a stream
exactly when, for every output coordinate \(j\), there is a prefix
length \(k\) such that \(r(\langle j,\alpha|k\rangle)>0\).
In the numerical assembly \(N\), the name of \(m\)
is the constant stream \(\nu_m(k)=m\) for all \(k\in\mathbb N\).
We use fixed effective encodings for finite words, natural numbers
and tuples. Thus a map from a subspace \(S\subseteq
B\), with inherited stream names, to \(B\) or \(\mathbb N\) is tracked
exactly when its carrier map is continuous. Indeed, a tracker reads
finitely many coordinates for each output. Conversely, represent numerical
outputs by their constant streams. For a finite word \(s\), let \([s]\)
be the set of streams extending \(s\). Set \(r(\langle j,s\rangle)=m+1\)
if \(S\cap[s]\) is nonempty and the \(j\)-th output coordinate is
constantly \(m\) there. Set all other entries of \(r\) to zero.
By continuity, each input in \(S\) has a prefix certifying each output
coordinate, so \(r\) tracks the map. The full embedding
of assemblies retains these maps in \(\mathcal E\).

\subsection{A stream-indexed family and its univalent classifier}

Give the bits \(0,1\) of \(C_2=\mathbb Z/2\) the stream names
\(\nu_0,\nu_1\), respectively. For each stage \(s\), use
a copy of \(C_2\) and let the transition \(d_{\alpha,s}\) to the next
copy be zero if \(\alpha(s)\ne0\), and the identity otherwise. Write
\(d_\alpha[s,t)\) for the finite composite, including the identity at
\(t=s\). Define \(M_\alpha\) on representatives \([s,a]\), \(a\in C_2\),
by the relation
\[
 (s,a)\sim_\alpha(t,b)\quad\Longleftrightarrow\quad
 \exists v\ge s,t\quad d_\alpha[s,v)(a)=d_\alpha[t,v)(b).
\]
Its total assembly has names \(\langle\alpha,s,a\rangle\) for
\((\alpha,[s,a])\). Transport to a common later stage proves
transitivity. To add two classes, transport to the larger input stage
and add there. This is independent of representatives by transport to
a common witnessing stage, and is tracked by finite prefix inspection.
The class \([0,0]\) gives zero and every class is self-inverse. This
constructs an abelian group \(M\) in the assembly slice over \(\mathbb B\).

Put \(P=\{\alpha:\forall k\ \exists s\ge k\ \alpha(s)\ne0\}\).
For \(\alpha\in P\), each representative reaches zero, so the projection
and zero section give a tracked isomorphism \(M_\alpha\cong0\).
If \(S\) is a bound with \(\alpha(t)=0\) for \(t\ge S\), a tracked
isomorphism \(C_2\to M_\alpha\) sends \(a\) to \([S,a]\). Its inverse
sends \([s,a]\) to \(d_\alpha[s,\max(s,S))(a)\). These formulas are
uniform in a supplied bound. The assembly of representatives has
countable fibres and total carrier of size \(|B|\). Its operations
and the finitely iterated bar constructions below have presentations
at the chosen size bound, since the nerve terms use finite products
and there are countably many simplicial degrees.

\begin{proposition}[The classifier and its universal family]
\label{prop:fn-classifier}
For \(n\ge2\), the family \(K_{\mathbb B}(M,n)\) has a univalent
completion \(u_n:\widetilde U_n\to U_n\), with code cover
\(c_n:\mathbb B\cover U_n\). Let \(a:\mathbb B\cover Q\) be the
effective quotient by the relation whose fibre at \((\alpha,\beta)\)
is the internal object of group isomorphisms \(M_\alpha\cong M_\beta\).
Then \(Q\) is a zero-type, \(M\) descends along \(a\) to an abelian
group \(A\) over \(Q\), and
\[
 U_n\simeq\sum_{q:Q}K(A_q,n+1),\qquad
 u_n\simeq\left(Q\xrightarrow{\;0\;}\sum_{q:Q}K(A_q,n+1)\right).
\]
In particular \(U_n\) has exact height \(n+1\). The codes
\(u_*=c_n(1^\omega)\) and \(v_n=c_n(0^\omega)\), for the constant
streams, have fibres \(\one\) and \(K(C_2,n)\), respectively.
\end{proposition}

\begin{proof}
Form the assembly \(H(\alpha,\beta)\) of group isomorphisms
\(M_\alpha\cong M_\beta\). Its names supply trackers for inverse
homomorphisms. The homomorphism and inverse equations define it by
finite limits in the assembly slice of function objects. These
objects are preserved by the source embedding. Each carrier group
is zero or cyclic of order two, so there is at most one isomorphism
over each pair. Thus \(H\) is a mono and, by identity, composition,
and inversion, an equivalence relation. Evaluation gives descent
data for \(M\) along its effective set quotient \(a\), yielding
\(A\to Q\) with \(a^*A\simeq M\).

In every parameter slice, an unpointed equivalence of the deloopings
decomposes into a translation in the target and a pointed equivalence.
The latter corresponds to an isomorphism of coefficients by
Lemma~\ref{lem:finite-bars}. Hence the fibre-equivalence
groupoid with objects \(\mathbb B\) has arrows
\((b,\phi)\) with \(\phi\in H(\alpha,\beta)\) and \(b\in K(M_\beta,n)\).
Composition sends \((b',\psi),(b,\phi)\) to
\((b'+K(\psi,n)b,\psi\phi)\). Delooping and translation give this
description naturally in the parameters. As in the simplicial
construction of univalent completion~\cite{VanDenBergMoerdijk2018Univalent},
realize the groupoid and its evaluation action on \(K_{\mathbb B}(M,n)\). Effectiveness
identifies paths between the image codes with their equivalences,
and descent along the code cover gives univalence.

To identify both realizations, put \(T=\sum_{q:Q}K(A_q,n+1)\)
and let \(z:Q\to T\) be zero. This is the relative bar cover.
The composite \(za\) is a cover. An arrow of its \v{C}ech groupoid
over \((\alpha,\beta)\) is a path in \(Q\), hence a \(\phi\in
H(\alpha,\beta)\), together with a path between the transported
zero points. The latter is an element \(b\in K(M_\beta,n)\).
Path composition is precisely the action formula above, with its
coherences induced by functoriality of transport and the loop-group
law. The pullback of \(z\) to this atlas has fibres \(K(M_\alpha,n)\), and
its transport is \(x\mapsto b+K(\phi,n)x\). Descent now gives
\(U_n\simeq T\) and identifies \(u_n\) with \(z\).
The base \(Q\) is a set, and the fibres of \(T\to Q\) are
\(n\)-connected \((n+1)\)-types. Consequently \(Q=\tau_{\le0}U_n\),
and at \(v_n\) the \((n+1)\)-fold loop object is \(C_2\). This
proves the exact bound and the descriptions of the marked fibres.
\qed
\end{proof}

\subsection{\texorpdfstring{\(G_\delta\)}{G\textdelta} and coanalytic equality loci at fixed truncation height}

The code $u_*$ names the terminal fibre. To measure the complexity
of recognizing it, vary the input stream and record the inputs at
which an identity with $u_*$ has a global witness.
For \(f:\mathbb B\to V\) and \(v:\one\to V\) define the external locus
\[
 L_v(f)=\{x\in B:\Map_{\mathcal E}(\one,(f(x)=v))\ne\varnothing\}.
\]
Give \(B\) its usual topology. A \(G_\delta\) set is a countable
intersection of open sets. A coanalytic subset of \(B\) is the
complement of the projection of a closed subset of \(B\times B\).
We use the boldface pointclasses, with arbitrary Baire parameters
available to the continuous reductions \cite{Moschovakis}.

\begin{theorem}[From \(G_\delta\) to coanalytic identities]
\label{thm:fn-recognition}
For \(n\ge2\), the loci \(L_{u_*}(f)\), as \(f:\mathbb B\to U_n\)
varies, are exactly the \(G_\delta\) subsets of \(B\). Put
\(V_n=U_n^{\mathbb B}\) and let \(w_n\) be constant at \(u_*\).
The loci \(L_{w_n}(f)\), for \(f:\mathbb B\to V_n\), are exactly
the coanalytic subsets of \(B\). Both \(U_n\) and \(V_n\) have exact height \(n+1\).
One such locus for \(V_n\) is the set of codes for well-founded trees,
which is complete for the coanalytic sets under continuous reductions.
\end{theorem}

\begin{proof}
By Proposition~\ref{prop:fn-classifier} and the fibre calculation,
\(L_{u_*}(c_n)=P\). Every map \(\mathbb B\to U_n\) lifts up to homotopy
through \(c_n\), since \(\mathbb B\) is partitioned projective.
Its locus is therefore a continuous inverse image of the \(G_\delta\)
set \(P\). To realize any \(G_\delta\) set, write it as an intersection
of open sets indexed by \(i\in\mathbb N\). Code each open set by an
enumeration of basic cylinders, allowing blank entries. At stage \(t\)
compute the largest \(k\le t\) such that each row \(i<k\) contains,
within its first \(t\) entries, a cylinder of word length at most \(t\)
containing the input \(x\). The resulting
sequence of integers is nondecreasing. Output a
one at every strict increase and a zero elsewhere. Infinitely many
ones are produced exactly when \(x\) belongs to every open set. This
is a single continuous procedure on the code and \(x\), since each
stage performs finitely many bounded tests.

Transpose a map \(f:\mathbb B\to V_n\) to a map
\(\mathbb B^2\to U_n\). The square is also partitioned projective,
so the transpose is homotopic to \(c_n\eta\) for a jointly continuous
\(\eta:B^2\to B\). For each external parameter \(x\),
\begin{equation}
 x\in L_{w_n}(f)\quad\Longleftrightarrow\quad
             \forall\beta\in B\quad\eta(x,\beta)\in P .
 \label{eq:fn-branches}
\end{equation}
Evaluation proves necessity. For sufficiency, the pullback of the
group \(M\) along \(\eta(x,-)\) is an assembly over \(\mathbb B\)
with singleton carrier fibres. Its projection and its tracked zero
section are inverse assembly maps. Delooping supplies a uniform
trivialization, hence an identity in the function type by univalence.
This uniformity is what allows the quantifier in
\eqref{eq:fn-branches} to range over external branches.

The complement of the right side is a projection of a closed set.
A witness consists of \(\beta\) and a number \(m\) after which every
entry of \(\eta(x,\beta)\) is zero. Encode \(m\) as the first entry
of the witness stream and \(\beta\) as its tail. Conversely, for a
closed \(C\subseteq B^2\), its open complement is \(G_\delta\).
Apply the first reduction on \(B^2\cong B\) to obtain \(\eta\) with
\(\eta(x,\beta)\in P\) exactly off \(C\). Currying \(c_n\eta\) gives
the complement of the projection of \(C\) by \eqref{eq:fn-branches}.
Exponentiation preserves the truncation upper bound and constant
functions split evaluation at \(0^\omega\), exhibiting \(U_n\) as a
retract of \(V_n\). Thus \(V_n\) has the same exact height.

For the explicit locus, a stream \(a\) codes the tree \(T_a\) of
finite words \(s\) on which every prefix \(t\preceq s\) satisfies
\(a(\ulcorner t\urcorner)=0\). On input \((a,\beta)\), at stage \(r\)
output one if some prefix of \(\beta\) of length \(\le r\) has left
\(T_a\), and zero otherwise. This supplies a tracked \(\eta\) in
\eqref{eq:fn-branches} recognizing the absence of an infinite branch.
For completeness, given closed \(C\subseteq B^2\), code for each \(x\)
the tree of words \(s\) such that
\(([x|{|s|}]\times[s])\cap C\ne\varnothing\), where brackets denote
cylinders. Membership of \(s\) depends on \(x|{|s|}\), so these codes
vary continuously. By closedness a branch through this tree is precisely
a \(\beta\) with \((x,\beta)\in C\). This gives the claimed reduction.
\qed
\end{proof}

\subsection{Tail bounds and double-negation shift}

The code $v_n$ for $K(C_2,n)$ describes the fibre of every eventually
zero stream. A tail bound gives the coefficient isomorphism uniformly,
and a tracker for the isomorphism in turn supplies a bound. To express
this correspondence, let \(F\) be the subassembly of eventually zero
streams, each named by itself. For \(\alpha\in |F|\) let \(\ell(\alpha)\) be
the least \(S\) with \(\alpha(t)=0\) for all \(t\ge S\). Give the
same carrier the refined names \(\langle\alpha,\ell(\alpha)\rangle\)
to obtain \(F_{\min}\), with forgetful mono \(j:F_{\min}\into F\).
In the internal logic write
\[
 J(\alpha)=\left\|\sum_{S:N}\prod_{t:N}
          (S\le t\longrightarrow \alpha(t)=0)\right\|_{-1},
 \qquad I(\alpha)=(c_n(\alpha)=v_n).
\]
Each stream in the carrier of \(F\) has a finite tail bound. The
family \(J\) expresses local existence of such a bound over a varying
stream. We compare it with the \emph{support} of \(I\to F\), the
image in \(F\) with fibre \(\|I(\alpha)\|_{-1}\) at \(\alpha\).

\begin{lemma}[The support of a code identity]\label{lem:fn-support}
Over \(F\), \(J\) is represented by \(j\), and
\[
 J(\alpha)\simeq(a(\alpha)=a(0^\omega))_{Q},\qquad
 I(\alpha)\simeq J(\alpha)\times K(C_2,n).
\]
For \(-1\le r<n\), the relative \(r\)-truncation of \(I\) is \(J\)
and the truncation map admits a section over \(J\).
\end{lemma}

\begin{proof}
Valid pairs \((\alpha,S)\), named by the pair itself, represent the
untruncated sum in the definition of \(J\). On a valid pair the
universal tail-equality proof is tracked by the numerical equality
test on \(t\) and the stream entry at \(t\). By scanning the first
\(S\) entries one can compute \(\ell(\alpha)\). Hence the assembly
of valid pairs maps by a split cover onto \(F_{\min}\), with section
inserting the least bound. Composing with \(j\) is its projection to
\(F\), so image factorization proves the first identification.

An isomorphism \(C_2\cong M_\alpha\) is uniformly obtained from
a bound by the formulas for \([S,a]\). Conversely, applying its
tracker to the nonzero generator returns a representative \([s,1]\)
of a nonzero class. A reset at any \(t\ge s\) would kill this class,
so \(s\) is a valid bound. Normalizing to the least bound gives
inverse tracked maps between the assembly of coefficient isomorphisms
and \(F_{\min}\), since the coefficient isomorphism is unique.
Effectiveness of \(H\) identifies this object with the indicated
identity in \(Q\). The translation factor in the fibre equivalences
adds \(K(C_2,n)\) for the full identity in \(U_n\).
The projection to \(J\) has \((n-1)\)-connected fibres and the zero
point supplies a section. Since \(J\) is a proposition, the stated
relative truncations follow. \qed
\end{proof}

For any type \(X\) in the current context, write \(\neg X=(X\to\zero)\).
Since \(\zero\) is a proposition, the universal property of propositional
truncation gives \(\neg X\simeq\neg\|X\|_{-1}\). In particular,
\(\neg I(\alpha)\simeq\neg J(\alpha)\) by Lemma~\ref{lem:fn-support}.

\begin{theorem}[Pointwise and uniform double negation for tail bounds]
\label{thm:fn-shift}
For the proposition family \(J\),
\[
 \prod_{\alpha:F}\neg\neg J(\alpha)\simeq\one,\qquad
 \neg\neg\left(\prod_{\alpha:F}J(\alpha)\right)\simeq\zero.
\]
The same two equivalences hold with \(I\) in place of \(J\).
\end{theorem}
\begin{proof}
For an assembly mono \(A\into X\) with carrier image \(S\), its
pseudocomplement is the inherited subassembly on \(|X|\setminus S\).
Indeed, a disjoint subobject pulls back on an assembly atlas to a map
whose carrier avoids \(S\), since any pair of elements with the same
image would give a point of the pullback. The same tracker thus
factors the atlas through the inherited subassembly. Factorization
through a mono descends along the atlas. Applying this calculation
twice shows that the double pseudocomplement is the inherited
subassembly on \(S\). For \(F_{\min}\into F\) it is all of \(F\),
giving the first equivalence. Over \(\mathbb B\) this likewise
identifies the double negation of internal eventual zero with \(F\).

Consider any continuous function \(|F|\to\mathbb N\). Its value
\(s\) at \(0^\omega\) is constant on a relative cylinder determined
by \(k\) initial zeros. The stream with a single one at stage
\(t=\max(k,s)\) lies in that cylinder and has least tail bound \(t+1>s\).
This gives, for each continuous candidate, a stream on which its
value falls strictly below the least bound. A global section of \(j\)
requires a continuous bound on every stream. The internal product of \(J\) is subterminal.
Every noninitial object of \(\mathcal E\) has a point, by choosing a
name for an element of an assembly atlas. Thus \(\prod_{\alpha:F}J(\alpha)
\simeq\zero\), proving the second equivalence. The support of \(I\) is
\(J\) by Lemma~\ref{lem:fn-support}, and any section of \(I\) projects
to a section of \(J\). The two assertions follow for \(I\) as well.
\qed
\end{proof}

\subsection{Sections under classical reflection}

Double-negation sheafification makes the least tail bound available
uniformly. For the universal family of Proposition~\ref{prop:fn-classifier},
the same reflection gives a family in spaces whose relative
truncations have contractible section space through degree $n-1$.
The nonzero top homotopy group of one base component distinguishes
the full section space from those of the lower truncations.
The identification of the reflector with global sections lets us
compute both section problems explicitly.

Let \(\Sp_{<\infty}\) denote the spaces of finite truncation height
small at an admitted cardinal. Write \(\Gamma=\Map_{\mathcal E}(\one,-)\).
The terminal partitioned assembly is projective for completed covers.
Refine a lifting problem by an assembly atlas and use projectivity
for the resulting source cover. Thus \(\Gamma\) preserves effective
epimorphisms. It preserves finite limits by representability, hence
sends an effective groupoid and its quotient to a \v{C}ech nerve of
an effective epimorphism and its quotient in spaces. Moreover
$\Gamma X$ has admitted size. This holds on assemblies by the
carrier formula. It is stable under effective groupoid realization:
choose one admitted inaccessible for the spaces of objects and arrows.
All nerve terms are finite pullbacks of these spaces and their
countable simplicial realization stays small there. Exact generation
gives the size bound for every $X$. Representability preserves its
finite truncation bound as well.

\begin{theorem}[The double-negation reflection of function realizability]
\label{thm:fn-classical}
There is an equivalence
\(\Sh_{\neg\neg}(\mathcal E)\simeq\Sp_{<\infty}\) under which
sheafification is \(\Gamma\). Its fully faithful right adjoint
\(D_{\mathrm{cl}}\) extends the codiscrete assembly functor on sets.
\end{theorem}

\begin{proof}
Use the realization of \(\mathcal E\) inside space-valued regular
sheaves on \(\Asm(\mathcal K_2)\). For a bounded space \(K\) set
\[
 D_{\mathrm{cl}}K(A)=\Map_{\Sp}(|A|,K),
\]
where \(|A|\) is the discrete carrier of the assembly. A cover of
assemblies is surjective on carriers and the carrier functor preserves
finite limits. Mapping its carrier \v{C}ech nerve into \(K\) therefore
verifies descent. The construction preserves finite limits. It also
preserves effective groupoid quotients when taken in the ambient
sheaf category. If \(R_\bullet\) presents \(K\), then
\(R_0^{|A|}\to K^{|A|}\) is surjective on components by external
choice, and finite-limit preservation identifies its \v{C}ech nerve.
An objectwise effective epimorphism of sheaves is effective epic,
so \(D_{\mathrm{cl}}K\) realizes the resulting groupoid.

For discrete \(S\) the formula represents the assembly \(\nabla S\)
in which every name realizes every element, since a constant name
tracks any function into it. Induct on the height \(h\) of \(K\).
For \(h\ge1\), choose a discrete set of representative points covering
its components. In the \v{C}ech nerve the fibres at strings of
vertices are products of path spaces, of height at most \(h-1\).
All its terms therefore have height at most \(h-1\). The chosen set
and nerve fit a common permitted size bound. Induction followed by
effective realization puts \(D_{\mathrm{cl}}K\) in \(\mathcal E\).

Evaluation at \(\one\) identifies \(\Gamma D_{\mathrm{cl}}K=K\).
Consequently application of \(\Gamma\) defines a natural comparison
\[
 \Map_{\mathcal E}(X,D_{\mathrm{cl}}K)\longrightarrow
 \Map_{\Sp}(\Gamma X,K).
\]
It is an equivalence for assemblies by the sheaf formula and Yoneda.
Both sides send effective groupoid quotients to limits, since
\(\Gamma\) preserves these quotients. Since the assemblies
generate \(\mathcal E\) under such realizations, the
comparison is an equivalence for every \(X\). This
establishes the adjunction, the invertible counit, and left exactness
of \(L=D_{\mathrm{cl}}\Gamma\).

The nucleus is determined by its dense monos. We show that a mono
\(m:X\into Y\) is dense for double negation exactly when \(\Gamma m\)
is an equivalence. A dense mono pulls back at every point of
\(Y\) to a dense subterminal, which is terminal, since \(\one\)
has only the empty and full subobjects. Conversely if \(\Gamma m\)
is an equivalence, a point of the pseudocomplement \(\neg_Y X\) would
lift to \(X\) and belong to the empty intersection. An object with
no point is initial by the assembly atlas argument above. Thus
the pseudocomplement is initial. The adjunction shows that each
\(D_{\mathrm{cl}}K\) is orthogonal to every dense mono. For a
double-negation sheaf \(Z\), the unit \(A\to\nabla|A|\) for an
assembly \(A\) is such a mono, being the identity on carriers.
Therefore \(\Map(LA,Z)\to\Map(A,Z)\) is an equivalence. This
extends to every \(A\in\mathcal E\) by generation under effective
quotients, since \(L\) preserves them. At \(A=Z\), extending the
identity makes \(Z\) a retract of \(LZ\), hence local. This identifies
the reflective image with the double-negation sheaves. \qed
\end{proof}

\begin{corollary}[Two section problems after reflection]
\label{cor:fn-reflected-sections}
Classical reflection inverts \(F_{\min}\to F\), and the identity
family \(I\) then has a section. In the reflected category the
universal family \(u_n\) has empty section space, while its relative
\(r\)-truncation has contractible section space for \(-1\le r<n\).
\end{corollary}
\begin{proof}
The global points of \(F_{\min}\to F\) form a bijection. Applying the
reflector to the product decomposition of \(I\) now supplies the
section given by zero. To calculate \(\Gamma u_n\), lift points of
\(Q\) through \(a\) by terminal projectivity. The coefficient fibre
calculation and effectiveness of \(H\) identify two lifts exactly
when both have infinitely many resets or both have finitely many.
Thus \(\Gamma Q\) is a discrete space with
two points. Finite-limit preservation
identifies the two fibres of \(\Gamma U_n\) over this pair, and
preservation of effective quotients computes their finite deloopings.
Under Theorem~\ref{thm:fn-classical} the reflected family is therefore
\[
 \{0,1\}\longrightarrow *\amalg K(\mathbb F_2,n+1),
\]
selecting the two basepoints. On the second component a section is a
contraction of \(K(\mathbb F_2,n+1)\). Its nonzero top homotopy group
therefore makes the section space empty. Every fibre is either
contractible or equivalent to \(K(\mathbb F_2,n)\),
so each relative truncation in the indicated range is terminal.\qed
\end{proof}

\section{Induction and extraction in modified realizability}
\label{sec:modified-semantics}

Using projective covers in modified realizability, we give a direct
proof of dependent induction for the ordinary natural numbers $N$ in
the higher completion $\mathcal M=\HH(\mathsf{Mod})$. In every higher
parameter context, the zero and successor data for a bounded family
over $N$ determine a contractible space of sections equipped with the
prescribed computation paths. 

For extraction we use contextual choice and the interpretation of
program indices in the closed subtopos of the nested numerical model
described in Section~\ref{sec:unified-theorem}. Suppose a closed proof
gives, for each input satisfying a stable numerical precondition,
local existence of a numerical witness with a bounded higher
certificate. We extract one program that halts on every input,
together with a coherent certificate for its output whenever the
precondition holds.

\subsection{Projective presentations and the natural-number eliminator}

We use the numerical presentation of $\mathsf{Mod}$. As in
Section~\ref{sec:unified-theorem}, we write a truth value as an
ordered pair $(A^p,A^a)$ of potential and actual realizer sets.
In the numerical convention for closed truth values, $0\in A^p$
and $A^a\subseteq A^p$.
A projective has a presentation by sets
$X\subseteq Y\xrightarrow{r} I\subseteq\mathbb N$, with $r$ surjective.
The equality of $y,y'\in Y$ has potential part $\{0,r(y)+1\}$ on the
diagonal and $\{0\}$ off the diagonal. Its actual part is
$\{r(y)+1\}$ for $y=y'\in X$ and empty elsewhere. These objects cover
the ordinary topos \cite{VanOosten1997}. Refining a zero-type atlas by
one of these projective covers gives a projective atlas for every
completed object. To lift a map from an original projective through a
higher cover, pull the cover back to that projective and cover the
pullback by an ordinary projective. The composite onto the original
projective is an ordinary cover and splits.

An arrow of named inclusions before reflection is a commuting diagram
on the potential carriers, actual carriers and numerical labels. Its
function on labels is partial recursive and defined on the source
label set. After reflection each arrow is represented by such a diagram
on a refinement
\[
 (X\subseteq r^{-1}(J)\to J)\longrightarrow(X\subseteq Y\to I),
\]
where $r(X)\subseteq J\subseteq I$ and a partial recursive test
defined throughout $I$ decides membership in $J$. The reflector
inverts this inclusion \cite{VanOosten1997}. The labels
in these diagrams are the unshifted $r(y)$. The corresponding
positive equality names in the closed presentation are $r(y)+1$.

Before reflection, products pair the labels. For parallel maps with
potential-carrier functions $g,h:Y\to Y_1$, their equalizer has
potential carrier $Y_e=\{y:g(y)=h(y)\}$, actual carrier
$X\cap Y_e$, and label map $r|_{Y_e}:Y_e\to r(Y_e)$. Indeed any
equalizing map factors on its carriers through these subsets with
the same tracker. After reflection two arrows out of one projective
can be represented on a common refinement by intersecting the two
decidable tests. The equalizer just constructed there is projective
before reflection, and its reflection is projective and computes the
desired equalizer by left exactness. The same argument for the
terminal diagram and products gives finite-limit closure. Finite
coproducts of projectives are projective by the lifting property.
The partitioned natural-number object before reflection has diagram
$(\mathbb N\subseteq\mathbb N\xrightarrow{\rm id}\mathbb N)$.
Its reflection is $N$ since the reflector preserves the terminal
object and transports the universal recursion property by adjunction.
In particular $\one$ and $N$ are projective.

Fix a higher context $\Gamma$. A displayed natural-number algebra
consists of a bounded family $E$ on $\Gamma\times N$, a section
$z:E(0)$, and maps $d_n:E(n)\to E(n+1)$, all in that context.
Their object of solutions in the slice over $\Gamma$ is
\[
 \mathsf{Sol}(E,z,d)=\sum_{h:\prod_{n:N}E(n)}
       (h(0)=z)\times\prod_{n:N}(h(n+1)=d_n(h(n))).
\]
Its construction by finite limits and dependent products commutes
with change of $\Gamma$.

\begin{theorem}[Induction with coherent computation paths]
\label{thm:mod-induction}
The structure map $\mathsf{Sol}(E,z,d)\to\Gamma$ is an equivalence.
Consequently $\Gamma\times N$ is homotopy-initial among pointed
endomorphism algebras in $\mathcal M/\Gamma$.
\end{theorem}

\begin{proof}
First pull the algebra back to an original projective atlas $R$ of
$\Gamma$, and choose an original projective atlas $q:Q\cover E_R$ of
the total family. Put $p:Q\to R\times N$. The zero and successor maps
on the total family lift by projectivity of $R$ and $Q$ to
$z':R\to Q$ and $t:Q\to Q$, with specified paths $qz'\simeq z$ and
$qt\simeq d q$. Their images in $R\times N$ give ordinary equalities
$pz'=(\mathrm{id}_R,0)$ and $pt=(\mathrm{id}_R\times s)p$, where
$s$ is successor. All maps between original zero-types here are
ordinary maps. In $\mathsf{Mod}/R$ use ordinary recursion with
initial value $z'$ and step $t$ to obtain $v:R\times N\to Q$.
Ordinary uniqueness of recursion gives $pv=\mathrm{id}$. Hence $qv$
together with the two chosen paths is a displayed solution over $R$.
The base $R\times N$ is a zero-type, so these paths lie over the
required base equalities. It follows that $\mathsf{Sol}\to\Gamma$ is
a cover.

The identity family of two solutions is itself a displayed algebra. For a solution
$h$, write $\beta^h_{\mathrm{zero}}:h(0)=z$ for its zero computation path and
$\beta^h_{\mathrm{succ},n}:h(n+1)=d_n(h(n))$ for its successor computation path
at $n$. For two solutions $h,h'$, the displayed family
$D(n)=(h(n)=h'(n))$ has zero
$\beta^h_{\mathrm{zero}}(\beta^{h'}_{\mathrm{zero}})^{-1}$ and step
\[
 a\longmapsto \beta^h_{\mathrm{succ},n}\,\operatorname{ap}_{d_n}(a)
                         (\beta^{h'}_{\mathrm{succ},n})^{-1},
\]
where concatenation is read from left to right. By the identity
rule for dependent pairs and function extensionality, the identity
type between the solutions $h$ and $h'$ is equivalent to the
solution object of this displayed algebra on $D$.
Indeed after specifying $a_n:h(n)=h'(n)$, compatibility of the
successor paths is precisely
$a_{n+1}\beta^{h'}_{\mathrm{succ},n}
=\beta^h_{\mathrm{succ},n}\operatorname{ap}_{d_n}(a_n)$,
and compatibility of the zero paths is
$a_0\beta^{h'}_{\mathrm{zero}}=\beta^h_{\mathrm{zero}}$.

We now induct on the finite relative truncation bound of $E$, allowing
arbitrary parameter contexts at each step. For proposition-valued $E$, the
solution object is a proposition and a cover, hence terminal in the
slice. At the next height apply the induction hypothesis to $D$ for
the two universal solutions over
$\mathsf{Sol}\times_\Gamma\mathsf{Sol}$. The diagonal of
$\mathsf{Sol}\to\Gamma$ is an equivalence. A monomorphism that is a
cover is an equivalence, completing the induction. Constant displayed
families give the assertion about pointed endomorphism algebras.
The eliminator thus includes its propositional computation paths and
is coherent under substitution.\qed
\end{proof}

\begin{corollary}[Induction for lists and finite trees]
\label{cor:mod-finitary}
For original projective labels $A_0,\ldots,A_m$, the ordinary
initial algebra $W$ of $F(X)=\coprod_{i=0}^m A_i\times X^i$
has dependent elimination into bounded higher families, with a
contractible object of displayed sections. In particular this holds
for lists and finite binary trees with projective labels.
\end{corollary}
\begin{proof}
Ordinarily $W$ consists of finite rooted ordered trees. In the
internal language of an elementary topos with $N$, code each finite
shape by its finite word of vertex arities in prefix order, with the
usual pending-slot count positive up to its last vertex and zero at
the end. Attach to each vertex of arity $i$ an element of $A_i$.
Single-root decomposition and recursion on the length of the word
give the constructor and its unique fold into any algebra. This
construction works with an arbitrary ordinary slice parameter $R$,
giving initiality of $R\times W$ there. Lists and binary trees use
the polynomials $\one+A\times X$ and $\one+A\times X^2$.

For a projective $R$ take a projective atlas $Q\to E_R$ of the total
displayed family on $R\times W$. The underlying object of $F_R(Q)$
is a finite coproduct of products of $A_i$ with $i$-fold fibre
products of $Q$ over $R$, using $R$ for the empty fibre product.
It is projective. Therefore the total displayed constructor lifts to
$F_R(Q)\to Q$ with its comparison path. Its projection to $R\times W$
is an ordinary algebra map because that base is a zero-type. The
ordinary fold $R\times W\to Q$ is a section of this projection by
initiality. As in Theorem~\ref{thm:mod-induction} this proves local
existence. Between two displayed folds the identity family has
constructor given by applying the displayed operation to the tuple
of paths between the inputs and composing with the two computation
paths. The identity rule for dependent pairs identifies its sections
with paths of folds. Induction on truncation height now proves
contractibility in every context.\qed
\end{proof}

\subsection{Extracting total programs under stable preconditions}

Write $\|-\|$ for propositional truncation in the current slice. A
proposition $P$ is stable when $\neg\neg P\to P$. From a local witness
under $P$, we obtain locally a numerical output chosen independently
of a proof of $P$, together with a certificate valid under $P$.

\begin{lemma}[Choice under a stable premise]
\label{lem:mod-stable-choice}
For a stable subobject $p:P\into\Gamma$, $\Pi_p$ preserves covers.
In particular $\|\Pi_pY\|\simeq\Pi_p\|Y\|$ for every bounded family
over $P$.
\end{lemma}
\begin{proof}
In the nested tripos the topology $k_1$ is represented by testing
actual inhabitation. Its restriction to the closed subtopos is
double negation. Indeed, the corresponding embedding of sets is the
composite of the double-negation inclusion into $\mathsf{Mod}$ with
the closed inclusion into the nested topos
\cite{VanOosten1997}.
Thus on modified truth values a uniform representative is
\[
 (A^p,A^a)\longmapsto
     (\mathbb N,\{n\in\mathbb N:A^a\ne\varnothing\}).
\]
A stable subobject of a named projective $X\subseteq Y\to I$ is
therefore represented by a subset $X'\subseteq X$ while retaining
$Y\to I$. Intersecting the double-negation predicate with projective equality
retains exactly those diagonal actual names at which the predicate
has an actual realizer, and retains all the original potential names. Pairing
with a fixed numeral in the double-negation representative and
projecting back implement this identification uniformly. For every
subset $X'\subseteq X$, the resulting named inclusion is projective.

Given a cover $Y\cover Z$ over $P$, test $\Pi_pY\to\Pi_pZ$ on a
map from an original projective $R$ to $\Pi_pZ$. Its adjoint has
domain $R\times_\Gamma P$, a stable subobject of $R$. Subobjects
and their logical operations on original bases agree with the
ordinary ones, so the preceding argument makes this domain
projective. Lift along the cover and transpose back. Projectives
cover the target and hence detect covers. Applying the result to
$Y\cover\|Y\|$ gives the assertion since $\Pi_p\|Y\|$ is
propositional. All operations commute with parameter pullback.\qed
\end{proof}

For a stable $P$ and an admitted family of higher certificates $A(m)$,
numerical Independence of Premise takes the form
\begin{equation}
 \bigl(P\to\|\textstyle\sum_{m:N}A(m)\|\bigr)
 \ \longrightarrow\
 \bigl\|\textstyle\sum_{m:N}(P\to A(m))\bigr\|.
 \label{eq:mod-higher-IP}
\end{equation}
The witness $m$ is chosen locally, and under $P$ it carries a certificate
in the bounded type $A(m)$. Theorem~\ref{thm:mod-extraction} uses this
choice at each input to extract from a closed proof of its premise one
program that halts on every input, with a coherent certificate wherever the
precondition holds. For ordinary predicates, van Oosten derives numerical
Independence of Premise with a negative antecedent from a principle for
$k_1$-closed antecedents and diagonal witness objects in the nested
topos~\cite{VanOosten1997}. The numerical object is diagonal because its
actual and potential self-equality names agree, and a stable $P$ is
equivalent to the negative proposition $\neg\neg P$.

To prove \eqref{eq:mod-higher-IP} with higher certificates, take an
original projective test $R=(X\subseteq Y\xrightarrow{r}I)$ carrying
a term of the premise. Its restriction to $P$ has domain the
projective $R'=(X'\subseteq Y\xrightarrow{r}I)$ of
Lemma~\ref{lem:mod-stable-choice}. Projectivity lifts the term over
$R'$ through the cover from $\sum_m A(m)$ to its propositional
truncation. This gives a numerical map $f:R'\to N$ and a section
of $A(f)$ over $R'$.

Represent $f$ on a refinement $r^{-1}(J)$ of $R'$, where
$r(X')\subseteq J$ and membership in $J$ is decidable on $I$.
The label map $\varphi$ of its representing diagram is partial
recursive and defined throughout $J$. Since the label map of $N$
is the identity, the value on $y\in r^{-1}(J)$ is
$\varphi(r(y))$. First run the test for $J$ and return
$\varphi(i)$ on $J$ and $0$ off $J$. This partial recursive
function is defined on all of $I$. It gives the label map and,
by composition with $r$, the carrier map of an arrow
$\bar f:R\to N$. Its restriction to $R'$ agrees with $f$ on
the representing refinement, hence in $\mathsf{Mod}$. Transporting
the section of $A(f)$ across this equality gives a term of
$P\to A(\bar f)$ over $R$. The pair with $\bar f$ proves the
truncated conclusion there. For any parameter carrying the premise,
such tests cover it, and the conclusion descends because it is a
proposition. This proves the contextual implication.

\begin{lemma}[Indices and computation traces]
\label{lem:mod-truncated-CT}
Fix an acceptable numbering of partial recursive functions and codes
for their finite halting computations. Use a primitive recursive
trace predicate $T(e,n,s)$ and output map $\operatorname{out}(s)$
with $\varphi_e(n)\mathord\downarrow=m$ exactly when some $s$ satisfies
$T(e,n,s)$ and $\operatorname{out}(s)=m$. The model $\mathcal M$ validates
\[
 \prod_{f:N^N}
 \left\|\sum_{e:N}\prod_{n:N}\sum_{s:N}
       T(e,n,s)\times(\operatorname{out}(s)=f(n))\right\|.
\]
\end{lemma}
\begin{proof}
Use the modified-assembly presentation of the double-negation-separated
objects of $\mathsf{Mod}$ \cite{VanOostenSimpson2000}.
An ordinary object $X$ is \emph{double-negation-separated} when its
diagonal is double-negation-closed. This says that equality in $X$ is
stable, meaning that $\neg\neg(x=y)\Rightarrow(x=y)$ holds internally
for all $x,y:X$.
Each carrier element $x$ has a nonempty set $|x|$ of actual names
inside a common nonempty set of potential inputs. A tracker is
defined on every potential input and sends actual names to actual
names of the output. In this presentation $|n|=\{n\}$ and the
potential set of $N$ is $\mathbb N$. Its exponential $N^N$ has
the total recursive functions as carrier, their indices as actual
names, and $\mathrm{Tot}=\{e:\varphi_e\text{ is total}\}$ as its
potential set. The separated objects form a full subcategory closed under finite
limits and exponentiation into a separated object. The exponential
here therefore agrees with the one in the topos. Images of
maps between separated objects are separated subobjects of their
codomains, so the image factorizations agree as well.

Let $I$ have carrier and potential set $\mathrm{Tot}$ and singleton
actual names $|e|=\{e\}$. The map $I\to N^N$, $e\mapsto\varphi_e$,
is tracked by the identity and is a cover because the actual names of
each function are the union of the actual names in its fibre and the
potential sets agree. On $I\times N$, least-stage search for a trace
terminates on every potential input $(e,n)$ precisely because that
domain requires $e\in\mathrm{Tot}$. The search returns the numeral
coding the first halting trace. Primitive recursive checking realizes
$T$ on this trace, and determinism identifies the recorded output
with the function value. These operations therefore give tracked witnesses for the displayed
formula after this cover of $N^N$. Its image over $N^N$ is all of
$N^N$. The zero-type inclusion into $\mathcal M$ preserves these
finite limits, dependent products on original families, and covers.
Pullback establishes the assertion for generalized $f$ as well.\qed
\end{proof}

\begin{theorem}[Total extraction with stable preconditions]
\label{thm:mod-extraction}
For stable numerical propositions $P(n)$ and an admitted bounded
family $C(n,m)$, there is a contextual implication
\[
\begin{aligned}
 &\prod_{n:N}\bigl(P(n)\to\|\textstyle\sum_{m:N}C(n,m)\|\bigr)\\
 &\quad\longrightarrow
 \left\|\sum_{e:N}\prod_{n:N}\sum_{s:N}
   T(e,n,s)\times\bigl(P(n)\to C(n,\operatorname{out}(s))\bigr)\right\|.
\end{aligned}
\]
For closed data and a closed term of the premise this supplies a
standard index for a program total on every input and a coherent
certificate on every input satisfying its precondition.
\end{theorem}
\begin{proof}
Apply \eqref{eq:mod-higher-IP} separately in the context of each $n$.
Since $N$ is projective and the original projectives are closed under
products, Corollary~\ref{cor:projective-choice} gives contextual
countable choice for bounded higher families over $N$. It gives local
existence of an $f:N^N$ together with a term of
$\prod_n(P(n)\to C(n,f(n)))$. The conclusion is a proposition, so
eliminate this truncation and the truncation of
Lemma~\ref{lem:mod-truncated-CT}. For each input transport the
certificate along the equality $f(n)=\operatorname{out}(s)$ of the
chosen trace and then form the displayed dependent products. Finally
take propositional truncation. Over a projective parameter one can
lift the resulting cover of witnesses to an actual section. For the
terminal parameter, the projective presentation identifies
$\Map(\one,N)$ with $\mathbb N$: the terminal named inclusion has one
actual point, a map chooses its numerical value, and the constant
program realizes every such choice. Thus each closed premise witness
yields a standard index for the extracted total program.\qed
\end{proof}

\section{Predicates, choice, and second identities in Krivine realizability}
\label{sec:krivine-semantics}

In the relative Boolean presentation of Krivine realizability, the
logical pole is a proper subterminal $U$ of the ordinary relative
topos. The ordinary tripos is the Booleanization for
$b_U(p)=((p\Rightarrow U)\Rightarrow U)$
\cite{VanOostenZou2016}.
For an admissible filtered presentation, Theorem~\ref{thm:bounded-reflection}
lifts this nucleus to the reflection
$a:\mathcal R\rightleftarrows\mathcal K:i$ on
$\mathcal R=\HH(\RT(A',A))$. Under $i$ the
false proposition in $\mathcal K$ corresponds to $U$ and the true
proposition to $\top$. Thus $\mathcal K$ is nondegenerate.
In the absolute $\mathcal K_1$ model every inhabited realizer set
contains a filter element, so its proper pole is $\bot$ and $b_U$ is
ordinary double negation.
Every predicate in $\mathcal K$ splits its parameter into two
complementary pieces. By choosing a bundle on each piece we can
record the predicate in the existence of a section. The same
decomposition will let us record it in an identity between two
self-equivalences of a fixed type.

For $V\into Z$ in the ambient model, put $U_Z=Z\times U$, and
let $\eta_Z:Z\to iaZ$ be the unit. The interpretation of target
negation in $\mathcal R/Z$ is
\begin{equation}
  \eta_Z^*i(\neg_{aZ}aV)=(V\Rightarrow U_Z).
  \label{eq:kr-relative-complement}
\end{equation}
Indeed the local predicates over $Z$ are the $b_{U_Z}$-fixed elements
of $\Sub(Z)$, with bottom $U_Z$. Relative negation is
$p\mapsto (p\Rightarrow U_Z)$, and triple relative negation equals
single relative negation. All finite cyclic groups below are the intrinsic
ones in the target. Set $T_n=K(C_2,n)$, with $T_0=C_2$, and write
$2=\one\amalg\one$ for the underlying binary object.

\subsection{Encoding a predicate with fixed base and fibre}

Two bundles with the same base and fibre let us encode a predicate
in their section objects. The product bundle supplies a section on
one piece of the parameter, and a cyclic extension gives an empty
section object on the complementary piece.

\begin{lemma}[The connecting map of the cyclic extension]
\label{lem:kr-cyclic}
For $n\geq1$ the extension
$0\to C_2\xrightarrow{1\mapsto2}C_4\xrightarrow{\bmod 2}C_2\to0$
induces a pointed map $\beta_n:T_n\to T_{n+1}$. Its homotopy fibre
over zero is $K(C_4,n)$ with its quotient map $p_n$ to $T_n$, and
its internal type of nullhomotopies is initial.
\end{lemma}
\begin{proof}
The cyclic extension gives the asserted fibre sequence by
Lemma~\ref{lem:principal-sections}. A nullhomotopy of $\beta_n$
gives a section of $p_n$. In every parameter context, subtract the value
of the section at zero to obtain a pointed section. Applying
this operation to a homotopy $p_ns\simeq\mathrm{id}$ also subtracts
the path at zero. The loop calculation then gives a group
homomorphism $C_2\to C_4$ splitting reduction modulo two.
Every possible value of the generator is killed by two and is
therefore $0$ or $2$ in $C_4$. Its reduction would have to be
both zero and one. This supplies internally a map from the
nullhomotopy type to the empty type, as claimed.\qed
\end{proof}

\begin{theorem}[A predicate as a bundle with fixed base and fibre]
\label{thm:kr-fixed-fibre}
For $S\into C$ in $\mathcal K$ and $n\geq1$ there is a principal
$T_n$-bundle $q_{S,n}$ over $C\times T_n$. With dependent product
along $C\times T_n\to C$, its full internal section object is
\[
                 \operatorname{Sec}_C(q_{S,n})
                 \simeq(\neg_C S)\times 2\times T_n
                 \quad\text{over }C.
\]
Two such bundles over the identity of the same base are equivalent precisely when
their defining predicates coincide. For $C=aZ$ and $S=aV$, the
bundle admits a section precisely when $V\leq U_Z$.
\end{theorem}
\begin{proof}
Over $C=S\amalg\neg_C S$, glue $\beta_n$ on $S$ to the zero map on
$\neg_C S$, giving a pointed family of maps $T_n\to T_{n+1}$
parametrized by $C$. Pull back the universal $T_n$-torsor
$\one\to T_{n+1}$ along this family. Over $S$ the resulting bundle
is $K(C_4,n)\to T_n$ and Lemma~\ref{lem:kr-cyclic} makes its
section object initial there. On $\neg_C S$ it is the product
bundle with section object
\[
   T_n^{T_n}\simeq T_n\times
       \underline{\Hom}_{\Ab}(C_2,C_2)\simeq T_n\times 2.
\]
Here Lemma~\ref{lem:finite-bars} separates the value at zero from
the pointed map, and a binary endomorphism is specified by the
image of its generator. Beck--Chevalley identifies both local
computations with restrictions of the section object. Gluing gives
the formula and shows that its propositional truncation is
$\neg_C S$, since the two other factors have points. Taking the
complement recovers $S$ from the bundle. Conversely $S$ specifies
the two pieces of the bundle, compatibly with pullback.
For $S=aV$ the section condition is $\neg_{aZ}aV=\one_{aZ}$,
which by \eqref{eq:kr-relative-complement} is $V\leq U_Z$.\qed
\end{proof}

\subsection{The first fibre degree at which choice detects the pole}

In the preceding construction the predicate determines a bundle
over a fixed index. For the choice test it determines the index
itself. Varying the fibre height then identifies the first degree at
which choice recovers the pole. To express the finite-height test of
Corollary~\ref{cor:projective-choice} internally, fix a univalent
universe $\mathcal U_{\leq k}$ of $k$-types.
For an index $Z$, the choice proposition is
\[
 \operatorname{Ch}_{k}(Z)=
   \prod_{F:Z\to\mathcal U_{\leq k}}
   \left(\left(\prod_{z:Z}\|F(z)\|\right)
             \longrightarrow
         \left\|\prod_{z:Z} F(z)\right\|\right).
\]
Here $\|X\|=\tau_{\leq-1}X$. The product over $F$ tests the families
classified by $\mathcal U_{\leq k}$ and defines a proposition in a
universe containing these products.
For a local proposition $P$, let $M_P$ be the
abelian group equal to $C_2$ on $P$ and to the trivial group on
$\neg P$, with underlying set $\one\amalg P$. The construction
is fibrewise in any parameter. Write $Z_{P,n}=K(M_P,n)$, $n\geq1$.

\begin{theorem}[The choice threshold]
\label{thm:kr-choice-threshold}
At any admitted test level, large enough in the second line to
include the based path family of $Z_{P,n}$,
\[
 \operatorname{Ch}_{k}(Z_{P,n})\simeq
 \begin{cases}
 \one,&-1\leq k\leq n-2,\\
 \neg P,&k\geq n-1 .
 \end{cases}
\]
For $P=aV$ with an ambient subterminal $V$, validity in the second
range is equivalent to $V\leq U$. At any one fixed $n$, these tests,
indexed by ambient truth values, determine the logical pole $U$.
\end{theorem}
\begin{proof}
Write $Z=Z_{P,n}$ and $z_0$ for its point. If $k\leq n-2$, the
univalent classifier of $k$-types is at most $(k+1)$-truncated,
while $Z$ is $(n-1)$-connected. Consequently evaluation at $z_0$
identifies maps $Z\to\mathcal U_{\leq k}$ with the classifier
itself, also in every parameter context. Each tested family is
therefore equivalent to the constant family on $F(z_0)$. Its
pointwise mere inhabitance provides $\|F(z_0)\|$, and
constant sections give the required truncated section.

For $k\geq n-1$, test the family $z\mapsto(z_0=z)$. Its fibres are
$(n-1)$-truncated and locally inhabited at each point. A section is
a contraction with centre $z_0$. Since $\isContr(Z)$ is a proposition,
local existence of a section implies contractibility. Taking $n$-fold
loops of a contractible $Z$ gives the trivial group $M_P$. Delooping
the trivial group proves the converse.
By its Boolean decomposition this condition is exactly $\neg P$.
Conversely on $\neg P$ the base is contractible, and evaluation at
its centre gives choice for every family. This proves the formulas
as equivalences of internal propositions. Now apply
\eqref{eq:kr-relative-complement} with $Z=\one$ to obtain
$i\operatorname{Ch}_{k}(Z_{aV,n})=(V\Rightarrow U)$ in the second
range. Thus the ambient truth values $V$ for which the choice test holds
form the principal downset of $U$. Its greatest element recovers $U$.\qed
\end{proof}

For instance, choice holds over $K(C_2,2)$ for all set-valued families.
Its based path family has groupoid fibres and locally inhabited paths
to every point. Sections of this family are contractions of $K(C_2,2)$,
so its nonzero second homotopy group makes the section type empty.
Thus groupoid-valued families distinguish choice over $K(C_2,2)$
from choice over a point at the first possible fibre degree.

\subsection{Totality as a second identity at one code}

The cyclic extension in Lemma~\ref{lem:kr-cyclic} also gives an
identity test at a single universe code. Fix $n\geq1$ and
choose a code $u$ for the type
$X=T_n\times T_{n+1}$. For a local predicate $S$ let $s_S$ be
the self-equivalence of $X$ given by
$(x,y)\mapsto (x,y+\beta_n(x))$ on $S$ and the identity on its
complement. Subtracting the same term gives its inverse. Both
pieces fix the basepoint and induce the identity on each based
homotopy group: $\pi_n$ comes from the unchanged first coordinate
and $\pi_{n+1}$ from the second coordinate over the first
basepoint. Univalence gives a loop $p_S:u=u$ in the parameter
context.

\begin{lemma}[Second identities of the shear]
\label{lem:kr-shear}
For $S\into C$, the identity type between
$p_S$ and the reflexivity loop at $u$ is
\[
    (p_S=\operatorname{refl}_u)
     \simeq (\neg_C S)\times T_{n-1}\times T_n\times 2
              \quad\text{over }C .
\]
\end{lemma}
\begin{proof}
On $S$ a homotopy from the shear to the identity, restricted along
$x\mapsto(x,0)$ and projected to the second factor, would contract
$\beta_n$ to the zero map. This gives the empty type by
Lemma~\ref{lem:kr-cyclic}. On $\neg_C S$, univalence identifies the second
identity type with the loop space of the object of self-equivalences of
$X$ at the identity map. Since being an equivalence is a proposition, this
is also the loop space of $X^X$ at the identity map.
Pointwise subtraction of the identity map translates the basepoint
to the zero map in this abelian group object. Exponentiation
preserves finite limits, giving
\[
 \Omega_0(X^X)\simeq(\Omega X)^X
          \simeq(T_{n-1})^X\times(T_n)^X .
\]
Since $\tau_{\leq n-1}X=\one$ and $\tau_{\leq n}X=T_n$, the first
factor is $T_{n-1}$ and the second is $T_n^{T_n}\simeq T_n\times2$.
These calculations hold after every base change and include
$n=1$. Disjoint gluing proves the claim.\qed
\end{proof}

To express totality by the shear, form the numerical predicate in the
ambient relative model and reflect it to a Boolean predicate. Let $N$
be the numerical assembly of the relative topos.
Assume that its standard numerals have filter names, that primitive
recursive tests have uniformly tracked true instances and empty
realizer sets at false standard instances, and that pairing,
witness formation and partial numerical computation have filter
trackers on their domains. In the ordinary realizability
interpretation of quantifiers an existential name includes a
witness and its realizer, and a universal name acts on each input
name. The ambient numbers carry their numerical realization relation
throughout this construction.

Kleene's first algebra with its full filter has these properties
by application of ordinary partial recursive indices. They also
hold for the relative second algebra with the filter of recursive
Baire functions. Represent the numeral $m$ by the constant Baire
function $\nu_m$. To convert to a Church numeral, read $m$ at
coordinate zero and construct the code for its $m$-fold iterator
by primitive recursion. Conversely apply a Church numeral to the
recursive successor on constant streams and to $\nu_0$. Successor
reads coordinate zero and returns the corresponding successor
stream, so iteration returns $\nu_m$, the name of $m$. Both conversions
have trackers in the recursive filter. In the prefix convention
for application, output coordinate $j$ is one less than the first
positive answer on a code for $j$ and a finite input prefix. Given a
program $e$, use the total recursive Baire function which on a
nonempty prefix $\sigma$ simulates $e$ on $\sigma(0)$ for
$|\sigma|$ steps. Return $q+1$ if it halts with output $q$, and
zero otherwise, including on the empty prefix and malformed codes.
On $\nu_m$ its application is $\nu_{\varphi_e(m)}$ when defined
and is undefined when the computation diverges. The construction
is effective in $e$. Pairing, bounded simulation, comparison and
the formation of a witness for a true decidable test are obtained
by this construction and recursive combinatory abstraction. Thus the
recursive filter supplies the trackers, and the ambient algebra
supplies all Baire functions as potential realizers.

\begin{theorem}[Second-identity completeness at a fixed code]
\label{thm:kr-second-identity}
Under these numerical hypotheses, for every fixed proper pole and
every $n\geq1$ there is a semantic family of loops $p$ at $u$ over
$aN$. At the reflected standard index $e$ it satisfies
\[
 (p_e=\operatorname{refl}_u)\simeq
 \begin{cases}
   T_{n-1}\times T_n\times2,&\varphi_e\text{ total},\\
   \zero,&\text{otherwise}.
 \end{cases}
\]
The standard indices with inhabited second identity form a
$\Pi^0_2$-complete set under computable many-one reductions.
All these loops induce identical actions on based homotopy groups.
\end{theorem}
\begin{proof}
Let $H(e,x,s)$ denote the primitive recursive test for termination
on $x$ within $s$ steps. In the ordinary \emph{ambient} relative
topos form
\[
    T(e)=\forall x:N\,\exists s:N\, H(e,x,s),\qquad V=\neg T\into N .
\]
At a standard total index $e$, least-stage search is total. A
filter tracker pairing the resulting numeral with the verification
of the bounded test therefore realizes the universal predicate,
giving $T(\bar e)=\one$. If $e$ diverges at $x_0$, the witness
set for $H(e,x_0,s)$ is empty for every $s$. Evaluation at a name
of $x_0$ therefore makes the entire set of potential universal names
empty, so $T(\bar e)=\zero$. Ambient negation interchanges these
two possibilities. Reflect the resulting subobject $V\into N$.
Left exactness preserves the pullback to each numeral and the left
adjoint preserves the initial object. Thus the Boolean predicate
$S=aV$ is false at total standard indices and true at the others.
By Lemma~\ref{lem:kr-shear}, the product $T_{n-1}\times T_n\times2$
supplies a point of the second identity at every total standard index.
At each other standard index the identity is initial, with empty global
point space because the pole is proper.

The displayed formula gives the $\Pi^0_2$ upper bound by the
numerical definition of $T$. For any recursive predicate $R$, the
index of the program on input $x$ searching for an $s$ with
$R(z,x,s)$ depends computably on $z$. This program is total
exactly when $\forall x\,\exists s\,R(z,x,s)$, proving hardness.
For each fixed pole, substituting the standard numerals into the same
semantic family gives all these loops. At a total
index the second identity type has exact height $n$,
since its factor $T_n$ has nonzero $\pi_n$.\qed
\end{proof}

\section{Grothendieck slices of unbounded localic rank with a common ordinary core}
\label{sec:grothendieck-ranks}

For the effective contexts $K(A_r,n)$ of
Section~\ref{sec:separation-specialization}, the loop space at the
basepoint determines the minimum dimension of a regular source for the
slice. Interpreting the same groups $A_r$ in sheaves of spaces on
countable assemblies gives a second dimension calculation. The least
dimension of a site presenting the entire Grothendieck slice is $n$,
and binary cohomology distinguishes the slices at each fixed $n$.

Let $\Asm_\omega$ be a small skeleton of the category of assemblies
over Kleene's first algebra with countable carrier. It is regular
and extensive. Indeed finite limits, regular images and disjoint
tagged finite sums in assemblies retain countable carriers; their
canonical factorizations and pullbacks are therefore computed in
this full subcategory. Let $J$ consist of the finite covering families
$\{U_i\to U\}$ for which $\coprod_i U_i\to U$ is a regular
epimorphism, including the empty family at the initial object.
Fix a Grothendieck universe $\mathbb U$ containing $(\Asm_\omega,J)$.
Taking spaces and sets in $\mathbb U$ gives a common size convention
for the sheaf categories and their bounded exact-completion comparison. Set
\[
 \mathcal G=\Sh_J(\Asm_\omega,\Sp),\qquad
 y:\Asm_\omega\longrightarrow\mathcal G .
\]

\begin{lemma}[The assembly site]
The families $J$ define a subcanonical pretopology. The sheaf Yoneda
embedding $y$ preserves finite limits, finite coproducts and covers.
The terminal object of $\mathcal G$ is projective.
\end{lemma}
\begin{proof}
The axioms for covers follow from pullback stability of regular
epimorphisms, extensivity, and stability under composition and finite
sums. A compatible family into any representable gives a map
$\coprod U_i\to W$ coequalizing the kernel pair of the cover onto
$U$. The unique induced map $U\to W$ gives subcanonicity. This
proves full faithfulness and finite-limit preservation of $y$.
Singleton regular epimorphisms are covers. A sheaf of spaces $F$
sends $0$ to $1$, and descent for the two summands of $U\amalg W$,
whose cross-intersections are empty, gives
$F(U\amalg W)\simeq F(U)\times F(W)$. Yoneda gives preservation of
finite coproducts. In particular $yD$ is the intrinsic binary group.

In a finite covering family of the terminal assembly some member has
a carrier element. Choosing one of its names gives a section of that
member. Every covering sieve of $1$ is therefore maximal. Since
the terminal sheaf is $y1$, global sections evaluate at $1$.
An effective epimorphism of sheaves is locally surjective on
components. For a point over $1$, restrict a local lift along the
section of a covering member to obtain a global lift up to homotopy.
This proves projectivity. Finite limits and such covers are preserved
by global sections, hence so are effective groupoid quotients, by the
\v{C}ech argument in Theorem~\ref{thm:effective-observers}.\qed
\end{proof}

Recall that an $\infty$-topos is $q$-localic if it admits a site on a
small $q$-category with finite limits; for $q\geq1$ such a category has
$(q-1)$-truncated mapping spaces. We call the least such $q$ its
\emph{localic rank}. All $A_r$ have countable carriers. For $r\geq0$
and $n\geq2$, set
\[
 \widetilde B_{r,n}=K_{\mathcal G}(yA_r,n),\qquad
 \mathcal G_{r,n}=\mathcal G/\widetilde B_{r,n}.
\]

\begin{theorem}[Localic rank and binary cohomology]
\label{thm:grothendieck-rank}
Let $p:\widetilde B_{r,n}\to\one$ be the projection. For $0\leq q\leq n-2$,
pullback along $p$ gives the equivalence
\[
 p^*:\mathcal G_{\leq q}\xrightarrow{\;\sim\;}(\mathcal G_{r,n})_{\leq q}.
\]
In particular the ordinary core is always
$\Sh_J(\Asm_\omega,\mathbf{Set})$. The localic rank of
$\mathcal G_{r,n}$ is exactly $n$, and its intrinsic binary coefficient
satisfies
\[
 H^n(\mathcal G_{r,n};p^*(yD))\cong\mathbb F_2^r .
\]
Two of these categories are equivalent precisely
when their parameter pairs $(r,n)$ agree.
For fixed $n$ their bases have equivalent global point spaces
$\Gamma_{\mathcal G}\widetilde B_{r,n}\simeq K(V,n)$.
\end{theorem}
\begin{proof}
The proof of Theorem~\ref{thm:relative-context-cohomology} applies in
$\mathcal G$ degree by degree. A classifier of $q$-truncated families
has height at most $q+1$, so descent along the $(n-1)$-connected base
identifies the low-type categories.
Projectivity identifies $\Gamma_{\mathcal G}K(M,k)$ with
$K(\Gamma_{\mathcal G}M,k)$ for an abelian zero-type $M$ and $k\geq1$,
hence positive-degree base cohomology vanishes. Taking $M=yD$ and
using full faithfulness and finite-limit preservation of $y$, the
remaining degree-$n$ group is
\[
 \Hom_{\Ab(\mathcal G_{\leq0})}(yA_r,yD)
 \cong \Hom_{\Ab(\Asm_\omega)}(A_r,D)
 \cong \mathbb F_2^r
\]
by the common-realizer argument of
Proposition~\ref{prop:separation-recovered}. The same global-sections
calculation gives $\Gamma_{\mathcal G}\widetilde B_{r,n}\simeq K(V,n)$,
since $\Gamma_{\mathcal G} yA_r\cong\mathbb F_2^r\oplus V\cong V$.

The ordinary site makes $\mathcal G$ $1$-localic and hence $n$-localic.
Slicing an $n$-localic topos over an $n$-truncated object is $n$-localic
\cite{LurieSAG}, giving the upper bound. For the lower bound, recall
that in a $q$-localic topos with $q\geq1$, sheafified representables
are $(q-1)$-truncated by left exactness of sheafification. Their
coproducts have the same truncation bound and cover every object,
since the site gives a generating family. Applied to the basepoint
object, Lemma~\ref{lem:general-basepoint-obstruction} in $\mathcal G$
then bounds the height of $\Omega\widetilde B_{r,n}$ by $q-1$.
This loop object is $K(yA_r,n-1)$ and has exact height $n-1$, since
$yA_r\ne0$ by full faithfulness, including for $r=0$. In particular
the slice has rank greater than $n-1$, since a site of any smaller
dimension is also an $(n-1)$-site. Both the localic
rank and the group law on the canonical binary coefficient are
invariant under equivalence, so the two computed invariants recover
$n$ and $r$ in the unmarked category.\qed
\end{proof}

To compare these slices with the bounded construction, first consider
the unsliced $\mathcal G$. Every object has a cover by a coproduct of sheafified
representables, which is a zero-type. The full subcategory
$\mathcal G_{<\infty}$ of finitely truncated objects is exact:
for a groupoid of such objects, a bound for its arrow and object
spaces bounds the diagonal of its quotient after the object cover,
as in Proposition~\ref{prop:input-completion}. Its zero-types are the
ordinary sheaves of sets. Hence bounded exact-completion recognition
gives
\[
 \operatorname{Ex}_\infty\bigl(\Sh_J(\Asm_\omega,\mathbf{Set})\bigr)
 \simeq \mathcal G_{<\infty}.
\]

The basepoint gives the sharp dimension $n$ in both settings. In the
bounded effective model, $n$ is the minimum dimension of a regular
source for exact completion. In the Grothendieck model, it is the
minimum dimension of a site for the whole slice. At each fixed $n$,
the Grothendieck slices share their ordinary sheaf topos and their bases
have equivalent global point spaces. The rank of degree-$n$ binary
cohomology recovers $r$, the dimension of the space of tracked
homomorphisms from $A_r$ to $D$.

\section{Conclusion}

The tripos-to-topos construction gives ordinary realizability a semantics
in sets and maps. Higher tripos theory extends this interpretation to 
families of types, where a path between classifying maps witnesses 
a uniform equivalence of families. In function realizability, this connects 
identity types to descriptive set theory. To each tree on the natural 
numbers we assign a map from Baire space to a univalent classifier. 
There is a global path from this map to the constant map at the 
terminal code precisely when the tree is well founded.

Choosing a section has a proof-theoretic interpretation. In modified
realizability, suppose a closed proof gives local existence of a witness
in an admitted bounded family whenever a stable numerical precondition
holds. We can then extract a program that halts on every input and
returns a certified witness on each input satisfying the precondition.
In Krivine realizability, choice of sections recovers the pole $U$ that
defines the Boolean localization. Fix $n\geq1$. For each ambient 
truth value $V$, we construct a pointed $n$-type on which choice 
for $(n-1)$-truncated families holds precisely when $V\leq U$, 
provided the test universe contains the family of paths from the basepoint.

Higher types also distinguish the roles of logic and context. At any
fixed degree $n\geq1$, the locus where sheafification makes the fibres
of one generic family contractible recovers the Lawvere--Tierney nucleus.
With the logic fixed, varying the context can still yield infinitely
many higher toposes with Hyland's effective topos as their category of
zero-types. Binary cohomology distinguishes these examples at each fixed
degree $n\geq2$, even when their bases have equivalent spaces of global
points. For the base $K(A_r,n)$, the degree-$n$ classes correspond to
tracked homomorphisms from $A_r$ to the two-element group. The rank
therefore counts the independent binary linear functionals that a
realizer can compute.

\appendix
\section{Tracked atlases and small universes}\label{app:source-presentations}

For relative and nested assemblies we construct the projective atlases,
section codes, and classifying families of
Proposition~\ref{prop:input-completion}.
Theorem~\ref{thm:source-verification} uses two bounds to encode a
family. The cardinal bound measures each atlas fibre over a projective
test and thus works uniformly for tests at every admitted size. The
truncation bound determines how many simplicial degrees the code needs.

\subsection{Relative and nested assemblies}

Let \(A\) be an ordered partial combinatory algebra. Application is
written by juxtaposition and associates to the left. If \(ab\) is
defined, then \(a'b'\) is defined and lies below \(ab\) whenever
\(a'\le a\) and \(b'\le b\). Fix a filter \(A'\) of global trackers,
closed under defined application and containing realizers of the
combinators \(k,s\). We use their usual ordered abstraction property:
lambda terms built from the indicated operations have uniform trackers
in \(A'\), with the value of an abstraction bounded above by the
value of the specified term whenever that term is defined. In
particular, there are uniform pairing, projection and finite case
operators. Write \(\langle a,b\rangle\) for pairing and take a single
bracket abstraction \(c\in A'\) for \(\lambda d\,r\,b.\,d\langle r,b\rangle\).
Its first two applications are total. Set \(\sigma(d,r)=c d r\);
in particular \(r'\le r\) implies \(\sigma(d,r')\le\sigma(d,r)\), and
\[
       \sigma(d,r)b\le d\langle r,b\rangle
       \quad\text{whenever the right hand side is defined}.
\]
In a discrete algebra the comparison is equality wherever the right
hand side is defined.

A \emph{relative assembly} consists of a set \(X\) and nonempty
downward-closed name sets \(\rho_X(x)\subseteq A\). A function
\(f:X\to Y\) is a morphism when a single \(t\in A'\) sends every name
of \(x\), by defined application, to a name of \(f(x)\).
A \emph{partitioned} assembly has a specified principal name
\(\nu_P(p)\), with \(\rho_P(p)=\mathord\downarrow\nu_P(p)\).

For the nested case, let \(A_\#\subseteq A\) be a sub-PCA of a
discrete PCA, containing common combinators and closed under
application whenever the ambient application of two of its elements is
defined. A \emph{two-level assembly} is an arrow of sets
\(\alpha_X:X_a\to X_p\), with nonempty name sets at each level
satisfying
\[
      \rho_p(y)\subseteq A,\qquad
      \rho_a(x)\subseteq A_\#\cap\rho_p(\alpha_X x).
\]
The subscripts \(p,a\) mean potential and actual. Morphisms are
commuting squares of functions with one tracker in \(A_\#\) acting at
both levels. Partitioned objects are named inclusions
\(P_a\subseteq P_p\) with a function \(\nu_P:P_p\to A\) and
\(\nu_P(P_a)\subseteq A_\#\). At both levels the names are
singletons, with the same name on the inclusion. Let \(\mathcal P\)
denote the replete class of the respective partitioned objects.
Whenever a calculation uses principal names for \(P\in\mathcal P\), we
choose a partitioned representative and transport the construction
along the chosen isomorphism to \(P\).

\begin{lemma}[Projective covers and section objects]\label{lem:src-projectives}
Both assembly categories are regular with stable disjoint finite
coproducts. Their partitioned objects form finite-limit-closed
projective covers, and both categories admit dependent products.
Their ordinary exact/regular completions are respectively the relative
and the nested realizability toposes.
\end{lemma}
\begin{proof}
Finite limits use products of names and restrictions to equalizer
carriers. A terminal partitioned object has a fixed global name.
Tagged disjoint unions give the coproducts, with a uniform case
operator tracking their universal property. Pulling back a tagged sum
commutes with the carrier decomposition and pairing gives trackers
for the two comparison functions. For a morphism \(e:E\to B\) the
image at each level has the carrier image and the union of the name
sets of its preimages. Thus \(e\) is a cover exactly when a single
global \(c\) satisfies
\[
   r\in\rho_B(b)\quad\Longrightarrow\quad
   cr\ \text{is defined and belongs to}\
         \bigcup_{e(x)=b}\rho_E(x).
\]
In the nested case this formula holds on both levels with the same
\(c\). Pairing shows that this factorization is stable under pullback.
A tracked map constant on the kernel pair descends as a carrier
function, with tracker \(r\mapsto d(cr)\) when \(d\) tracks the given
map. This verifies regularity.

Every relative assembly is covered by the partitioned assembly of
pairs \((x,r)\), \(r\in\rho_X(x)\), with principal name \(r\).
In the nested case take as potential carrier the disjoint union of
the actual incidences \((x,r)\), \(r\in\rho_a(x)\), and the potential
incidences \((y,r)\), \(r\in\rho_p(y)\). The first summand is the
actual carrier. Both have name \(r\), and their potential images
are \(\alpha_Xx\) and \(y\), respectively. The identity is a cover
tracker. Given a map \(P\to B\) tracked by \(t\), choose, for each
\(p\), a preimage through \(e\) named by \(c(t\nu_P(p))\).
For a named inclusion first choose actual preimages at actual
points, retain their potential images there, and then choose lifts
on the remaining potential carrier. The composite \(r\mapsto c(tr)\)
tracks the lift. Monotonicity handles every smaller name in the
ordered case. These are external choices.

For \(Z\to Y\to X\), the relative section object over \(x\) consists
of carrier sections \(s:Y_x\to Z\) admitting a tracker \(e\in A\)
on all names of their inputs. Give \(s\) the downward closure of
the names \(\langle r,e\rangle\) with \(r\in\rho_X(x)\) and \(e\)
as above. In the nested category a potential section is such a
section of the potential fibre. An actual section over \(x\in X_a\)
consists of a potential section over \(\alpha_Xx\) and an actual
section over \(x\), commuting with the structure arrows, with a
common tracker \(e\in A_\#\). Its image forgets the actual section.
At both levels names pair a base name with an appropriate tracker.
Evaluation applies the second component to the argument. For a
section parameterized by \(T\to X\) with base tracker \(h\) and
evaluation tracker \(d\), the transpose has tracker
\(r\mapsto\langle hr,\sigma(d,r)\rangle\). At actual names both
components lie in the sub-PCA. At an arbitrary parameter name
\(\sigma(d,r)\) can range over \(A\). The pointwise carrier description
gives the inverse adjunction, and all function sets involved fit below
a common larger inaccessible.

For the ordinary identifications, give a set \(I\) the codiscrete
assembly \(\nabla I\), with all possible names at every point, using
\(I\xrightarrow{1}I\) in the nested case. Subobjects of \(\nabla I\)
are exactly predicates consisting of name sets \(U_i\) in the
relative case, or pairs \((U_{i,p},U_{i,a})\) with
\(U_{i,a}\subseteq A_\#\cap U_{i,p}\) in the nested case, modulo
uniform mutual entailment. The carrier of
each component is its nonempty support. To see that every mono has
this form, its component functions are injective: two preimages can
be tested from a singleton carrying the pair of their names, at the
level in question. For a potential comparison in the nested case the
test has empty actual carrier. For an actual comparison its singleton
is present at both levels with the same paired actual name. The two
projections track the two parallel maps in either test. One such
subobject is contained in another precisely when a global tracker
witnesses the corresponding entailment uniformly in \(i\in I\).

Every partitioned object embeds in
the codiscrete object on its (potential) carrier. Consequently every
object of the exact completion is the quotient of a partial equivalence
relation on a codiscrete carrier: cover it by a partitioned object
and extend the kernel pair on that subobject by the empty predicate
off its support. Conversely, restricting any such relation to its
support and taking its effective quotient gives an object of the
completion. Pulling a graph back along the quotient covers gives a
relation on the two supports. It respects both partial equalities,
is total on the source support, and is single-valued up to target
equality. Conversely, such a relation descends to a graph, and
equivalent predicates give the same graph. In the ordered relative
case the predicates are downsets
with entailment tracked in $A'$ \cite{VanOostenZou2016}.
In the nested case the two sets of names give the potential and actual
predicates of Maschio and Streicher \cite{MaschioStreicher2015}. With these predicates
the objects and arrows just described are precisely those of the
tripos-to-topos construction
\cite{HylandJohnstonePitts1980}.\qed
\end{proof}

\subsection{Fibre bounds and bounded Kan presentations}

Write \(\E=\Ex_\infty\C\) for the completion of either source.
The projective covers and section data of
Lemma~\ref{lem:src-projectives} give the finite descent and section
reconstruction of Lemma~\ref{lem:coherent-sections}. In particular,
the ambient regular-sheaf category computes dependent products in
\(\E\). At a selected inaccessible \(\kappa>|A|\), put \(\mathcal S_{\kappa,-1}=\{\text{all
monos of }\E\}\). For \(n\ge0\), an \(n\)-truncated \(Y\to X\) belongs
to \(\mathcal S_{\kappa,n}\) if its diagonal belongs to
\(\mathcal S_{\kappa,n-1}\) and every \(P\to X\), \(P\in\mathcal P\),
admits a cover \(D\cover Y\times_XP\) with \(D\in\mathcal P\) and
all carrier fibres of \(D\to P\) of cardinality \(<\kappa\).
In the two-level case we count potential fibres, which already
bound the actual fibres of named inclusions.

\begin{lemma}[Closure of the small families]\label{lem:src-closure}
These classes are pullback stable, cover local, increasing in both
indices, and exhaustive. Each level contains all monos and is closed
under composition, relative diagonals, and finite limits in slices.
An effective groupoid of objects small over a base has a small
quotient at the same level if the quotient has the indicated relative
truncation bound.
\end{lemma}
\begin{proof}
Original projectives remain projective in the completion: cover the
source of a cover over \(P\) by an original object and split the
resulting original cover. Thus every projective test lifts through
a cover of its base. Testing at such lifts gives locality, and
precomposing tests gives pullback stability, in each case by
induction on the diagonal. A subobject of an original \(P\) is
original: cover it by an original object and take the image of the
composite into \(P\). Its partitioned incidence cover above has at
most \(2|A|\) points per potential point (at most \(2\max(1,|A|)\)
in all cases). The mono therefore lies at level zero. Its diagonal
is an isomorphism. Induction gives inclusion of consecutive levels,
and every atlas at one cardinal bound is admitted at each larger bound.
Conversely an \(m\)-truncated map already at level \(n\ge m\) lies at
level \(m\), by using the same atlas and lowering its diagonal's
level inductively.

For a composite, first cover \(Y_P\) by a small partitioned atlas
\(D\), and then cover \(Z\times_Y D\) by a small partitioned atlas
over \(D\). Their composite covers \(Z_P\), with small fibres by
regularity of \(\kappa\). The diagonal of \(Z/X\) factors through
\(Z\times_YZ\), first by the diagonal of \(Z/Y\), then by a
pullback of the diagonal of \(Y/X\). Induction proves composition
closure. Finite limits in a slice follow from products and diagonals.
In particular every map between small objects of a slice is small:
its graph is a pullback of the target diagonal and the subsequent
projection is a pullback of the source's structure map.

For the groupoid \(R_\bullet\) with quotient \(Q\) over \(X\), an
atlas of \(R_0\) also covers \(Q\). The diagonal of \(Q/X\) pulls
back along \(R_0\times_XR_0\cover Q\times_XQ\) to
\(R_1\to R_0\times_XR_0\). This map is small by the graph argument
and has the truncation bound of the diagonal. Lower its level and
descend. For degree \(-1\) the conclusion is the truncation
assumption itself. Finally, for any bounded \(Y\to X\), choose
partitioned covers \(P_0\cover X\) and \(D_0\cover Y_{P_0}\).
Choose a selected cardinal above their total sizes. Every projective
test lifts to \(P_0\), and the pullback of \(D_0\) along that lift
is a partitioned atlas with the same fibre bound. Increase the
bound finitely many times to include the successive diagonals.
This proves exhaustiveness.\qed
\end{proof}

A \emph{Kan presentation} of \(Y\to P\) here consists of a simplicial
diagram of partitioned assemblies over \(P\), chosen tracked sections
of its horn maps, and an equivalence over \(P\) from its realization
in regular sheaves to \(Y\). For \(m\geq1\) and \(0\leq i\leq m\),
the horn \(\Lambda_i^m\subseteq\Delta^m\) is the union of the
codimension-one faces other than the \(i\)-th. A horn section assigns
to every compatible tuple of these faces an \(m\)-simplex restricting
to that tuple. In an \(r\)-coskeletal diagram, for \(m>r\), every
compatible assignment on the \(r\)-skeleton of \(\Delta^m\) extends to
a unique \(m\)-simplex. Thus in an \((n+1)\)-coskeletal presentation
a chosen filler in dimension \(n+2\) is specified by its missing
\((n+1)\)-face together with the given horn.

\begin{lemma}[Finite Kan presentations]\label{lem:src-kan}
For \(P\in\mathcal P\) and \(n\ge0\), every
\(Y\to P\) in \(\mathcal S_{\kappa,n}\) has a Kan presentation by
partitioned objects with carrier fibres \(<\kappa\). Levels through
\(n+1\), their face and degeneracy operators and chosen horn sections
through dimension \(n+2\) determine an \((n+1)\)-coskeletal
presentation realizing \(Y/P\). Conversely such a presentation
realizes to an object of \(\mathcal S_{\kappa,n}\).
\end{lemma}
\begin{proof}
Build a split simplicial hypercover \(D_\bullet\to Y\). At degree
\(k\) choose a partitioned cover \(N_k\) of the augmented matching
object with fibres of size \(<\kappa\) over \(P\), and put
\[
 D_k=\coprod_{\theta:[k]\twoheadrightarrow[j]}N_j .
\]
The identity surjection supplies the new summand and its given
boundary. Degeneracies act on summands by precomposition of
surjections. On a proper-surjection summand, faces are determined
by factoring the composite with a coface into a surjection followed
by an injection and using the previously specified faces. Uniqueness
of these ordinal factorizations and the matching condition give the
simplicial identities. The matching object is the
finite limit of the boundary in the slice over \(Y\), hence small
over \(P\) by Lemma~\ref{lem:src-closure}. There are finitely many
degenerate summands in each degree, indexed by the proper
surjections out of that ordinal. Tagged coproducts of partitioned
objects are coproducts in the completion as well. Indeed, mapping
from the sum and taking the product of the mapping spaces from
the summands agree on original targets. Both preserve finite limits
and covers by projectivity, hence effective groupoid quotients by
effectivity of the image of each quotient's \v{C}ech nerve. They
therefore agree on the full exact completion. This justifies the
split matching construction there. Let \(\mu=\operatorname{card}(|P|)\)
in the relative case and \(\mu=\operatorname{card}(P_p)\) in the nested
case. Every degree has total carrier size at most \(\max(1,\mu)\kappa\)
throughout the external countable recursion.

For a horn of dimension \(m\), attach first the missing
\((m-1)\)-face along its boundary and then the \(m\)-simplex.
Passing to limits in the slice over \(Y\) gives two consecutive
covers, by the matching conditions, including \(D_0\cover Y\)
when \(m=1\). The finite face category of the horn is contractible,
so its limit agrees with the underlying horn object in the slice
over \(P\). This horn object is a finite limit of partitioned
objects over \(P\). Projectivity splits the composite cover, and
full faithfulness gives a tracked section. Choose such sections.
For every test assembly \(T\to P\), these sections supply horn fillers
in the simplicial set \([m]\mapsto\Hom_{\C/P}(T,D_m)\), so it is Kan.

For a Kan complex the map to its \((n+1)\)-coskeleton preserves
homotopy groups through \(n\): representatives and homotopies in
those degrees only use simplices through degree \(n+1\).
Every boundary above degree \(n+1\) extends in the coskeleton, so
all higher groups vanish. The horns up to degree \(n+1\) retain
their chosen fillers. A horn in degree \(n+2\) needs exactly its
missing \((n+1)\)-face, supplied by an original filler. In higher
degrees the horn and simplex have identical \((n+1)\)-skeleta.
These facts apply objectwise, and left exact sheafification
preserves the truncation calculation. A hypercover induces an
equivalence after every finite truncation (finite hypercover
descent, as in Lemma~\ref{lem:coherent-sections}), so its
coskeletal realization here is \(Y\).

For the converse we prove a stronger assertion. Over a partitioned
base \(P\), let \(K_\bullet\) be any simplicial diagram of partitioned
objects with fibres of size \(<\kappa\) at each degree and chosen
tracked sections of all horn maps. If its ambient realization
\(Q\to P\) is \(r\)-truncated, it belongs to
\(\mathcal S_{\kappa,r}\). We induct on \(r\) for all such diagrams
and bases. For \(r=-1\) the realization is the image of the vertex assembly and
hence a mono in the completion.

For \(r\geq0\) form the simplicial
path object with endpoints in the vertices,
\[
  L_\bullet=K_\bullet^{\Delta^1}
     \times_{K_\bullet^{\partial\Delta^1}}
              (K_0\times_P K_0).
\]
Here a finite cotensor is obtained in degree \(m\) by the finite
limit on the nondegenerate simplices of
\(\Delta^m\times\Delta^1\), or of its boundary-endpoint analogue.
The relative prism horn extends by successive pushouts of horn inclusions
(the standard prism triangulation), hence by finitely many chosen
fillers of \(K\). The objectwise Kan path-space formula followed by
left exact sheafification identifies \(|L_\bullet|\) with
\(K_0\times_Q K_0\). Here the inclusion extended to fill a prism is
\[
 (\Lambda_i^m\times\Delta^1)\cup
       (\Delta^m\times\partial\Delta^1)
       \ \hookrightarrow\ \Delta^m\times\Delta^1 .
\]
The two end faces prescribe the endpoint pair, so the resulting
filler lies over that same pair. The relative truncation of the
realization over \(K_0\times_PK_0\)
is one lower. This endpoint base is partitioned. In each degree
the path object is a finite limit of partitioned objects with small
fibres, and the prism fillers just constructed give its tracked horn
sections over the endpoint base. The strengthened induction hypothesis
therefore places this kernel pair at level \(r-1\). Its
iterated fibre products give the \v{C}ech nerve of
\(K_0\cover Q\). Its effective quotient in \(\E\) realizes \(Q\).
For partitioned levels with the given fibre bounds, the finite
limits in this argument are partitioned with small fibres. The
smallness assertion therefore descends for the diagonal from the
cover \(K_0\times_PK_0\cover Q\times_PQ\); the vertex cover supplies
the atlas condition after every projective base change. This proves
the converse and the bounded realization used below. In this way the
realization and projective-resolution arguments for ordinary Kleene
assemblies \cite{Maruyama2026} extend to relative and nested assemblies.
The relative horn sections have trackers in the filter, and each nested
horn section has a tracker acting at both levels, by the projective
lifting property of Lemma~\ref{lem:src-projectives}.\qed
\end{proof}

\subsection{Verifying admissibility for relative and nested assemblies}

\begin{theorem}[Verification of the assembly presentations]
\label{thm:source-verification}
Suppose the carrier of \(A\) lies below some selected inaccessible.
The relative assembly presentation of a filtered ordered PCA and the
two-level assembly presentation of a discrete sub-PCA as specified
above satisfy the admissibility hypotheses of
Proposition~\ref{prop:input-completion}. For each \(\kappa>|A|\) in
the hierarchy and \(n\ge0\), there is a weak classifier for
\(\mathcal S_{\kappa,n}\) on a partitioned base of carrier size at
most \(\kappa\). The resulting univalent hierarchy contains all monos
and satisfies
\[
 f\in\mathcal S_{\kappa,N},\quad g\in\mathcal S_{\kappa,r},\quad
 -1\le r\le N,\quad N\ge0
 \quad\Longrightarrow\quad
                 \Pi_f g\in\mathcal S_{\kappa,r}
\]
for composable \(g,f\). In particular the construction applies to
relative numerical and function realizability, to their absolute
instances \(\mathcal K_1,\mathcal K_2\), and to nested numerical
realizability with \(A_\#=A=\mathcal K_1\).
\end{theorem}
\begin{proof}
\emph{Codes for families.} Fix \(n\) and \(\kappa\). A potential code \(v\) consists of ordinal
carriers \(B_0,\ldots,B_{n+1}<\kappa\), principal-name functions
\(\nu_k:B_k\to A\), face and degeneracy functions, horn sections
through \(n+1\), and missing-face operations for the horns of
dimension \(n+2\). The operators satisfy the truncated simplicial
identities and the horn section equations, including the boundary
conditions for the missing faces. The chosen fillers and their
trackers are part of a code. In particular the same underlying
diagram occurs with each admissible choice. Equip these data with a tuple
\(q_v\) of trackers in \(A\) for all those operations on the
specified names. Horn names here are finite tuples of the names
of their faces, using the iterated finite-limit encoding. In the
ordered case each tracker acts on all names in the indicated
principal downsets. Admissibility of these data is
an external set-theoretic condition. A relative code base \(V\)
has all such codes as carrier and gives \(v\) the principal name
\(q_v\).

In the nested case, an enhancement \(H\) of \(v\) is a choice of
subsets \(B_{k,a}\subseteq B_k\) whose names lie in \(A_\#\).
For an enhancement, the operators and the chosen horn fillers preserve
the subdiagram, and the components of \(q_v\) belong to \(A_\#\). Put
\[
 V_a=\{(v,H):H\text{ enhances }v\},\qquad
 V_p=V_a\amalg\{v:v\text{ is a potential code}\}.
\]
The structure map is the inclusion of the actual carrier as the first
summand of the potential carrier. The potential copy and
each enhanced copy have the same name \(q_v\), so a
single name can represent several enhancements.
Over an enhanced copy use \(B_k\) as potential carrier and
\(B_{k,a}\) as actual carrier of the universal degree \(k\). Over
a potential-only copy take \(B_k\) on the potential side.

In either case a universal simplex \((v,b)\), with the enhancement
retained in its base when present, has name
\(\langle q_v,\nu_k(b)\rangle\). Uniform global trackers perform
the projections, extract the appropriate operator from \(q_v\),
apply it to the names of the indicated simplex or horn, and pair
the result again with \(q_v\). The tracked section equations
are equations of carrier functions. In the ordered case the
projection inequalities yield outputs in the required downsets.
In the nested case the common trackers preserve the enhancements.
Taking the coskeleton and applying Lemma~\ref{lem:src-kan} gives a
family \(p_{\kappa,n}:E_{\kappa,n}\to V\) in
\(\mathcal S_{\kappa,n}\). Each fixed finite tuple of ordinals
below \(\kappa\) supports fewer than \(\kappa\) possible name
assignments, functions, subsets and tracker tuples, by the strong
limit and regularity conditions and \(|A|<\kappa\). There are at
most \(\kappa\) such ordinal tuples. Thus both \(V\) and every
universal degree have total size at most \(\kappa\) and belong
to the next selected size stratum.

To classify \(Y\to X\), first take a projective cover \(P\cover X\)
and the presentation of Lemma~\ref{lem:src-kan} over \(P\).
Externally identify each (potential) fibre carrier with an ordinal,
transporting its principal names, functions, and fillers. At an
actual parameter take the corresponding enhanced copy, since each
partitioned actual fibre is a subset of the potential one.
The original structure maps and chosen sections have a finite tuple
of global trackers, common to all the fibres. When an operator over
\(P\) needs a parameter name, a simplex or a vertex of the input
horn supplies one through its tracked projection to \(P\); horns
used here are nonempty. Thus a single tuple \(q\) tracks the
relabelled fibre operations. The map \(P\to V\) sending each
parameter to its code is tracked by the constant \(q\), at both
levels where applicable. If \(\pi_k\) tracks projection of the
original level to \(P\), the comparison to the pulled-back universal
level is tracked on input names \(r\) by
\[
  r\longmapsto\langle\pi_k r,\langle q,r\rangle\rangle .
\]
Two successive second projections track the inverse. The two maps
are inverse on carriers, preserve all the named operations, and hence recover
the family on realization. Thus the local class \(\mathcal M(p_{\kappa,n})\)
generated by \(p_{\kappa,n}\) is exactly \(\mathcal S_{\kappa,n}\).

\emph{Codes for sections.} Work first over a
projective test \(P\to X\); write \(Z\to Y\to P\) for the
base-changed maps. Put \(m=r+1\), allowing \(m=0\). Resolve \(Y\)
by a partial semisimplicial projective hypercover \(U_{\le m}\) over
\(P\), and choose a relative resolution with matching covers
\[
 V_k\cover D_k\times_{M_kD}M_kV,\qquad
 D_k=Z\times_YU_k,\qquad 0\le k\le m .
\]
Here \(M_k\) is the finite boundary limit over \(P\), with
\(M_0D=M_0V=P\). As all matching objects are finite homotopy limits of
objects small over \(P\) at level \(N\), projective atlases make
the \(U_k,V_k\) partitioned with fibres \(<\kappa\).

Form a partitioned object \(W\to P\). For a base point \(b\), write
\(U_k(b),V_k(b),W(b)\) for the carrier fibres over \(b\), taking
potential carriers in the nested case. The points of \(W(b)\) consist
of strict semisimplicial sections \(s_k:U_k(b)\to V_k(b)\) over \(U_k(b)\), with a tuple
\((e_0,\ldots,e_m)\in A^{m+1}\) tracking the component maps on
their input names. A point over an actual \(b\) is actual when
all its trackers lie in \(A_\#\) and its functions carry the actual
subdiagram into the actual subdiagram. Its actual functions are
then fixed by restriction. Use the principal name
\(\langle\nu_P(b),e_0,\ldots,e_m\rangle\). For each \(b\),
\[
  |W(b)|\ \le\ |A|^{m+1}
       \prod_{k=0}^{m}|V_k(b)|^{\,|U_k(b)|}<\kappa .
\]
The strong limit property of \(\kappa\) bounds the cardinal powers in this estimate.

Application of \(e_k\) gives a tracked strict family of evaluations
on \(U_k\) with values in \(V_k\). Compose into \(D_k\). The finite
descent equivalence for sections of an \(r\)-truncated map
(Lemma~\ref{lem:coherent-sections}) turns these compatible evaluations
into a section over \(W\times_PY\). Conversely, take any section
parameterized by a completed \(T\to P\), and cover \(T\) by
a partitioned \(T'\). The objects \(T'\times_P U_k\) are
projective. Inductively lift along the relative matching covers,
using the earlier lifts on faces and the given section into \(D_k\).
This gives strict component functions into \(V_k\); equations
between maps of original assemblies are literal because their
mapping spaces are discrete. The relative matching maps retain the
coherences into \(D_\bullet\). If \(d_k\) are trackers of the
lifts and \(h\) tracks \(T'\to P\), the representing map
\(T'\to W\) is tracked by the abstraction
\[
    t\longmapsto
          \langle ht,\sigma(d_0,t),\ldots,\sigma(d_m,t)\rangle .
\]
More explicitly, at a point \(t'\) of \(T'\) with image \(b\) in \(P\), use its principal
name \(t=\nu_{T'}(t')\) to choose the displayed tuple. For another
name \(v\le t\), we have \(hv\le ht\le\nu_P(b)\) and
\(\sigma(d_k,v)\le\sigma(d_k,t)\) for each \(k\). The resulting tuple
therefore names the same point of \(W\). At actual names the specialized
trackers belong to \(A_\#\). The descent equivalence identifies the
represented section with the given one, including its comparison
over \(Y\). Applying this construction to the universal section over
\((\Pi_f g)_P\) shows that \(W\to(\Pi_f g)_P\) is a cover.

Induct now on \(r\). The dependent product of a mono is monic.
For \(r\ge0\) let \(S=\Pi_f g\to X\), which is \(r\)-truncated
by preservation of finite limits by the right adjoint on slices.
The small covers \(W\) give its atlas condition. Over
\(B=S\times_X S\), evaluation gives two sections of \(g\) over
\(Y_B\). Their identity family \(I\to Y_B\) is a pullback of
\(\Delta_g\), hence lies at level \(r-1\). The adjunction,
including its action on mapping spaces, identifies
\(\Delta_{S/X}\) over \(B\) with \(\Pi_{Y_B/B}I\). Apply the
induction hypothesis to the pullback \(Y_B\to B\), which belongs to level \(N\).
This proves the diagonal condition.

\emph{Codes for propositions.} The ordinary subobject classifier
\(\Omega\) classifies monos on zero-type parameters. As in the
ordinary-core argument of Proposition~\ref{prop:input-completion},
characteristic maps descend along a cover of any parameter by a
zero-type. Agreement on its kernel pair can be tested after a further
cover by a zero-type, and all higher agreements are unique because
\(\Omega\) is a set. Thus \(\Omega\) classifies all monos in
\(\E\). Pull back
\(\one\into\Omega\) along a partitioned cover of \(\Omega\) to
obtain the required weak family on an original base. The preceding
construction, Lemmas~\ref{lem:src-projectives} and
\ref{lem:src-closure}, and the product bound verify all data of
Proposition~\ref{prop:input-completion}. Its univalent reconstruction
turns the tracked families on these partitioned bases into the closed
hierarchy of universes in the completion.\qed
\end{proof}
\end{document}